\documentclass[%
 amsmath,amssymb,
 aps, physrev,
]{revtex4-2}

\usepackage{makecell}
\usepackage{enumitem}
\usepackage{graphicx}
\usepackage{dcolumn}
\usepackage{bm}

\begin{document}

\preprint{APS/123-QED}

\title{\textbf{Revisiting the mixing length scaling in pressure-gradient turbulent boundary layers via a symmetry approach} 
}%

\author{Wei-Tao Bi}
 \email{Contact author: weitaobi@pku.edu.cn}
\affiliation{Department of Mechanics, School of Mechanics and Engineering Science}%
\affiliation{State Key Laboratory for Turbulence and Complex Systems, Peking University, Beijing 100871, China}%

\date{\today}

\begin{abstract}
A century after Prandtl’s mixing length hypothesis, full-profile scaling of the mixing length in pressure-gradient turbulent boundary layers (PG TBLs) remains debated, especially for adverse pressure gradients (APGs). This work presents a symmetry-based analytical model for the mixing length in equilibrium APG TBLs by extending the structural ensemble dynamics theory and coupling a two-layer total shear stress model. The framework unifies the inner layer, logarithmic region, half-power-law transition zone, and wake region with an invariant K\'{a}rm\'{a}n constant ($\kappa=0.45$). A critical Clauser parameter $\beta_c\approx6.2$ is identified, above which the logarithmic layer shrinks and transitions to the half-power-law scaling. The wake-region mixing-length parameter $\lambda$ is analytically formulated, and the viscous sublayer and buffer layer thicknesses are determined self-consistently without ad hoc fitting. With only one finite-Reynolds-number correction parameter determined by the maximum shear stress, the model accurately predicts full profiles of mixing length, mean velocity, and Reynolds shear stress, validated against extensive numerical and experimental data ($Re_\theta=1250\sim50980$, $\beta=0\sim39$). This work provides a unified, physically consistent framework for mixing-length scaling in PG TBLs and clarifies the transition mechanism from the log law to the half-power law under strong APG. It also enables assessment of the invariance of the logarithmic law and K\'{a}rm\'{a}n constant using the full-profile scaling law of the mixing length.

\end{abstract}

\maketitle


\section{Introduction} \label{sec:Intro}
\noindent The mixing length ($\ell_m$) was introduced by Prandtl a century ago \cite{prandtl1925ausgebildete}, marking a major milestone in turbulence research \cite{Voyage}.  Prandtl proposed $\ell_m$ as the turbulent analog of the molecular mean free path in the kinetic theory of gases. Although this analogy is not exact \cite{Bradshaw1974,Heisel2020}, $\ell_m$ remains a key length scale characterizing momentum transport by turbulent eddies. In pipe flows and zero-pressure-gradient (ZPG) turbulent boundary layers (TBLs), Prandtl \cite{prandtl1932} postulated $\ell_m\mbox{=}\kappa y$, where $\kappa$ is the K\'{a}rm\'{a}n constant and $y$ denotes the wall-normal distance. This linear scaling, known as Prandtl's mixing-length hypothesis, leads to the celebrated logarithmic law of the wall for the mean velocity in the overlap region, representing one of the most crucial achievements in turbulence theory and modeling \cite{Voyage}. 

Extending Prandtl's mixing-length hypothesis to predict the full mean-velocity profile in canonical and complex wall-bounded turbulent flows has been a long-standing challenge. For canonical flows (pipes, channels, ZPG TBLs, etc.), van Driest \cite{van1956turbulent} proposed a damping function to account for the deviation of $\ell_m$ from Prandtl's linear scaling in the near-wall region. In the wake region (except near the center of pipes and channels), $\ell_m$ is often taken as a constant fraction of the boundary-layer thickness \cite{Schlichting}.

Compared to ZPG TBLs, pressure gradient (PG) TBLs are substantially more complex, and ongoing debates persist regarding their mixing length and mean velocity profiles \cite{JoTreview2024}. To avoid additional complications from flow history effects, we focus here on equilibrium PG TBLs, where the mean flow is determined solely by the local Reynolds number and PG \cite{bradshaw1967turbulence}. However, even in such simplified flow configurations, ambiguities and disagreements remain \cite{devenport2022equilibrium}.

A central question is whether the log law for the mean velocity and the linear scaling of the mixing length remain valid in the overlap region of PG TBLs \cite{resilience}. As noted by Sk{\aa}re and Krogstad \cite{Skare1994exp}, it is now necessary to distinguish the K\'{a}rm\'{a}n constant in the log law of mean velocity (denoted by $\kappa_u$) from that in Prandtl's mixing-length hypothesis (denoted by $\kappa_\ell$), since $\kappa_\ell$ is sensitive to PG. In the overlap region of adverse pressure gradient (APG) TBLs, Townsend \cite{townsend1961equilibrium} retained Prandtl's mixing-length hypothesis and assumed a linear profile for the total shear stress (TSS) to derive the so-called generalized law of the wall \cite{Schlichting} (or extended wall law \cite{Knopp2021}). This law recovers the log law in the ZPG case, but indicates its breakdown under APG conditions. In the zero-skin-friction limit, it reduces to the well-established half-power (square-root) law first derived by Stratford \cite{stratford1959prediction}. Therefore, the generalized law of the wall notably captures a smooth transition from the log law to the half-power law as the PG parameter (denoted by $p^+$) increases from zero to infinity. However, the validation of the law is weakened by two main limitations: (1) uncertainties in determining the location of the overlap region under APG conditions, and (2) empirical evidence suggesting that $\kappa_\ell$, $B$ (the intercept), and $\alpha$ (the coefficient in the linear TSS) vary with $p^+$ and Reynolds number, making the application of the law essentially a data-fitting procedure. 

In contrast, the logarithmic velocity profile is widely regarded as significantly more robust compared to the linear mixing-length profile in the overlap region \cite{ColesHirst1968,Galbraith1977,resilience}. If the log law is preserved for PG TBLs as in Coles' law of the wake \cite{coles1956law}, the mixing length in the overlap region must be scaled as $\ell_m\mbox{=}\sqrt{\tau^+}\kappa_u y$ ($\tau^+$ denotes TSS in wall units) \cite{Reeves1974,Galbraith1977}, indicating that Prandtl's mixing-length hypothesis is invalid under nonzero PG conditions. However, this revised scaling does not hold in general, as it can lead to nonphysical, imaginary mixing-length values in certain favorable PG (FPG) TBLs \cite{JoTreview2024}. In addition, the inevitable breakdown of the log law as $p^+\to\infty$ indicates that the $\sqrt{\tau^+}$-scaling cannot always be obeyed.  

To improve near-wall predictions under PG conditions, numerous studies in the 1970s (as summarized by Granville \cite{Granville1989}) modified van Driest's damping model by empirically incorporating PG effects into the damping parameter $A$. Broadly, these efforts followed one of two approaches \cite{Granville1989}: either directly applying Prandtl's mixing-length hypothesis or scaling it with $\sqrt{\tau^+}$ to preserve the log law for the mean velocity. In contrast, the $\ell_m$ profile in the wake region remains largely independent of $y$ under APG conditions, with the ratio $\ell_m/\delta$ (called Prandtl's mixing-length parameter \cite{Thomas1989}, where $\delta$ is the boundary-layer thickness) influenced by the APG \cite{Cantwell2022}. Moreover, this constant-$\ell_m$ regime expands and progressively encroaches upon the overlap region as the APG intensifies \cite{devenport2022equilibrium}. These behaviors still require appropriate quantification based on solid physical foundations. 

After a long period of limited activity \cite{Bernard2003,Buschmann2005,Pirozzoli2014} mainly due to the rise of computational fluid dynamics, research on mixing-length modeling has advanced significantly in the past decade. Using a Lie-group method, She et al. \cite{she2010new,she2017quantifying} recently established the so-called structural ensemble dynamics (SED) theory for wall turbulence. The SED derived a smooth analytical expression for the entire profile of $\ell_m$ in canonical wall-bounded turbulence. For ZPG TBLs, the expression is
 \begin{equation}
  \ell_m^+=\frac{\kappa {y_s^+}^2}{y_b^+}\left(\frac{y^+}{y_s^+}\right)^{3/2}\left[1+\left(\frac{y^+}{y_s^+}\right)^4\right]^{1/8}\left[1+\left(\frac{y^+}{y_b^+}\right)^4\right]^{-1/4}\frac{1-r^4}{4(1-r)},
  \label{eq:SEDmodel}
\end{equation}
where the superscript plus denotes normalization by wall units, $y_s^+\approx9.7$ and $y_b^+\approx41$ represent the thicknesses of the viscous sublayer and buffer layer, respectively, and $r\mbox{=}1-y/\delta$. Equation (\ref{eq:SEDmodel}) captures a four-layer structure for $\ell_m$: in the wake region ($y_b^+{\rm \ll} y^+{\rm \le} \delta^+$), $\ell_m\approx\frac{\kappa}{4} \delta$, a constant; in the log-law region ($y_b^+{\rm \ll} y^+{\rm \ll}\delta^+$), $\ell_m\mbox{=}\kappa y$, following Prandtl's hypothesis; in the buffer layer ($y_s^+{\rm \ll} y^+{\rm \ll} y_b^+$), $\ell_m{\rm \propto} y^2$, consistent with van Driest's damping scaling; and in the viscous sublayer ($y^+{\rm \ll} y_s^+$), $\ell_m{\rm \propto} y^{3/2}$, satisfying the near-wall constraint. In equation (\ref{eq:SEDmodel}), $\kappa$ is a global parameter, enabling the determination of the K\'{a}rm\'{a}n constant using the complete profile of the mixing length or mean velocity, thereby avoiding uncertainties associated with locating the log layer. She et al. reported $\kappa=0.45$, which is independent of flow geometry and Reynolds number (if above moderate values) \cite{she2017quantifying}. This finding contradicts conventional observations \cite{MarusicPoF2010} and thus requires further elucidation, although it aligns with the theoretical preferences for simplicity and invariance. Additionally, extending equation (\ref{eq:SEDmodel}) to PG TBLs remains a significant challenge due to the interplay of the Reynolds number and PG effects.

A crucial point in the construction of equation (\ref{eq:SEDmodel}) is the formulation of scaling crossover between adjacent layers through a Lie-group-derived ansatz: $[1+(y/y_0)^\zeta]^{\Delta/\zeta}$, where $y_0$ is the crossover scale, $\Delta$ is the scaling-exponent increment after the crossover, and $\zeta$ is a positive integer characterizing the crossover steepness ($\zeta=4$ appears to be a preferred value \cite{she2017quantifying,Bi2025}). This ansatz has been widely used to describe crossover scaling phenomena in statistical physics and complex systems \cite{Gluzman1998}. Examples in turbulence include the K\'{a}rm\'{a}n spectrum \cite{Karman1948} and the Klebanoff intermittency function \cite{Klebanoff1954}. The mathematical foundation of the ansatz has been investigated by various authors, including Gluzman and Yukalov, who used an algebraic self-similar renormalization method \cite{Gluzman1998}, She et al., who used the Lie group method \cite{she2017quantifying}, and Li et al., who used a Bernoulli differential equation \cite{LiRong2025}. She et al. \cite{she2017quantifying} postulated that the ansatz is a universal functional form for describing dilation-symmetry-breaking from one scaling law to another. 

Cantwell et al. \cite{Cantwell2019,Cantwell2022} proposed an alternative mixing-length formulation for canonical wall turbulence and PG TBLs. Using the crossover scaling ansatz, their model blends a modified van Driest expression for the inner layer with a constant mixing length for the wake region, expressed as
 \begin{equation}
  \ell_m^+=\kappa y^+\left[1-e^{-(y^+/A)^m}\right]\left[1+\left(\frac{y^+}{b\delta^+}\right)^n\right]^{-1/n},
  \label{eq:Cantwellmodel}
\end{equation}
where $\kappa$, $A$, $m$, $b$, and $n$ are empirical parameters optimized to match velocity profiles across different flow geometries, Reynolds numbers, and PG conditions. The model yields a unified and highly accurate description of the mean velocity profile from the wall to the outer edge in both channel flows and boundary layers \cite{Cantwell2022}. However, the framework relies on a stress balance that assumes a linear TSS distribution, which is valid in pipes and channels but generally not in TBLs, particularly under strong APGs. Although parameter flexibility compensates for this discrepancy, the model's apparent universality remains partly empirical. The model's limitations have inspired follow-up research to improve it \cite{ShuXu2025,MaChen2026}. 

Notably, both equation (\ref{eq:SEDmodel}) and equation (\ref{eq:Cantwellmodel}) adhere to Prandtl's mixing-length hypothesis in the overlap region, which is against the $\sqrt{\tau^+}$-scaling. The latter, however, has gained more support from recent direct numerical simulations (DNS) on channel flows \cite{Pirozzoli2014,XuYJ2025}.   

In summary, after a century of research, the mixing length remains a key quantity in understanding and modeling the mean-field properties of wall turbulence. Although significant advances have been made in canonical flows and PG TBLs during the past decade, unresolved questions persist regarding the scaling of the mixing length, the invariance and precise value of the K\'{a}rm\'{a}n constant, and the resilience and breakdown of the log law under APG conditions. Notably, an analytical description of the complete profile of mixing length in equilibrium PG TBLs remains an open challenge, let alone non-equilibrium flows.

This study aims to develop a full-profile formulation for mixing length in equilibrium APG TBLs, guided by the following hypotheses.
\begin{itemize}
\item The formulation reduces to equation (\ref{eq:SEDmodel}) in the ZPG limit. Using equation (\ref{eq:SEDmodel}) as the baseline model is due to its apparent elegance in capturing the multilayer behavior of TBLs. 
\item Before its breakdown, the log law remains valid in the overlap region. This hypothesis requires the $\sqrt{\tau^+}$-scaling in equation (\ref{eq:SEDmodel}). 
\item The crossover scaling ansatz effectively captures the scaling transitions (dilation-symmetry-breakings) induced by the APG.
\item The formulation is consistent with established qualitative and quantitative knowledge of equilibrium APG TBLs.
\end{itemize}
Based on these hypotheses, we construct a compact (no adjustable parameters), symmetry-based, full-profile formulation for the mixing-length distribution in equilibrium APG TBLs. By combining this formulation with a recently developed symmetry-based TSS model \cite{ZhengBi2025}, we derive accurate predictions for mean velocity and Reynolds shear stress throughout the APG range. Key new findings include a formulation of the mixing-length parameter and the identification of a critical Clauser PG parameter, above which the log-law region begins to shrink and eventually transitions to the half-power-law region. The study also provides insight for the scaling of the mixing length in canonical internal flows (pipes and channels), as these flows are subject to nonzero PG.

This paper is structured as follows. Section \ref{sec:theory} develops the theoretical framework for predicting the complete mean velocity profile in equilibrium APG TBLs, using symmetry-based formulations for the TSS and mixing-length profiles. Section \ref{sec:valid} validates the model assumptions and predictions against experimental and numerical data. Section \ref{sec:discuss} addresses the achievements and limitations of the theory, and suggests future developments. Section \ref{sec:conclude} provides concluding remarks.

\section{Theory}\label{sec:theory}
\subsection {Theoretical framework for predicting mean velocity in PG TBLs}\label{subsec:theoframe}
\noindent We focus on incompressible, attached, and statistically two-dimensional (2D) TBLs. The streamwise mean momentum equation for a 2D incompressible planar TBL is given by
 \begin{equation}
  u^+\frac{\partial u^+}{\partial x^+}+v^+\frac{\partial u^+}{\partial y^+}  = -p^++\frac{\partial^2 u^+}{\partial {y^+}^2}-\frac{\partial \left<{u'u'}\right>^+}{\partial x^+}-\frac{\partial \left<{u'v'}\right>^+}{\partial y^+},
  \label{eq:x_momentum}
\end{equation}
where the superscript plus denotes normalization by wall units, angle brackets denote Reynolds averaging, primes denote fluctuations, $x$ and $y$ represent the streamwise and wall-normal Cartesian coordinates, $u$ and $v$ represent the streamwise and wall-normal mean velocity components, $p^+\mbox{=}\nu/(\rho u_\tau^3){\partial P}/{\partial x}$, $\rho$ is the density, $P$ the mean static pressure, $\nu$ the kinematic molecular viscosity, and $u_\tau$ the friction velocity. The wall-normal integration of (\ref{eq:x_momentum}) yields
\begin{align}
  \tau^+ &\equiv\frac{\partial u^+}{\partial y^+}-\left<{u'v'}\right>^+ \nonumber\\
&=1 +\int_{0}^{y^+}{\left(u^+\frac{\partial u^+}{\partial x^+}+v^+\frac{\partial u^+}{\partial y^+}+p^+ +\frac{\partial \left<{u'u'}\right>^+}{\partial x^+}\right){\rm d}y^+},
  \label{eq:tau}
\end{align}
where $\tau^+$ is the dimensionless TSS. Within the integral on the right-hand side of (\ref{eq:tau}), the advection terms (i.e., the first two terms) and the streamwise PG ($p^+$) provide the dominant contributions to the $\tau^+$ profile. During incipient separation and in detached TBLs, the evolution of the streamwise turbulent kinetic energy plays a limited role \cite{devenport2022equilibrium}.

Modeling the Reynolds shear stress with the mixing length as $-\left<u'v'\right>^+\mbox{=}\left(\ell_m^+\partial u^+/\partial y^+\right)^2$, the mean velocity profile can be derived by further integrating (\ref{eq:tau}) to obtain
 \begin{equation}
  u^+ =  \int_{0}^{y^+} {\frac{{ - 1 + \sqrt {1+4{\tau ^{\rm{ + }}}{{ {\ell_{m}^+}}^2}} }}{{2{{ {\ell_{m}^+}}^2}}}\mathrm{d} {y^{\rm{ + }}}}.
  \label{eq:u_model}
\end{equation}
To predict the mean velocity using equation (\ref{eq:u_model}), both $\ell_{m}^+$ and $\tau^+$ require modeling, which are presented next.

\subsection {Two-layer defect scaling of total shear stress in equilibrium APG TBLs}\label{subsec:TSS_model}
\noindent Recently, Zheng et al. \cite{ZhengBi2025} proposed an explicit symmetry-based formulation for the TSS profile in equilibrium APG TBLs, expressed as
 \begin{equation}
  \tau^+=1-{y^*}^{3/2}+P_0^+\left\{\left[{1 + {\left(\frac{y_p^*}{y^*}\right)^4}} \right]^{-1/4}-{y^*}^{3/2}\right\},  
  \label{eq:tau_APG}
\end{equation}
where $y^*\mbox{=}y/\delta$, $\delta$ is the boundary-layer thickness, $P_0^+$ represents an APG-induced characteristic shear stress, and $y_p^*\mbox{=}y_p/\delta$ denotes a characteristic thickness induced by the APG. Equation (\ref{eq:tau_APG}) describes a two-layer structure for $\tau^+$ in equilibrium APG TBLs. In the near-wall region ($y{\rm \ll} y_p$), equation (\ref{eq:tau_APG}) reduces to $\tau^+\approx1+(P_0^+/y_p^+)y^+$, corresponding to a linear profile. Near the boundary-layer edge ($y{\rm \gg} y_p$), it approaches $\tau^+\mbox{=}(1+P_0^+)\left(1-{y^*}^{3/2}\right)$, which extends the $3/2$ defect law, $\tau^+\mbox{=}1-{y^*}^{3/2}$, proposed by Chen et al. for ZPG TBLs \cite{chen2016analytic}, to the equilibrium APG case. Therefore, $P_0^+$ quantifies the shear stress overshoot induced by APG. The near-wall and near-edge expressions are smoothly connected by the crossover scaling ansatz $\left[{1 + {\left(y_p^*/y^*\right)^4}} \right]^{-1/4}$, indicating that the APG induces a characteristic thickness $y_p^*$ that separates the boundary layer into near-wall and near-edge regions in terms of the shear stress distribution.

Drawing on a large body of numerical simulations and experimental data, Zheng et al. \cite{ZhengBi2025} determined that $P_0^+\mbox{=}\frac{3}{2}\beta$ and $y_p^*\mbox{=}0.491\left[3\beta/(2+3\beta)\right]^{2/3}$. Here, $\beta$ denotes the Clauser PG parameter \cite{clauser1954turbulent}, defined as $\beta\mbox{=}\frac{\delta_1}{\rho u_\tau^2}\frac{\partial P}{\partial x}$, where $\delta_1$ is the displacement thickness. The maximum shear stress and its wall-normal location are given by $\tau_{\rm max}^+\mbox{=}1+\frac{3}{4}\beta$ (first proposed by Sk{\aa}re and Krogstad \cite{Skare1994exp}) and $y_{\rm max}^*\mbox{=}0.456\left[3\beta/(2+3\beta)\right]^{2/3}$, respectively. However, it should be noted that the maximum shear stress predicted by $\tau_{\rm max}^+\mbox{=}1+\frac{3}{4}\beta$ is achieved if only the Reynolds number is sufficiently large (the critical Reynolds number increases with increasing $\beta$). Otherwise, $\tau_{\rm max}^+\mbox{=}1+\sigma_\tau\beta$ with $\sigma_\tau<3/4$. For limited Reynolds numbers, $\sigma_\tau$ can be measured from empirical data through $\sigma_\tau\mbox{=}(\tau_{\rm max}^+-1)/\beta$. A numerical experiment on the TSS model gives a finite-Reynolds-number correction for the relationship between $P_0^+$ and $\beta$: $P_0^+=\sigma_p\beta$, where $\sigma_p\approx\frac{3}{2}-1.85(\frac{3}{4}-\sigma_\tau)$. 

\subsection {Multilayer scaling of mixing length in equilibrium APG TBLs}\label{subsec:mixinglength_model}
\noindent We continue to propose a symmetry-based formulation for the mixing-length profile in equilibrium APG TBLs. As will be shown, the construction of the formulation is dictated by the multilayer properties of TBLs, symmetry constraints, and established knowledge, leaving little room for arbitrary choices.

\subsubsection {The logarithmic-law region} \label{subsubsec:log}
\noindent In the log-law region of ZPG TBLs, $\ell_m\mbox{=}\kappa y$ is well-established. We aim to extend this linear scaling to include the PG effect. As addressed by Sethna \cite{Sethna2022}, the renormalization group theory predicts universal scaling functions for relations involving two or more parameters. If $Z$ depends on $X$ and $Y$, then $Z\propto X^a\mathcal{F}(X/Y^b)$, where $a$ and $b$ are universal scaling exponents, and $\mathcal{F}$ is a universal function. Based on this assertion, we propose that $\ell_m\mbox{=}\kappa y \mathcal{F}\left(\frac{y \partial P/\partial x}{\rho u_\tau^2}\right)=\kappa y \mathcal{F}(p^+y^+)$ in the log-law region of APG TBLs. To determine the universal function $\mathcal{F}$, we recall the classic scaling of $\ell_m\mbox{=}\sqrt{\tau^+}\kappa y$ proposed to preserve the log law for mean velocity in the overlap region \cite{Granville1989}. In this region, $\tau^+\approx1+\alpha p^+ y^+$, where $\alpha$ is a coefficient to account for a reduction in effective PG due to mean flow inertia \cite{townsend1961equilibrium,Skare1994exp}. Therefore, in the log-law region,
 \begin{equation}
  \ell_m\mbox{=}\kappa y\sqrt{1+\alpha p^+ y^+}.  
  \label{eq:ellm_log}
\end{equation}
The $\sqrt{\tau^+}$-scaling can also be understood within the SED framework, which postulates that a new physical factor (here, the APG) deforms the multilayer structure of a wall-bounded flow (such as changing characteristic thicknesses) or introduces a dilation-symmetry-breaking that can be formulated using a crossover scaling ansatz \cite{chen2018quantifying,Bi2025}. Here, $\sqrt{\tau^+}=[1+(\frac{y^+}{1/\alpha p^+ })^1]^{0.5/1}$ is the ansatz to describe the APG-induced dilation-symmetry-breaking to $\ell_m^+$ in the log-law region.

To be consistent with the near-wall asymptote of the TSS model in section \ref{subsec:TSS_model}, we replace $\alpha p^+$ in equation (\ref{eq:ellm_log}) with $P_0^+/y_p^+$, to obtain
 \begin{equation}
  \ell_m\mbox{=}\kappa y\sqrt{1+\frac{P_0^+}{y_p^+}y^+}.  
  \label{eq:ellm_log_1}
\end{equation}
Substituting the relationships of $P_0^+$ and $y_p^*$ with $\beta$ into equation (\ref{eq:ellm_log_1}) yields
 \begin{equation}
  \ell_m\mbox{=}\kappa y\sqrt{1+\frac{1.5\beta}{0.491} \left(\frac{2+3\beta}{3\beta}\right)^{2/3}y^*}.  
  \label{eq:ellm_log_2}
\end{equation}
In equation (\ref{eq:ellm_log_2}), we apply the high-Reynolds-number limit $P_0^+=1.5\beta$, instead of using $P_0^+=\sigma_p\beta$. This setting simplifies the modeling of $\ell_m^+$, based on the assumption that finite-Reynolds-number effects are important for TSS but negligibly influence the $\sqrt{\tau^+}$-scaling of $\ell_m^+$. Indeed, we have checked using the measured $\sigma_p$ to model $\ell_m^+$, and found that the mean-velocity predictions are generally no better (and sometimes worse) than those obtained by keeping $\sigma_p=3/2$. 

Equation (\ref{eq:ellm_log_2}) is the present model for $\ell_m$ in the log-law region of equilibrium APG TBLs, employing $\beta$ as the governing similarity parameter. The model undermines the linear scaling of $\ell_m$ when the APG is strong, but preserves the log law of the wall with an invariant K\'{a}rm\'{a}n constant. When $\beta=0$, the model reverts to the ZPG expression, i.e., Prandtl's mixing-length hypothesis. When $\beta\to +\infty$, however, $\ell_m$ incorrectly becomes infinitely large, thus necessitating a mechanism for the breakdown of the log law, which is discussed in section \ref{subsubsec:half}.   

\subsubsection {The wake region} \label{subsubsec:wake}
\noindent Within the wake region, the mixing length should keep constant wall-normally, since the wake flow becomes more and more like a mixing layer when the APG increases \cite{devenport2022equilibrium}. Therefore, $\ell_m=\lambda \delta$, with $\lambda$ dependent on APG. In view of the mixing-layer nature of the wake flow, we introduce a PG parameter $G$ and propose $\lambda=\lambda(G)$, where $G=\frac{\delta \partial P/\partial x}{\tau_{\rm max}}=\frac{\beta}{\tau_{\rm max}^+\delta_1^*}$. Since $\delta_1^*\propto \left(\frac{3\beta}{2+3\beta}\right)^{2/3}$ \cite{ZhengBi2025} and $\tau_{\rm max}^+=1+\frac{3}{4}\beta$ (assuming the Reynolds number is large), we redefine the PG parameter $G$ as
 \begin{equation}
  G=\frac{\frac{3}{4} \beta}{1+\frac{3}{4}\beta}\left(\frac{2+3\beta}{3\beta}\right)^{2/3}.  
  \label{eq:G}
\end{equation}
Consequently, as $\beta$ increases from zero to infinity, $G$ increases from zero to unity. We use $\lambda_0$ to denote $\lambda$ at $\beta=0$, and $\lambda_\infty$ to denote $\lambda$ at $\beta=\infty$. Then a model of $\lambda$ can be proposed as:
 \begin{equation}
  \lambda=\lambda_0-(\lambda_0-\lambda_\infty)f(G),  
  \label{eq:lambda}
\end{equation}
where $f(0)=0$ and $f(1)=1$. At the moment, $f$ is determined by fitting empirical data. As shown in Section \ref{subsubsec:ML_parameter}, $f=\sqrt{G}$ describes the data rather well.

The SED theory predicted $\lambda_0=\kappa/4=0.1125$ (note $\kappa=0.45$ in the SED theory). Here, we estimate $\lambda_\infty$ based on established knowledge on the mean velocity profile in equilibrium APG TBLs. As is well-known, the mean velocity profile in equilibrium APG TBLs can be well described using Coles' law of the wake \cite{coles1956law,White2005}, which is given by
\begin{equation}
  u^+=\frac{1}{\kappa}{\rm ln}y^++B+\frac{2{\rm \Pi}}{\kappa}{\rm sin}^2\left(\frac{\pi}{2}y^*\right),
  \label{eq:U_wake_law}
\end{equation}
where $\kappa=0.41$, $B=5.0$, and ${\rm \Pi}$ is Coles' wake parameter. Das \cite{Das1987,White2005} established an empirical relationship between $\beta$ and ${\rm \Pi}$: $\beta=0.42\Pi^2+0.76\Pi-0.4$, which correlates hundreds of data points from the 1968 Stanford Conference \citep{ColesHirst1968}. Equating the velocity gradients ${\rm d}u^+/{\rm d}y^*$ from equation (\ref{eq:U_wake_law}) at $y_{\rm max}^*$ and from $\sqrt{1+\frac{3}{4}\beta}/\lambda$ yields the following equation
\begin{equation}
  \frac{1}{\kappa y_{\rm max}^*}+\frac{\pi\Pi}{\kappa}{\rm sin}\left(\pi y_{\rm max}^*\right)=\frac{\sqrt{1+\frac{3}{4}\beta}}{\lambda},
  \label{eq:Pi_model}
\end{equation}
where $y_{\rm max}^*\mbox{=}0.456\left[3\beta/(2+3\beta)\right]^{2/3}$. For sufficiently large $\beta$, we have $y_{\rm max}^*\to0.456$ and $\lambda\to\lambda_\infty$. In this limit, $\beta$ becomes a parabolic function of $\Pi$, as suggested by Das. At large $\Pi$ the leading-order term is $\frac{4}{3}\frac{\pi^2}{\kappa^2}\lambda_\infty^2{\rm sin}^2(0.456\pi)\Pi^2$, which according to Das' empirical relation is equal to $0.42\Pi^2$. Solving for $\lambda_\infty$ gives $\lambda_\infty\approx0.074$. 

Consequently, we construct an explicit model for the mixing-length parameter $\lambda$ in the wake region. This represents a physically grounded refinement over the turbulence model of Escudier, who set $\lambda=0.09$ \cite{Schlichting}, and over the recent observations of Shu and Xu, who found $\lambda\approx0.08$ in APG TBLs \cite{ShuXu2025}. It is worth pointing out that the finite-Reynolds-number effect may affect the prediction accuracy of equation (\ref{eq:lambda}), owing to the deviation from $\tau_{\rm max}^+=1+\frac{3}{4}\beta$ when the Reynolds number is limited.

\subsubsection {The half-power-law region} \label{subsubsec:half}
\noindent Now, we aim to match the expressions of $\ell_m$ in the log-law and wake regions. To maintain consistency with equation (\ref{eq:SEDmodel}), the mixing length in the outer region of APG TBLs is postulated to obey the following defect power law:
 \begin{equation}
  \ell_{m}=\lambda \delta \left(1-r^m\right),
  \label{eq:ell_out_APG}
\end{equation}
where $r=1-y^*$. In ZPG TBLs, $m=4$ according to the SED theory. In equilibrium APG TBLs, $m$ should vary with the strength of APG for the following reasons. Observations indicate that the wake region expands toward the wall as the APG intensifies \cite{devenport2022equilibrium}, implying an increase in $m$ with the strengthening APG. Furthermore, in the limit as $r{\rm \to}1$, equation (\ref{eq:ell_out_APG}) simplifies to $\ell_{m}\approx m\lambda y$. For consistency with equation (\ref{eq:ellm_log}), the product $m\lambda$ must increase with increasing APG. This requires $m$ to increase because equation (\ref{eq:lambda}) shows that $\lambda$ decreases as APG increases.

Equation (\ref{eq:ell_out_APG}) displays a Prandtl-type linear scaling close to the wall. Therefore, to ensure a match with equation (\ref{eq:ell_out_APG}), equation (\ref{eq:ellm_log_2}) should be revised to
 \begin{equation}
  \ell_m\mbox{=}\kappa y\sqrt{1+\frac{1.5\beta}{0.491} \left(\frac{2+3\beta}{3\beta}\right)^{2/3}y^*\left[1+{\left(\frac{y^*}{y_{\rm log}^*}\right)^4}\right]^{-1/4}},  
  \label{eq:ellm_log_3}
\end{equation}
where $y_{\rm log}^*$ denotes the upper edge of the log-law region. With the crossover scaling ansatz introduced, $\ell_m$ retrieves the linear scaling to match equation (\ref{eq:ell_out_APG}) when $y^*>y_{\rm log}^*$. The ansatz represents the second dilation-symmetry-breaking induced by APG, describing a transition from the log-law to the half-power-law scaling, as discussed below.

The matching condition yields the following equation
 \begin{equation}
  m\lambda=\kappa \sqrt{1+\frac{1.5\beta}{0.491} \left(\frac{2+3\beta}{3\beta}\right)^{2/3}y_{\rm log}^*}.  
  \label{eq:match}
\end{equation}
For ZPG TBLs, $0.15\delta$ is generally accepted as the upper bound of the log-law region \cite{Marusic2013Rapid}. If this bound stays constant in APG TBLs, equation (\ref{eq:match}) indicates that $m\to\infty$ as $\beta\to\infty$, which implies that the wake region engulfs the entire boundary layer when the flow approaches the separation point. To avoid this nonphysical state, we should introduce a mechanism for the breakdown of the log-law region, which is obtained by introducing the third dilation-symmetry-breaking, as below
 \begin{equation}
  y_{\rm log}^*=0.15\left[1+\left(\frac{\beta}{\beta_c}\right)^4\right]^{-1/4}.  
  \label{eq:ylog}
\end{equation}
In the crossover scaling ansatz in equation (\ref{eq:ylog}), $\beta_c$ denotes a critical Clauser PG parameter above which the log-law region shrinks and eventually disappears as $\beta\to\infty$, representing a ``progressive breakdown'' of the log law as proposed by Galbraith et al. \cite{Galbraith1977}. This progressive breakdown is particular: below $\beta_c$, $y_{\rm log}$ remains at $0.15\delta$; above $\beta_c$, $y_{\rm log}=0.15\delta \beta_c/\beta$. When $\beta$ is infinitely large, equation (\ref{eq:match}) becomes
 \begin{equation}
  m_\infty\lambda_\infty=\kappa \sqrt{1+\frac{3/2}{0.491} 0.15\beta_c}.  
  \label{eq:match_1}
\end{equation}
Therefore, we introduce an upper bound for $m$ (i.e., $m_\infty$) through $\beta_c$.

The matching condition indicates that there is a layer with linear-scaling mixing length in between the log-law region and the wake region in APG TBLs. In this layer, if the APG is not minor, the Reynolds shear stress is approximated by $-\left<{u'v'}\right>^+=P_0^+ y^+/y_p^+$, and the mixing length simplifies to $\ell_m^+=\kappa y^+\sqrt{P_0^+ y_{\rm log}^+/y_p^+}$. Therefore, $\frac{{\rm d}u^+}{{\rm d}y^+}=\frac{1}{\kappa \sqrt{y_{\rm log}^+} \sqrt{y^+}}$, predicting a half-power law for the mean velocity. The coexistence of the log law and an overlying half-power law for the mean velocity in TBLs subjected to strong APGs was identified early by Perry and Schofield \cite{Perry1973}, and more recently proposed by Knopp et al. \cite{Knopp2021,Knopp2022} and Ma et al. \cite{MaPoF2024-2}. Here, it is rediscovered through a symmetry-based argument on the mixing length. With the shrinkage of the log-law region as $\beta\to\infty$, the half-power law eventually dominates the overlapping region, as predicted by Stratford \cite{stratford1959prediction}.

The remaining question is how to determine $\beta_c$ (or $m_\infty$). We consider the asymptotic condition of the mean velocity gradient in the half-power-law region as $\beta\to\infty$, which is 
 \begin{equation}
  \frac{{\rm d}u^+}{{\rm d}y^+}=\frac{1}{\kappa \sqrt{0.15\delta^+\beta_c/\beta}\sqrt{y^+}}.  
  \label{eq:dudy_half}
\end{equation}
Since $u_\tau$ is no longer an appropriate velocity scale, we change to using $u_p\equiv(\frac{\nu}{\rho}\frac{\partial P}{\partial x})^{1/3}$ as the velocity scale and $y_p=\nu/u_p$ as the length scale \cite{stratford1959prediction,skote2002direct}. Substituting the relationship $\beta=\delta_1^+u_p^3/u_\tau^3$ into equation (\ref{eq:dudy_half}) yields
 \begin{equation}
  \frac{{\rm d}u^{(p)}}{{\rm d}y^{(p)}}=\frac{\sqrt{\delta_1^*}}{\kappa \sqrt{0.15\beta_c}\sqrt{y^{(p)}}},
  \label{eq:dudy_half1}
\end{equation}
where $u^{(p)}=u/u_p$ and $y^{(p)}=y/y_p$. The solution of equation (\ref{eq:dudy_half1}) is the half-power law for the mean velocity \cite{stratford1959prediction}: $u^{(p)}=C\sqrt{y^{(p)}}+D$, where 
 \begin{equation}
  C=\frac{2\sqrt{\delta_1^*}}{\kappa \sqrt{0.15\beta_c}},
  \label{eq:dudy_half2}
\end{equation}
and $D$ is the integration constant. Equation (\ref{eq:dudy_half2}) is a refinement of the classic proposal of $C=2/\kappa$ \cite{stratford1959prediction}.

As proposed earlier, $\alpha p^+=P_0^+/y_p^+$. Since $p^+=\beta/\delta_1^+$, we have $\delta_1^*=\frac{\alpha \beta}{P_0^+}y_p^*=\frac{\alpha }{1.5}y_p^*$ (assuming the Reynolds number is large). As $\beta\to\infty$, $y_p^*\to0.491$. Granville specified that the coefficient $\alpha$ is 0.9 \cite{Granville1989}, which is consistent with the Sk{\aa}re and Krogstad datasets, which have high Reynolds numbers and a rather large $\beta$ ($\sim 20$) \cite{Skare1994exp}. Consequently, $\delta_1^*\to0.295$ as $\beta\to\infty$. The ``universal constants'' $C$ and $D$ have been reported to vary significantly in different studies \cite{coleman2017}. Through a DNS of a zero-skin-friction Couette-Poiseuille flow, which arguably provides the best chance of success in confirming the half-power law, Coleman et al. recently gave a compact estimate: $C=2.25\sim2.5$ and $D=-2.2$ \cite{coleman2017}. Here, we employ $C=2.5$, since it is also the lower bound in the estimation ($C=2.5\sim4.9$) cited in Schlichting and Gersten's book \cite{Schlichting}. Using $\kappa=0.45$ from the SED, we derive from equation (\ref{eq:dudy_half2}) that $\beta_c\approx6.2$. Then, $m_\infty\approx12$ according to equation (\ref{eq:match_1}).

\subsubsection {The inner region} \label{subsubsec:inner_ellm}
\noindent The formulation of the near-wall mixing-length profile is motivated as follows. Numerous studies have observed that APG tends to reduce the thicknesses of the viscous sublayer and buffer layer, resulting in a lower log-law intercept \cite{devenport2022equilibrium}. This effect has previously been captured by adjusting the damping parameter $A$ in the van Driest model \cite{Granville1989}. Here, we assume that $\ell_m^+$ in the near-wall region retains the analytical form given by the SED (equation [\ref{eq:SEDmodel}]), with the characteristic thicknesses $y_s^+$ and $y_b^+$ depending on the APG. Consequently, the near-wall mixing-length model is written as
 \begin{equation}
  \ell_m^+=\frac{\kappa {y_s^+}^2}{y_b^+}\left(\frac{y^+}{y_s^+}\right)^{3/2}\left[1+\left(\frac{y^+}{y_s^+}\right)^4\right]^{1/8}\left[1+\left(\frac{y^+}{y_b^+}\right)^4\right]^{-1/4}.
  \label{eq:ell_inner}
\end{equation}

In the SED theory, $y_s^+$ is a measure of the thickness of the viscous sublayer. Specifically, it is defined as the midpoint between the viscous sublayer and the buffer layer. At this location, $p^+$ replaces $\beta$ as the governing PG parameter. Therefore, $y_s^+$ is a function of $p^+$. In contrast to $y_s^+$, Nickels \cite{Nickels2004} proposed a critical thickness $y_c^+$ for the sublayer ``edge'', above which the flow becomes unstable. Assuming $Re_c\equiv\frac{u_cy_c}{\nu}$ (where $u_c\equiv\sqrt{{\tau|_{y=y_c}}/{\rho}}$) is a universal constant for all wall-bounded flows, he derived a cubic equation for $y_c^+$:
 \begin{equation}
  p^+(y_c^+)^3+(y_c^+)^2-Re_c^2=0,
  \label{eq:yc}
\end{equation}
where $Re_c=12$. The physically relevant solution for $y_c^+$ is the smallest positive root of equation (\ref{eq:yc}). Because $y_s^+$ and $y_c^+$ are highly relevant, we can calculate $y_s^+$ from $y_c^+$ using the following equation:
 \begin{equation}
  y_s^+=\frac{9.7}{12}y_c^+,
  \label{eq:ys}
\end{equation}
where 9.7 and 12 are the reference values of $y_s^+$ and $y_c^+$ under ZPG, respectively.

$y_b^+$ represents the midpoint between the buffer layer and the log-law layer. Its magnitude significantly influences the intercept of the log law. Because by definition $y_b^+$ is the lower edge of the overlap region between the inner and outer flows, in principle, it is dictated by $p^+$ and modulated by $\beta$ and $\delta^+$, exhibiting a complex nature. So far, we have not identified the functional form of $y_b^+(p^+,\beta,\delta^+)$ (since $p^+$ is related to $\beta$ and $\delta^+$, the number of independent variables is probably two). Rather than introducing an ad hoc empirical closure, $y_b^+$ is determined self-consistently by matching the present velocity profile to Nickels’ universal inner-layer scaling at $y^+=4y_c^+$, which represents a robust and well-validated inner-layer reference \cite{Nickels2004}. At this location, Nickels’ model yields the asymptotic relation: $u^+\approx y_c^++2.24\sqrt{1+p^+y_c^+}$ (Appendix \ref{Append_Nickels}). By equating this target value to the integrated velocity profile derived from the current mixing-length model (equation [\ref{eq:ell_APG}]), we obtain the following equation 
 \begin{equation}
  y_c^++2.24\sqrt{1+p^+y_c^+} =  \int_{0}^{4y_c^+} {\frac{{ - 1 + \sqrt {1+4{\tau ^{\rm{ + }}}{{ {\ell_{m}^+}}^2}} }}{{2{{ {\ell_{m}^+}}^2}}}\mathrm{d} {y^{\rm{ + }}}}.
  \label{eq:yb}
\end{equation}
Consequently, $y_b^+$ is uniquely determined without additional fitting parameters. This self-consistent procedure ensures that $y_b^+$ automatically adapts to both APG and finite-Reynolds-number effects, preserving the analytical and symmetry-based foundation of the present framework.

\subsubsection {The full-profile model of mixing length}
\noindent Combining the above formulations, the full-profile model of $\ell_m^+$ in equilibrium APG TBLs is written as
 \begin{align}
  \ell_m^+=&\frac{\kappa {y_s^+}^2}{y_b^+}\left(\frac{y^+}{y_s^+}\right)^{3/2}\left[1+\left(\frac{y^+}{y_s^+}\right)^4\right]^{1/8}\left[1+\left(\frac{y^+}{y_b^+}\right)^4\right]^{-1/4}\nonumber\\
&\times\sqrt{1+\frac{1.5\beta}{0.491} \left(\frac{2+3\beta}{3\beta}\right)^{2/3}y^*\left[1+{\left(\frac{y^*}{y_{\rm log}^*}\right)^4}\right]^{-1/4}}\frac{1-r^m}{m(1-r)},
  \label{eq:ell_APG}
\end{align}
where $\kappa=0.45$ as proposed by the SED, $y_s^+$ is derived from equation (\ref{eq:ys}), $y_b^+$ is derived from equation (\ref{eq:yb}), $m$ is derived from equation (\ref{eq:match}), and $y_{\rm log}^*$ is derived from equation (\ref{eq:ylog}). 

In summary, this section presents a symmetry-based analytical model to predict the full mixing-length profile in equilibrium APG TBLs. The model extends the ZPG formulation by She et al. \cite{she2017quantifying} to describe the equilibrium APG effects on the mixing-length distribution. Combining this mixing-length model with the TSS model developed by Zheng et al. \cite{ZhengBi2025}, a theoretical prediction is derived for the full profile of mean velocity in equilibrium APG TBLs, which maintains the standard logarithmic law of the wall and reproduces the half-power law in the overlap region. The current models are physically sound, explicit, and compact, containing only one empirical parameter ($\sigma_\tau$ or $\sigma_p$) that accounts for the finite-Reynolds-number effect and can be easily measured through the maximum shear stress.

\section{Validation}\label{sec:valid}
\subsection{Database}\label{subsec:databases}
\noindent To validate the current models, we collected a database of equilibrium APG TBLs from the literature. As shown in Table \ref{tab:cases}, the datasets span a wide range of $Re_\theta$ ($1250\sim50980$) and $\beta$ ($0\sim39$), representing the current state of the art in DNS, Large eddy simulation (LES), and experiments. Cases B, P, and SV are near-equilibrium, while the others have been finely tuned to be in equilibrium. In the SV cases \cite{Sanmiguel2020exp}, only the mean velocity profiles have been presented, with no information on the TSS or Reynolds shear stress profiles. Because their Reynolds numbers are rather high and the APGs are rather limited, we empirically set $\sigma_p=\frac{3}{2}$ to test the robustness of the current models.  

\begin{table}
\caption{\label{tab:cases} Datasets of the equilibrium APG TBLs used for validating the current theory.}
\begin{ruledtabular}
\begin{tabular}{lcccccc}
      \makecell{Case Name\footnotemark[1]} & Method  & $Re_\theta$ & $p^+$  &  $\beta$ & $H$  & $\sigma_p$ \\[3pt]
\hline
     S \cite{Schlatter2009dns} & DNS  & 2400 & 0 & 0 & 1.41 & $3/2$  \\
     EA \cite{Eitel-Amor2014} & LES   & 6600 & 0 &0 & 1.36 & $3/2$ \\
     LS \cite{Lee2008} & DNS  & \makecell{1410, 1250\\1250, 1350} & \makecell{0, 0.0027\\0.007, 0.016} & \makecell{0, 0.25\\0.73, 1.68} & \makecell{1.47, 1.55\\1.62, 1.84} & \makecell{$3/2$, 0.78\\1.48, 1.95} \\
     L \cite{Lee2017direct} & DNS   & \makecell{1320, 1605\\2180, 2840} & \makecell{0, 0.0088\\0.023, 0.082} & \makecell{0, 0.73\\2.2, 9.0} & \makecell{1.45, 1.56\\1.72, 1.98} & \makecell{$3/2$, 1\\1.18, 0.79} \\
     B \cite{Bobke2017les} & LES & 2870, 3360 & 0.006, 0.011 & 1, 2 & 1.56, 1.68 & 1.32, 1.57 \\
     P \cite{Pozuelo2022les} & LES  & 8700 & 0.0033 &1.4 & 1.5 & 1.45 \\
     SV \cite{Sanmiguel2020exp} & Exp.   & 9060, 19780 & 0.0026, 0.0013 & 1.11, 1.12 & 1.53, 1.44 & $3/2$, $3/2$ \\
     SK \cite{Skare1994exp} & Exp.   & 50980 & 0.0126 & 21.2 & 1.998 & 1.41 \\
     K \cite{Kitsios2017} & DNS  & 3440, 10000 & 0.005, 0.11 & 1, 39 & 1.47, 2.45 & 1.5, 0.85 \\
\end{tabular}
\end{ruledtabular}
\footnotetext[1]{Case names used in the text and figures are formed by combining the initials of the lead authors' names (as summarized in the table) with the $\beta$ value of each case (e.g., ``LS0.25'' denotes the case from Lee and Sung \cite{Lee2008} with $\beta=0.25$).}
\end{table}

\subsection{Total shear stress profile}\label{subsec:TSS}
\begin{figure}
    \centering
    \includegraphics[width=0.46\linewidth]{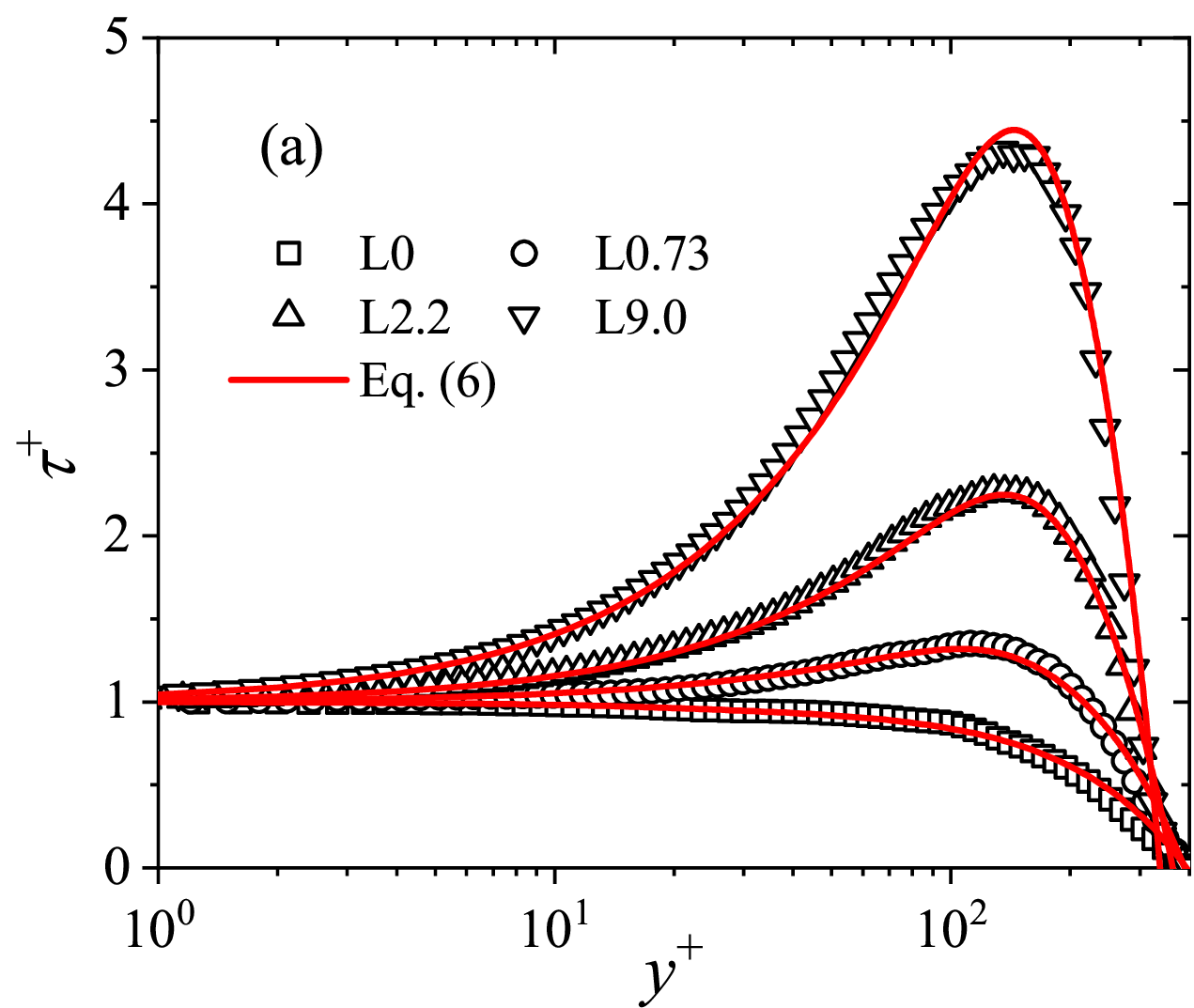}
    \includegraphics[width=0.49\linewidth]{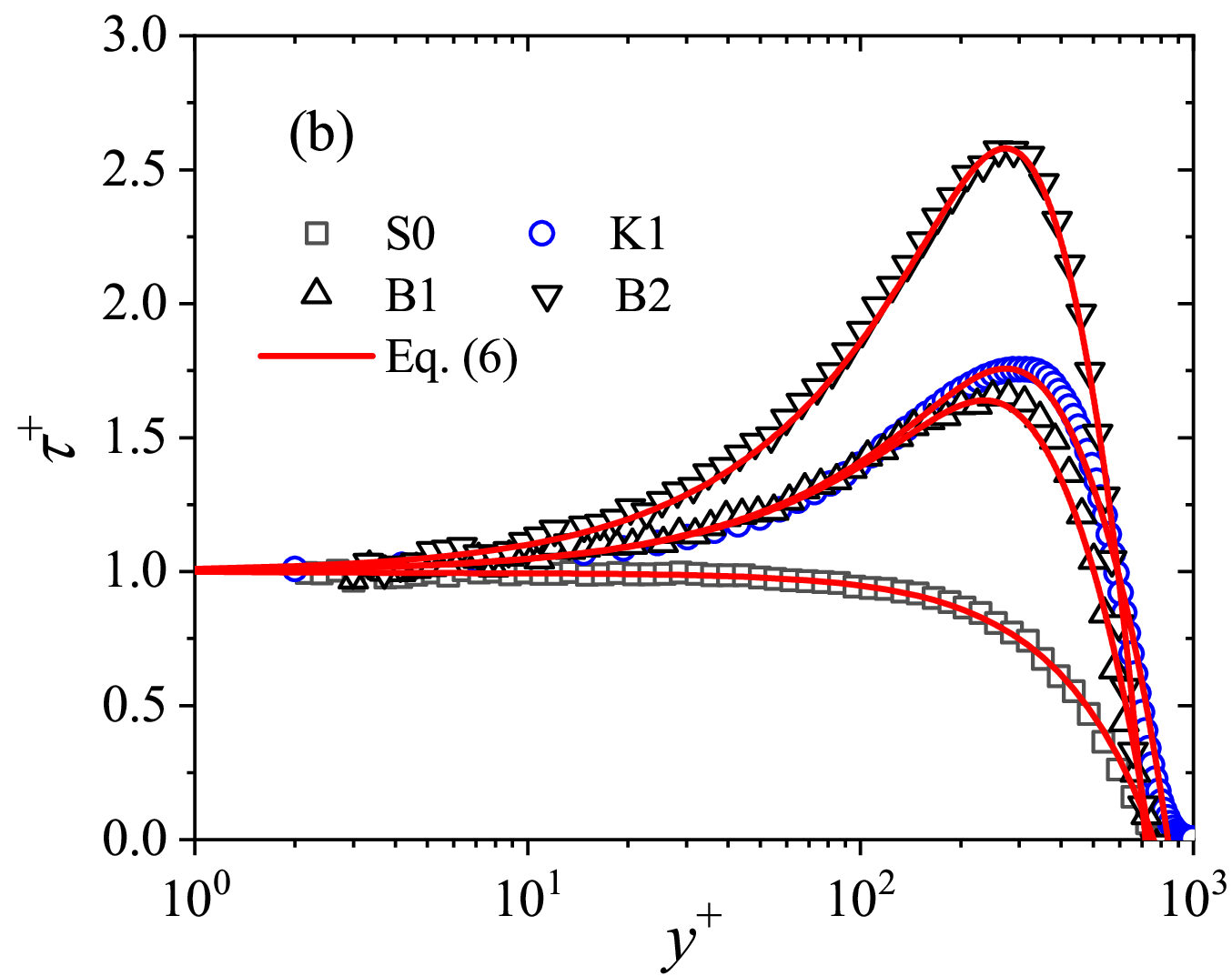}
     \includegraphics[width=0.49\linewidth]{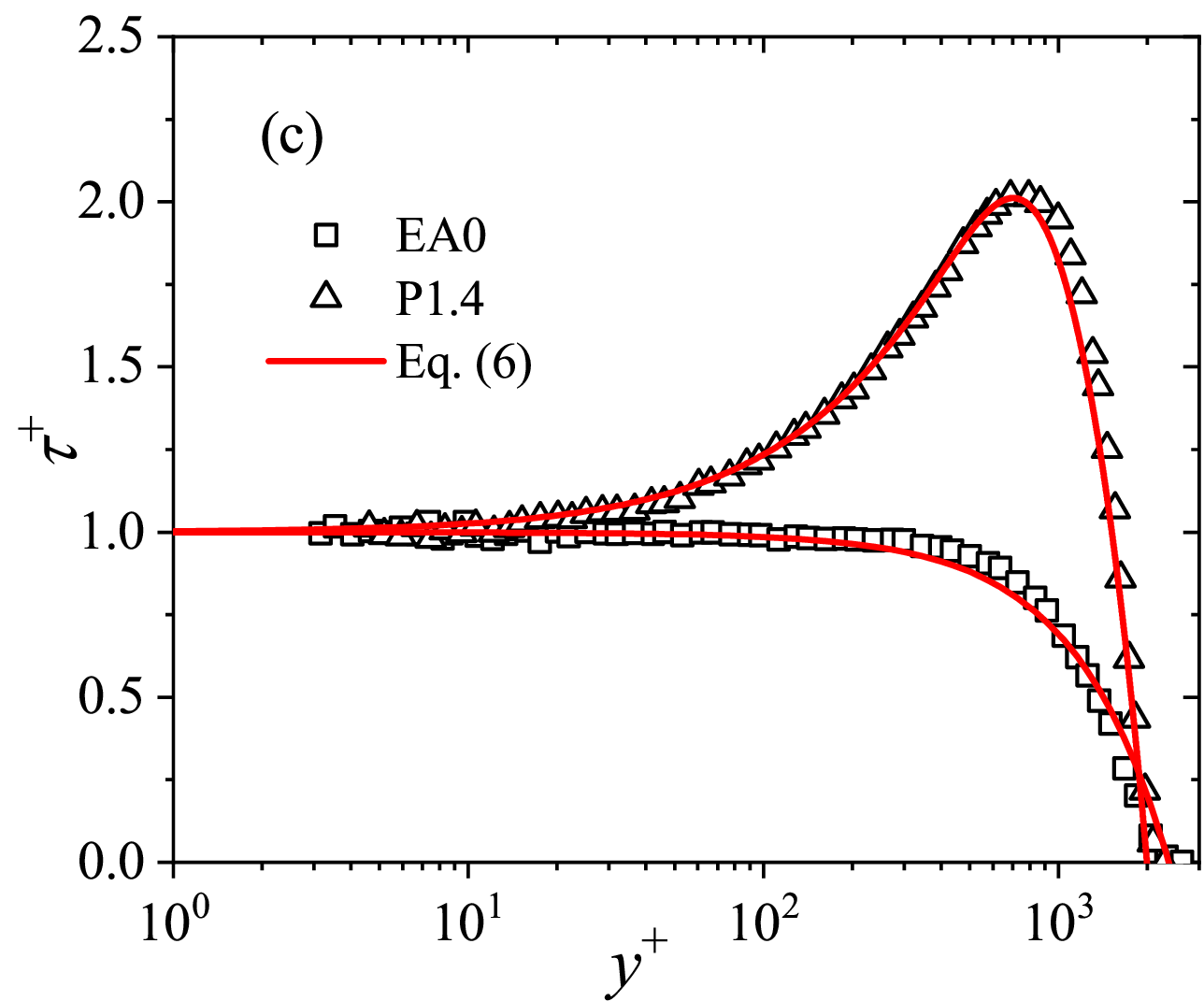}
    \includegraphics[width=0.49\linewidth]{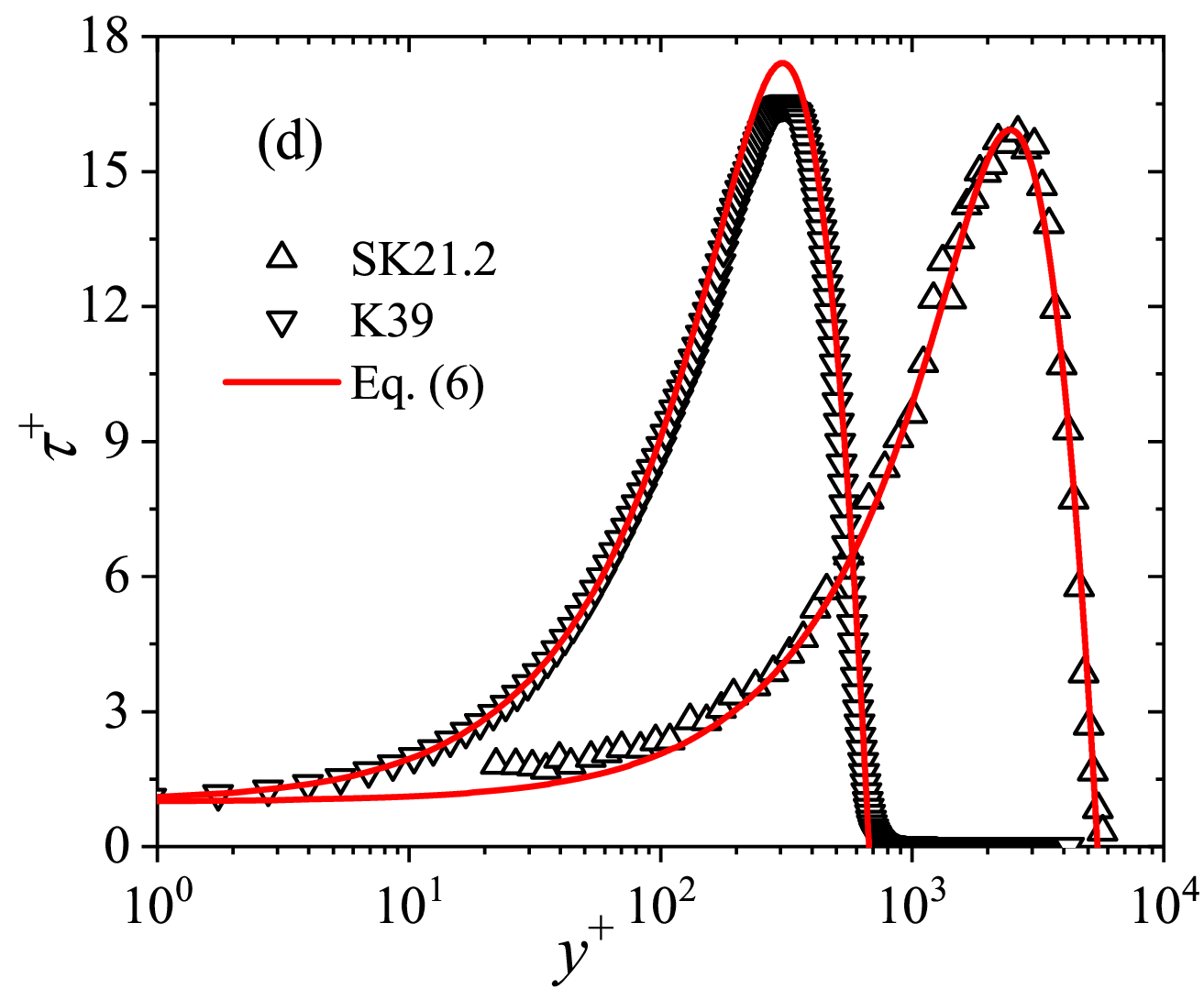}
  \caption{Profiles of TSS in equilibrium APG TBLs. Symbols represent data from the experiments and numerical simulations in Table \ref{tab:cases}. Solid lines denote predictions from the $\tau^+$ model (Eq. [\ref{eq:tau_APG}]), with the single empirical parameter $\sigma_p$ estimated from the peak shear stress and given in Table \ref{tab:cases}.}
  \label{fig:tau_result}
\end{figure}

In Fig. \ref{fig:tau_result}, we validate the TSS model (Eq. [\ref{eq:tau_APG}]) using the datasets in Table \ref{tab:cases}. To predict $\tau^+$ through equation (\ref{eq:tau_APG}), the only empirical parameter $\sigma_p$ is determined by using the peak shear stress data, as detailed in section \ref{subsec:TSS_model}; its value for each case is provided in Table \ref{tab:cases}. As shown in Fig. \ref{fig:tau_result}, the predictions of the $\tau^+$ model exhibit excellent agreement with all numerical and experimental data.  

\begin{figure}
    \centering
    \includegraphics[width=0.48\linewidth]{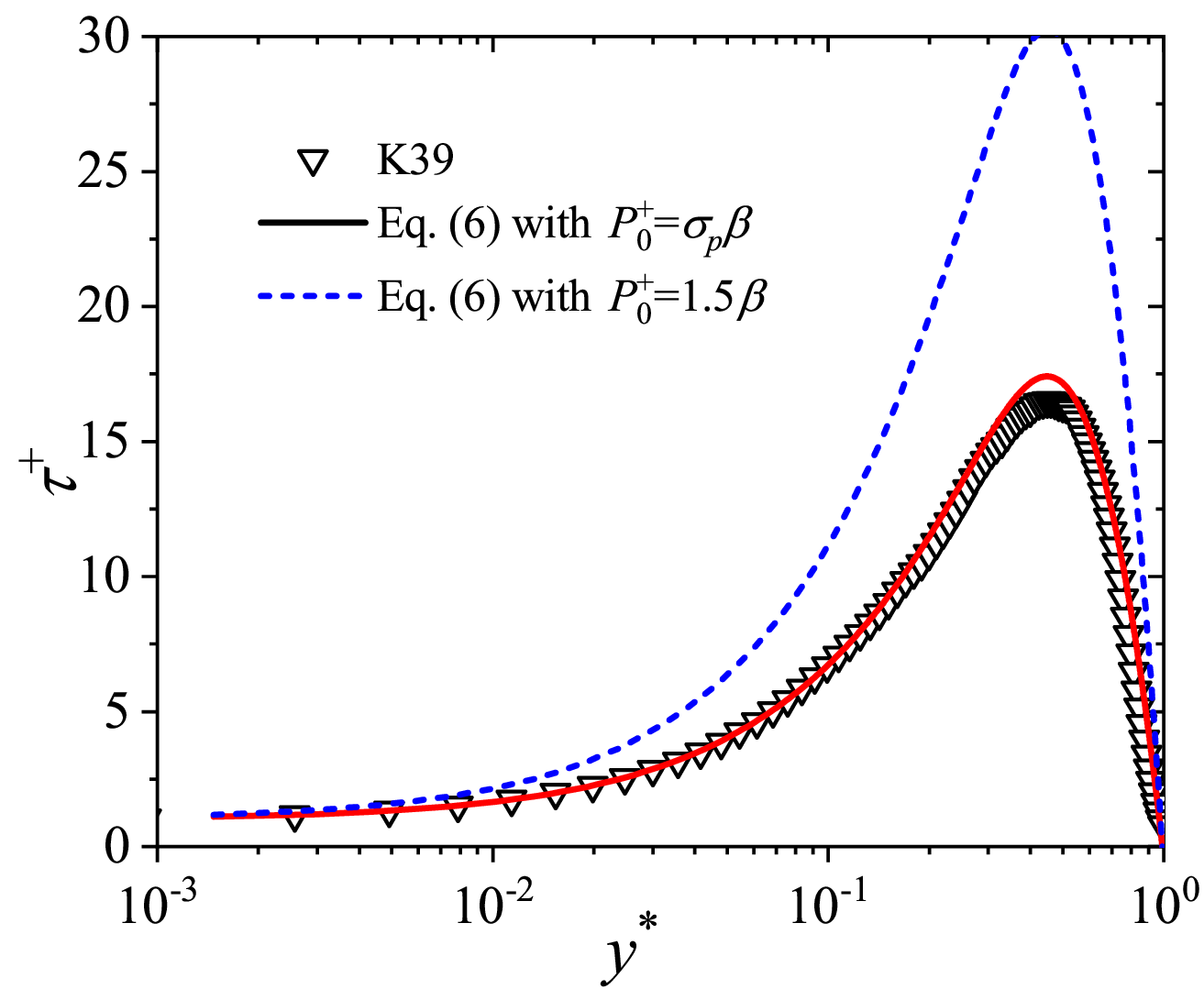}
    \includegraphics[width=0.48\linewidth]{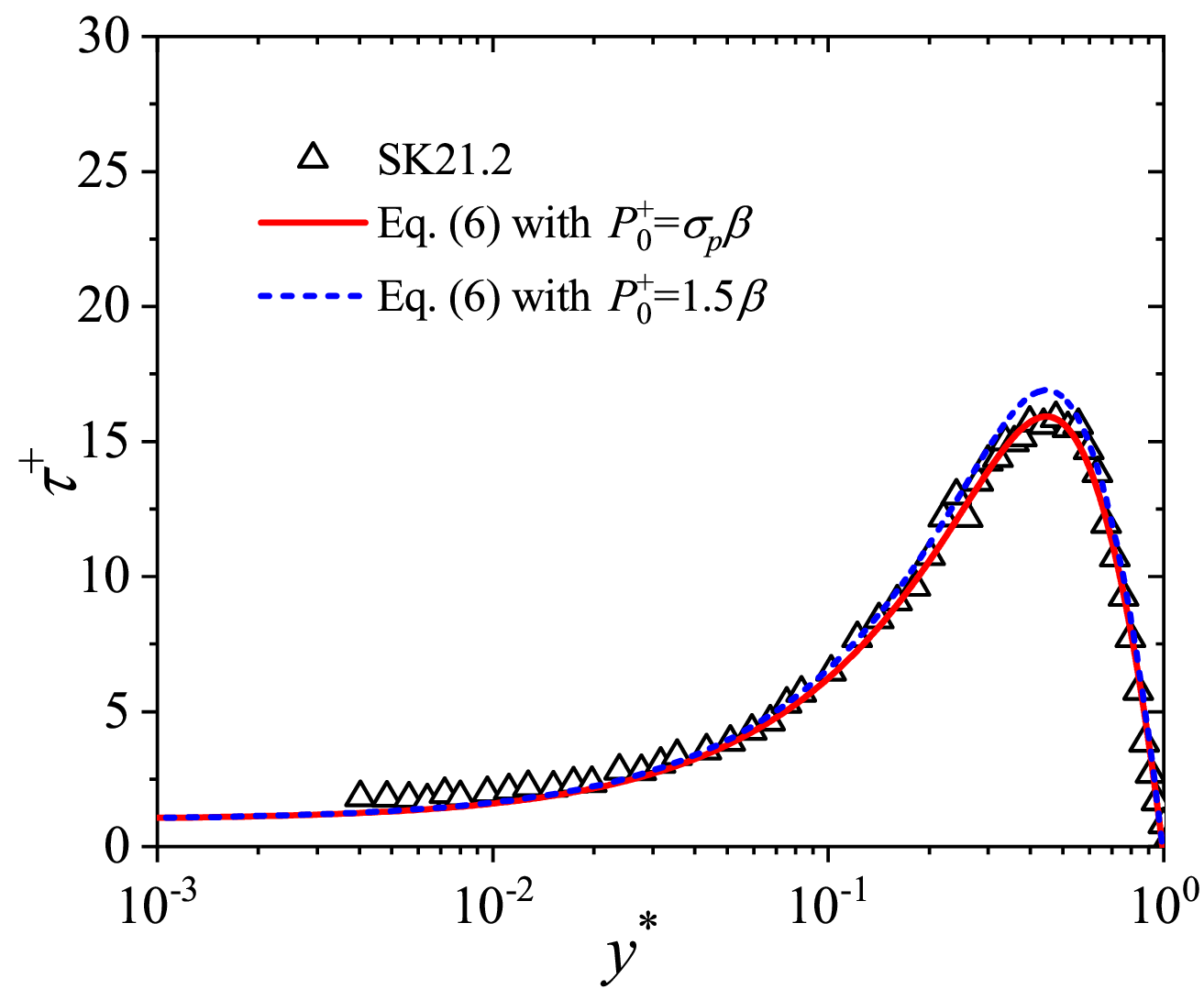}
    \\(a)\quad\quad\quad\quad\quad\quad\quad\quad\quad\quad\quad\quad\quad\quad\quad\quad\quad\quad\quad(b)\\
 \caption{Profiles of TSS for Cases (a) K39 and (b) SK21.2 in the outer coordinate $y^*$. Solid lines denote predictions from the $\tau^+$ model (Eq. [\ref{eq:tau_APG}]), with $\sigma_p$ estimated from the peak shear stress. Short-dashed lines illustrate the same model with $\sigma_p=3/2$, representing the high-Reynolds-number limit.}
 \label{fig:beta20_tau_3/2}
\end{figure}

Figure \ref{fig:beta20_tau_3/2} further illustrates, via short-dashed lines, the model's behavior when $\sigma_p=\frac{3}{2}$ (i.e., the high-Reynolds-number limit). For the K39 case, this high-Reynolds-number prediction significantly overestimates the result obtained with the empirical $\sigma_p$, indicating that the turbulence remains substantially underdeveloped in K39 compared to the high-Reynolds-number scenario with the same APG ($\beta=39$). In contrast, the SK21.2 case, which also features a strong APG ($\beta=21.2$) but is of a high Reynolds number, does not show this discrepancy.

It is worth noting that, in Eq. (\ref{eq:tau_APG}), the expression $y_p^*\mbox{=}0.491\left[3\beta/(2+3\beta)\right]^{2/3}$ was derived based on $\sigma_p=\frac{3}{2}$ \cite{ZhengBi2025}. Fig. \ref{fig:beta20_tau_3/2} thus reveals that the finite-Reynolds-number effects have little impact on the shape of the $\tau^+$ profile, which is governed by $y_p^*$, but markedly affect the maximum value of $\tau^+$, which is determined by $P_0^+$.  

The variation of $\sigma_p$ indicates that, for each $\beta$, there is a critical Reynolds number, which increases with increasing $\beta$, and above which $\sigma_p$ becomes saturated to $3/2$. A plausible hypothesis is that the transition is governed by a critical $p^+$, as the increase in the Reynolds number for a constant-$\beta$ flow decreases $p^+$. However, as shown in Tab. \ref{tab:cases}, the relationship between $\sigma_p$ and $p^+$ is complicated. For now, the dependence of $\sigma_p$ on the Reynolds number and APG remains an open question that warrants further investigation.  

\subsection{Key model parameters}\label{subsec:model_parameters}

\subsubsection{Mixing-length parameter $\lambda$}\label{subsubsec:ML_parameter}
\begin{figure}
    \centering
    \includegraphics[width=0.5\linewidth]{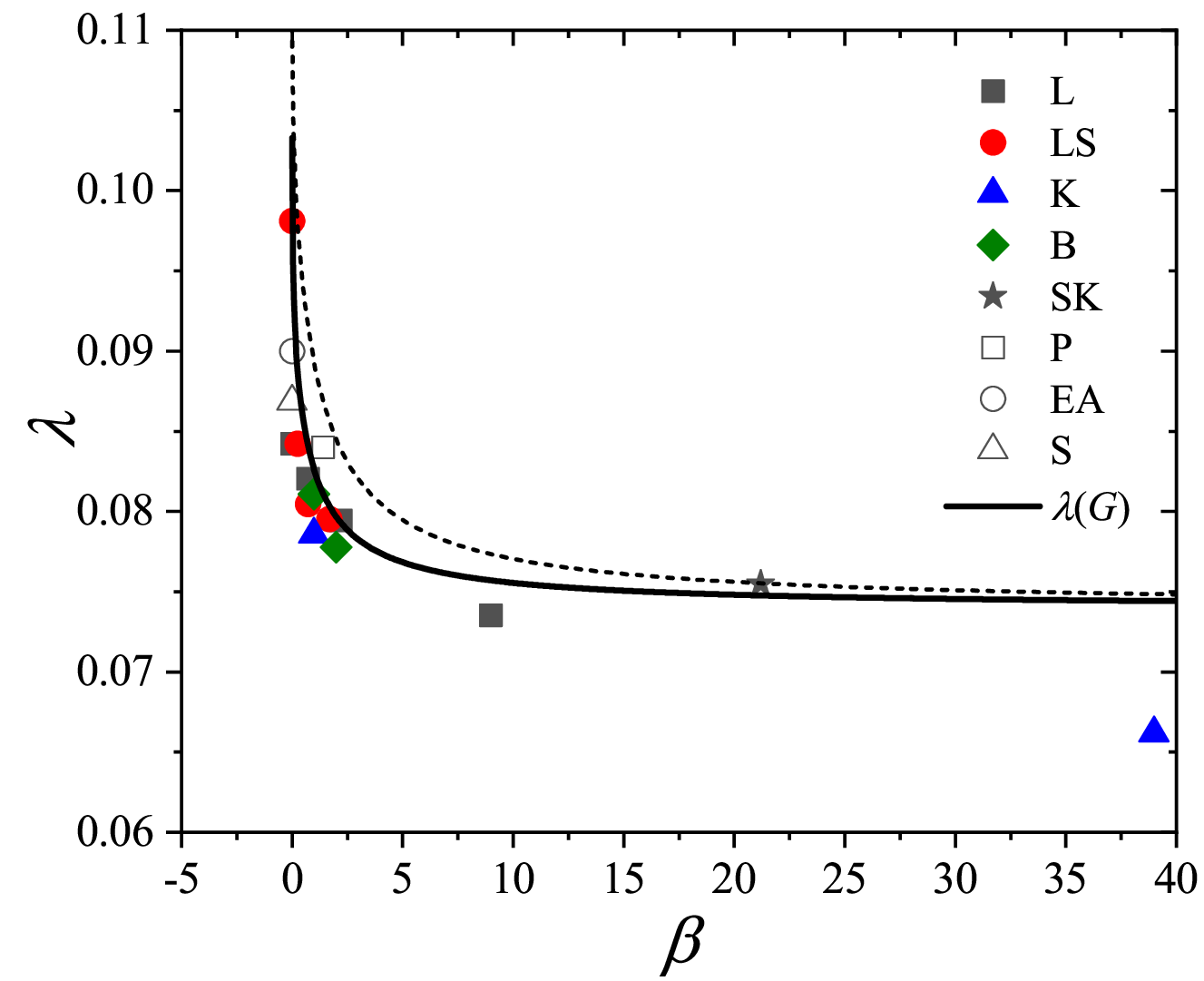}
 \caption{Variation of the mixing-length parameter $\lambda$ with $\beta$. Symbols denote the data points measured from the datasets listed in Table \ref{tab:cases}. The solid line represents the prediction from the current model (equation [\ref{eq:lambda}] with $f(G)=\sqrt{G}$). The short-dashed line corresponds to the prediction from equation (\ref{eq:lambda}) with $f(G)=G$.}
 \label{fig:lambda_beta}
\end{figure}

\noindent The wake region features a constant mixing length characterized by the mixing-length parameter $\lambda$. Figure \ref{fig:lambda_beta} validates the current model for $\lambda$ (equation [\ref{eq:lambda}]) using the datasets listed in Table \ref{tab:cases}. To measure $\lambda$ from the numerical and experimental data, we compute the ratio of the average mixing length within the wake region (where $\ell_m$ remains approximately constant) to the boundary-layer thickness. As shown in Fig. \ref{fig:lambda_beta}, equation (\ref{eq:lambda}) with $\lambda_0=0.1125$, $\lambda_\infty=0.074$, and $f(G)=\sqrt{G}$ accurately captures the variation of $\lambda$ with $\beta$. For comparison, the prediction that employs $f(G) = G$ exhibits a clear deviation from the data at moderate values of $\beta$. The theoretical rationale for $f(G)=\sqrt{G}$ deserves further investigation. 

In particular, the data points for Cases K39 and L9 lie substantially below the model prediction. Both cases feature relatively large $\beta$ values but comparatively low Reynolds numbers, which lead to underdeveloped turbulence within the wake region, as indicated also by the significantly smaller values of $\sigma_p$ reported in Table \ref{tab:cases}.

\begin{figure}
    \centering
    \includegraphics[width=0.5\linewidth]{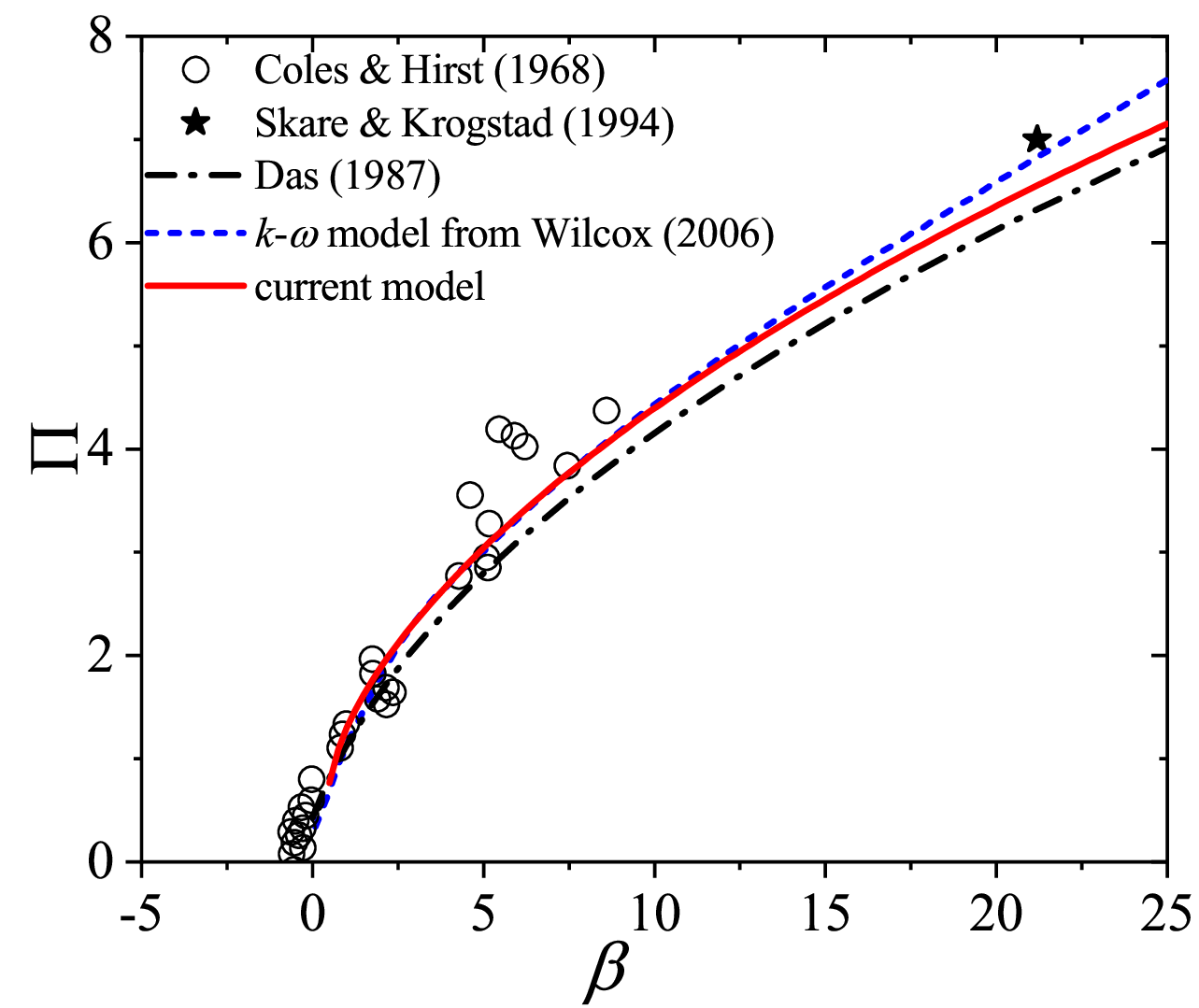}
 \caption{Variation of Coles' wake parameter $\Pi$ with $\beta$. Symbols denote experimental data points. The solid line represents the prediction of the current model (equation [\ref{eq:Pi_model}]). The short-dashed line denotes the prediction of the $k-\omega$ turbulence model from Wilcox's book \cite{Wilcox2006}. The dash-dotted line shows the prediction of Das' empirical correlation \cite{Das1987}.}
 \label{fig:pi_beta}
\end{figure}

Figure \ref{fig:pi_beta} illustrates the dependence of the Clauser wake parameter $\Pi$ on $\beta$. Data points are sourced from Coles and Hirst \cite{ColesHirst1968} and the experiment featuring large $\beta$ and high Reynolds numbers conducted by Sk{\aa}re and Krogstad \cite{Skare1994exp}. The current prediction of the $\Pi-\beta$ relationship (equation [\ref{eq:Pi_model}]) is consistent with the data points, as well as predictions derived from the $k-\omega$ model presented in Wilcox's book \cite{Wilcox2006} and the empirical correlation proposed by Das \cite{Das1987,White2005}. 

\subsubsection{Critical Clauser parameter $\beta_c$ and outer-layer exponent $m$}\label{subsubsec:beta_c}
\begin{figure}
    \centering
    \includegraphics[width=0.5\linewidth]{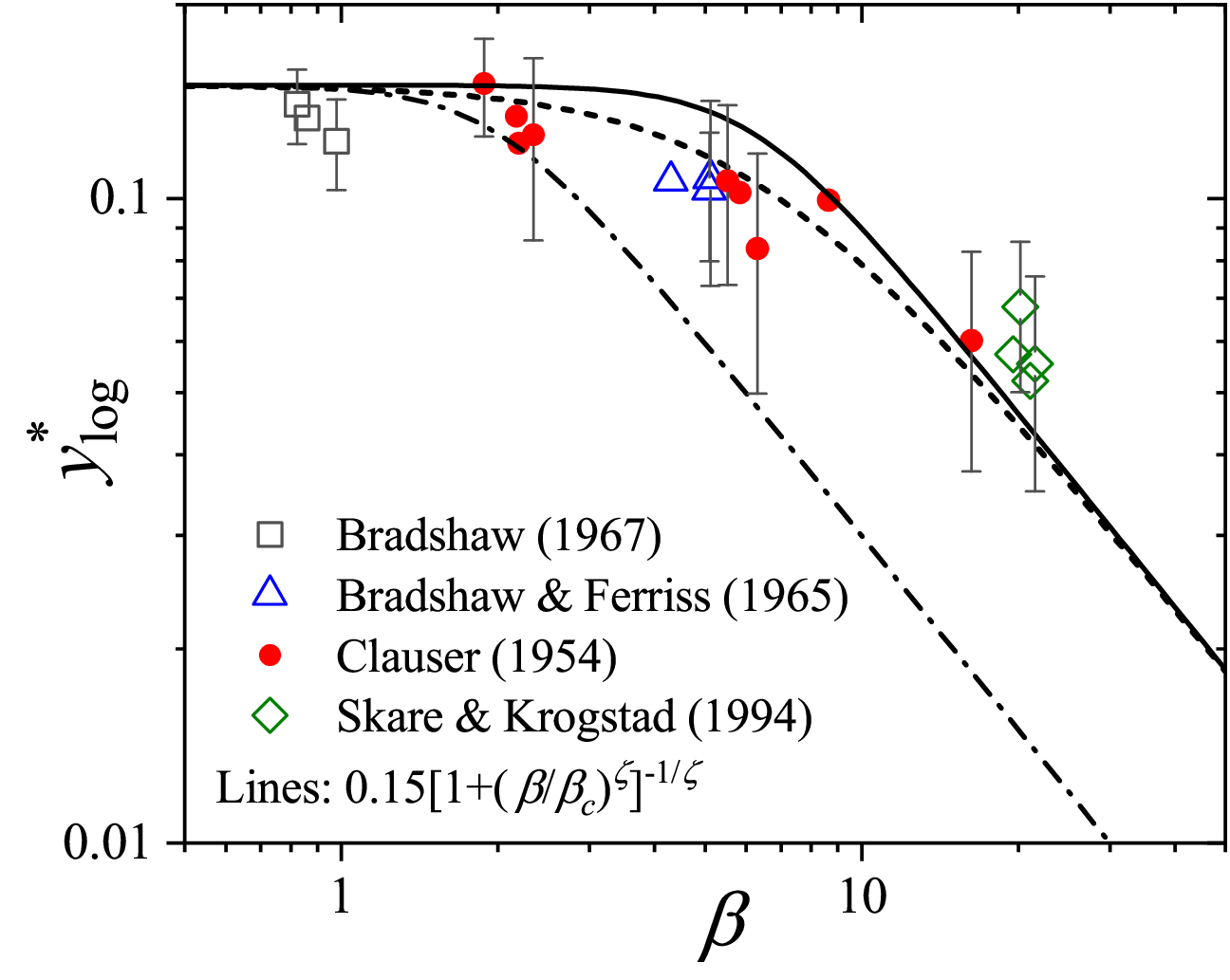}
 \caption{Variation of $y_{\rm log}^*$ with $\beta$ in equilibrium APG TBLs. Data are acquired from Figure 3 in \cite{Knopp2022}. Solid line: $\beta_c=6.2$, $\zeta=4$; short-dashed line: $\beta_c=6.2$, $\zeta=2$; dotted-dashed line: $\beta_c=2$, $\zeta=4$.}
 \label{fig:ylogstar_beta}
\end{figure}

In this study, we identify the critical Clauser parameter $\beta_c\approx6.2$, above which the logarithmic layer shrinks and transitions to the half-power-law scaling. Specifically, the upper boundary of the log-law region is formulated by $\beta_c$ through equation (\ref{eq:ylog}). This formulation is validated in Fig. \ref{fig:ylogstar_beta}, in which the data points are acquired from Knopp \cite{Knopp2022} and restricted to equilibrium APG TBLs (Bradshaw \cite{bradshaw1967turbulence,Bradshaw1965}, Clauser \cite{clauser1954turbulent}, Sk{\aa}re and Krogstad \cite{Skare1994exp}) for consistency with the present theory. As addressed by Knopp \cite{Knopp2022}, the data have been determined by visual identification of the log-law range from measured velocity profiles, since locating the exact upper bound of the log region is difficult and ambiguous, especially under APGs. The present model (equation [\ref{eq:ylog}]) aligns well with the data within measurement uncertainty, yielding a mean relative deviation of less than $15\%$. The agreement is slightly improved if the crossover steepness $\zeta=2$ rather than 4, but the improvement is marginal relative to the data uncertainty.

The data are also compared with the prediction using $\beta_c=2$. As presented in Section \ref{subsubsec:half}, in deriving $\beta_c$ from equation (\ref{eq:dudy_half2}), both $\alpha$ and $C$ are subject to significant scatterings presented in the literature, resulting in uncertainty in $\beta_c$. For example, Knopp \cite{Knopp2022} proposed $C=2/K$ with $K=0.45$, which leads to $\beta_c\approx2$ when $\alpha=0.9$. As shown in Fig. \ref{fig:ylogstar_beta}, this smaller $\beta_c$ significantly underestimates $y_{\rm log}^*$ for the strong APG case of Sk{\aa}re and Krogstad \cite{Skare1994exp}, confirming that $\beta_c=6.2$ is physically more reasonable. 

Notably, motivated by similarity and scaling arguments, Knopp \cite{Knopp2022} proposed an empirical correlation to describe the reduction of the log-law region, which is $y_{\rm log}^*=1.68{\delta^+}^{-1/2}{p^+}^{-1/5}$. Because $p^+\propto\beta/\delta^+$ for large $\beta$, Knopp's model predicts $y_{\rm log}^*\propto\beta^{-1/5}{\delta^+}^{-3/10}$ for strong APGs, in contrast to the current prediction of $y_{\rm log}^*\propto\beta^{-1}$ that does not carry explicit Reynolds-number dependence. We argue that, since the upper bound of the log region lies in the matching zone between the inner and outer layers, Reynolds-number dependence is a weak higher-order effect and should not be present as explicit scaling. More research is needed to clarify this issue.

\begin{figure}
    \centering
    \includegraphics[width=0.5\linewidth]{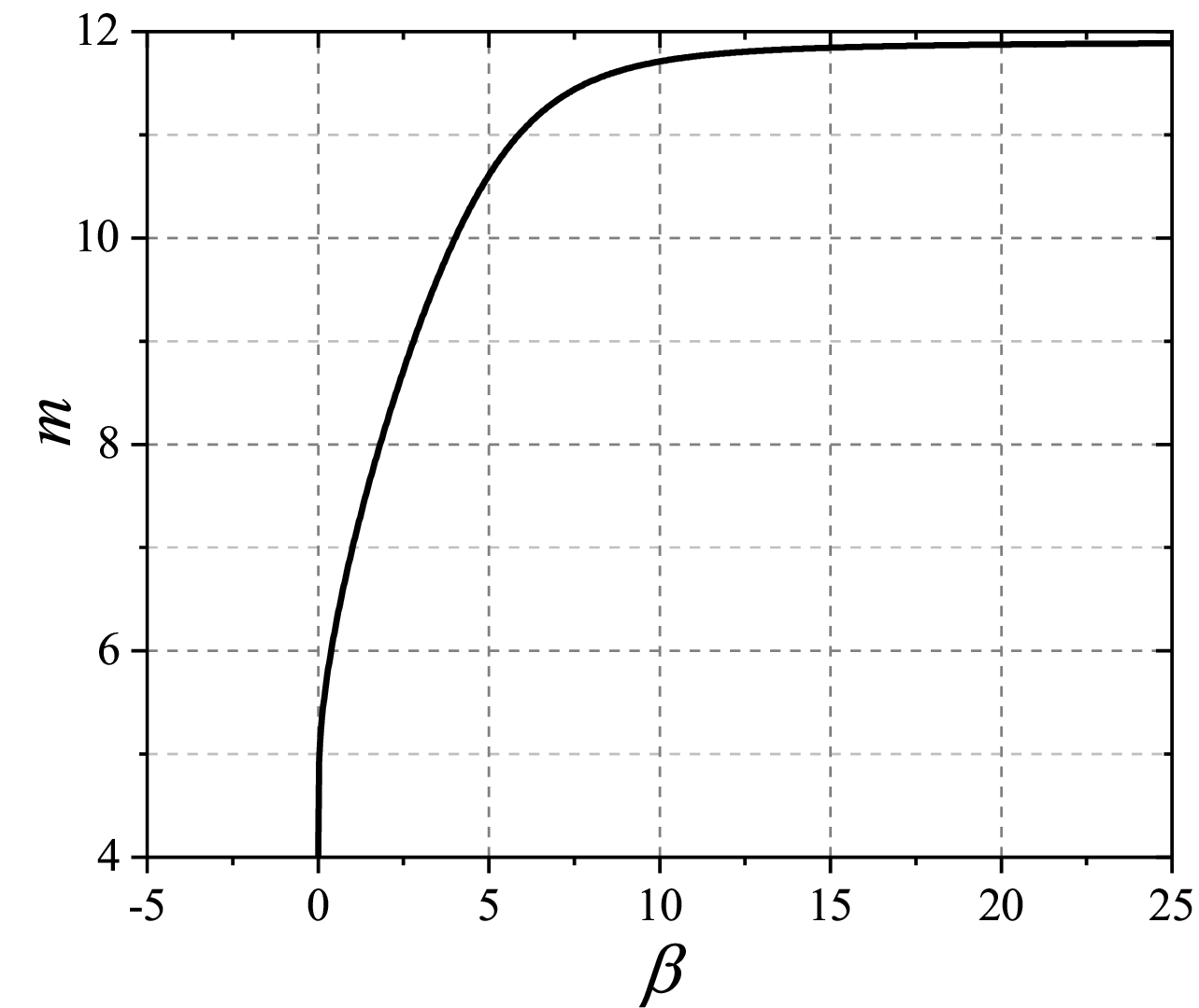}
 \caption{Variation of $m$ with $\beta$ as predicted by equation (\ref{eq:match}) for equilibrium APG TBLs.}
 \label{fig:m_beta}
\end{figure}

The mixing length profile in the outer region is formulated with a defect scaling law (equation [\ref{eq:ell_out_APG}]), in which the exponent $m$ is described by equation (\ref{eq:match}). Fig. \ref{fig:m_beta} illustrates the variation of $m$ with $\beta$. For strong APGs (e.g., $\beta>10$), $m$ quickly becomes saturated to $m_\infty$, which is slightly smaller than 12, as predicted by equation (\ref{eq:match_1}) with $\beta_c=6.2$ and $\lambda_\infty=0.074$.

The intersection of the linear asymptotic mixing-length scaling $\ell_m=m\lambda y$ and the constant wake-region scaling $\ell_m=\lambda \delta$ yields $y=\delta/m$, which serves as both the upper boundary of the half-power-law region and the lower boundary of the wake region (denoted by $y_{\rm wake}$). As $\beta$ increases, $y_{\rm wake}$ approaches $\delta/m_\infty$, which means that the upper bound of the half-power law region tends to a constant fraction of $\delta$, in agreement with Knopp \cite{Knopp2022}, who reported $y_{\rm wake}=0.21\delta$.
The present prediction is notably smaller than Knopp's value. This discrepancy stems from different definitions of $y_{\rm wake}$. Knopp empirically set $y_{\rm wake}=0.21\delta$ because the half-power law, fitted to data up to $y=0.15\delta$, accurately described the mean velocity even up to $0.21\delta$.

\subsubsection{Inner-layer thicknesses $y_s^+$ and $y_b^+$}\label{subsubsec:ys_yb}
\begin{figure}
    \centering
    \includegraphics[width=0.43\linewidth]{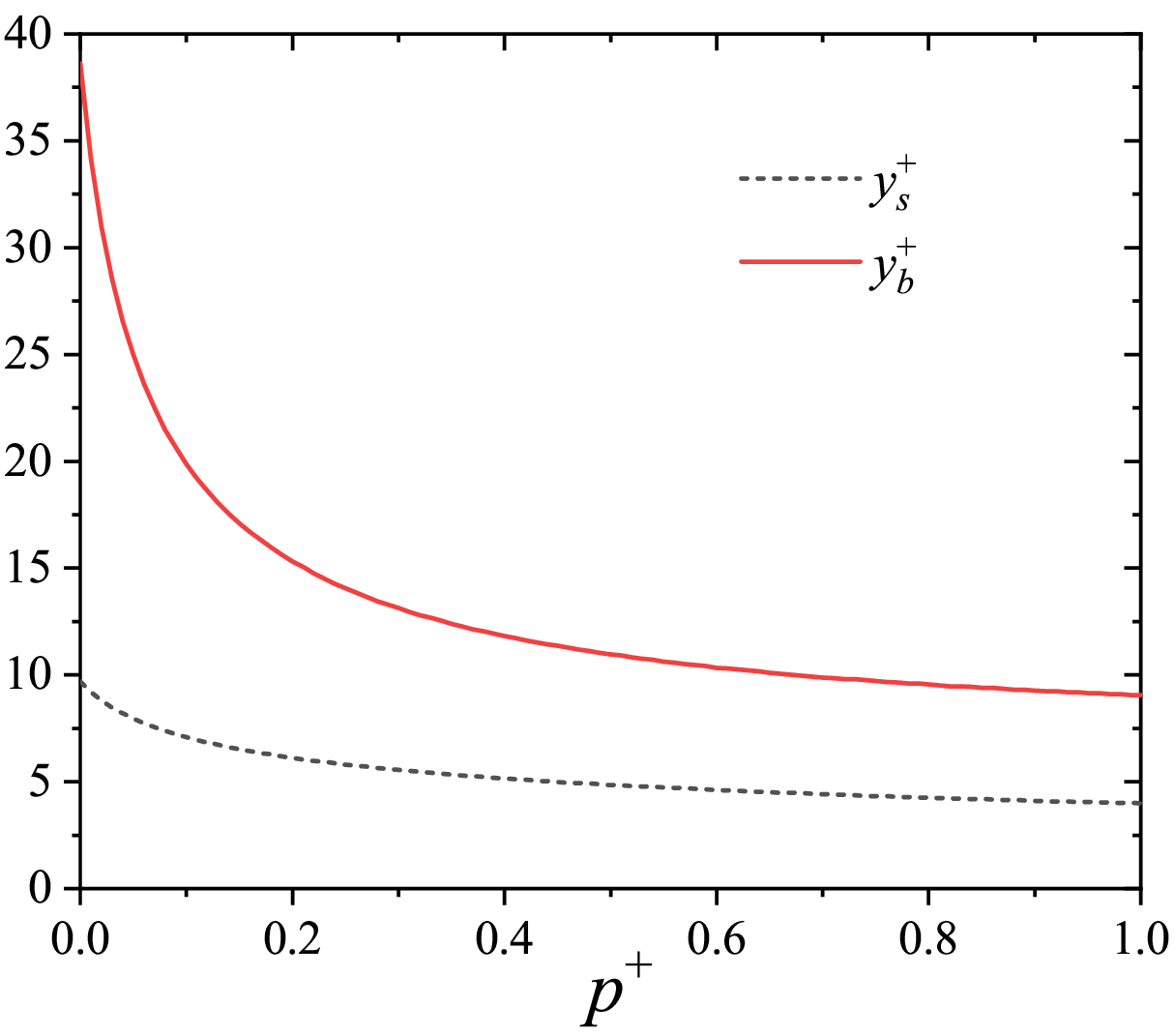}
    \includegraphics[width=0.48\linewidth]{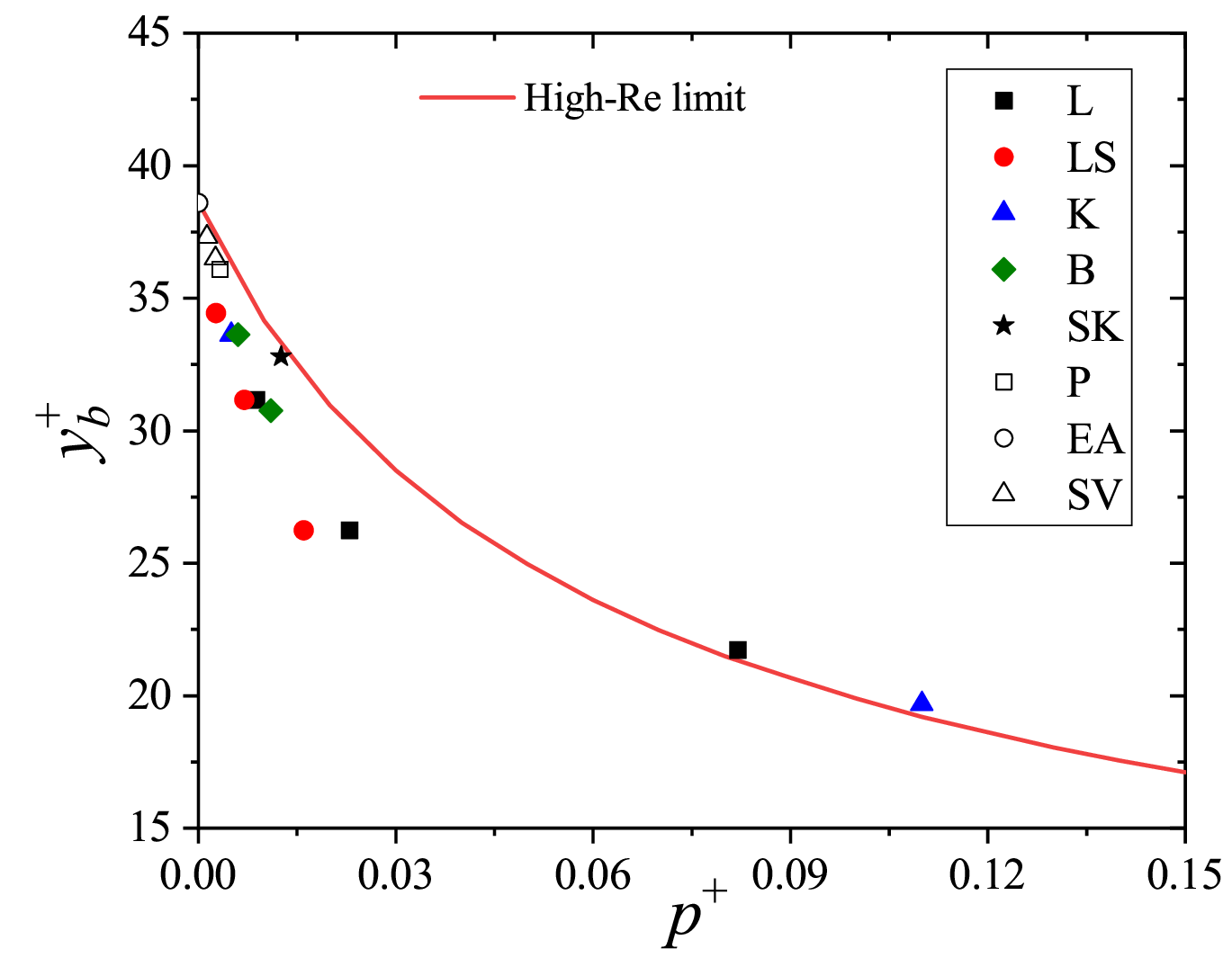}
    \\(a)\quad\quad\quad\quad\quad\quad\quad\quad\quad\quad\quad\quad\quad\quad\quad\quad\quad\quad\quad(b)\\
  \caption{(a) Variations of $y_s^+$ and $y_b^+$ with $p^+$, predicted by equation (\ref{eq:ys}) and equation (\ref{eq:yb}), respectively, for equilibrium APG TBLs. $y_b^+$ is predicted for the high-Reynolds-number limit. (b) $y_b^+$ predicted by equation (\ref{eq:yb}) for the datasets in Table \ref{tab:cases}, compared to the prediction in the high-Reynolds-number limit (solid line).}
 \label{fig:ysyb}
\end{figure}

The relationship between the viscous-sublayer thickness $y_s^+$ and $p^+$ is predicted by equation (\ref{eq:ys}) and illustrated in Fig. \ref{fig:ysyb}(a). As $p^+$ increases, $y_s^+$ decreases monotonically.  

As discussed in Section \ref{subsubsec:inner_ellm}, $y_b^+$ is more complex than $y_s^+$. However, when the Reynolds number is sufficiently large (i.e., $y_{\rm log}^+\gg4y_c^+$, and $\frac{1-r^m}{m(1-r)}\approx1$ at $r=1-4y_c^+/\delta^+$), within the integration range $0<y^+<4y_c^+$ in equation (\ref{eq:yb}), the mixing length reduces to
 \begin{equation}
  \ell_m^+=\frac{\kappa {y_s^+}^2}{y_b^+}\left(\frac{y^+}{y_s^+}\right)^{3/2}\left[1+\left(\frac{y^+}{y_s^+}\right)^4\right]^{1/8}\left[1+\left(\frac{y^+}{y_b^+}\right)^4\right]^{-1/4}\sqrt{1+\alpha p^+y^+},
  \label{eq:ell_inner_highRe}
\end{equation}
and the TSS reduces to $\tau^+=1+\alpha p^+y^+$ (which should be viewed as an equivalent TSS since $\tau^+=1+p^+y^+$ in the wall vicinity), where $\alpha\approx0.9$ \cite{Granville1989}. Therefore, in the large-Reynolds-number limit, $y_b^+$ uniquely depends on $p^+$ as $y_s^+$ does. In this limit, the variation of $y_b^+$ with $p^+$, shown in Fig. \ref{fig:ysyb}(a), is analogous to that of $y_s^+$, but there is no simple (such as linear) relationship between $y_b^+$ and $y_s^+$. Notably, both $y_s^+$ and $y_b^+$ decrease most rapidly with increasing $p^+$ in the vicinity of $p^+=0$, indicating that TBLs are sensitive to APG under the ZPG condition. At $p^+=0$, $y_b^+\approx39$, which is only slightly smaller than the SED prediction of $y_b^+=41$, meaning that Nickels' mean-velocity scaling is approximately consistent with the SED theory for ZPG TBLs.  

It should be pointed out that Nickels validated his mean-velocity scaling only within $-0.02<p^+<0.06$, which is much narrower than the range of $p^+$ presented in Fig. \ref{fig:ysyb}(a). The present study shows that Nickels' scaling remains valid at least up to $p^+=0.11$, as validated by the $\beta=39$ case from Kitsios et al. \cite{Kitsios2017}, even under finite-Reynolds-number conditions. However, for larger $p^+$, the variation of $y_b^+$ shown in Fig. \ref{fig:ysyb}(a) should be assessed with high-Reynolds-number experimental data.

In Fig. \ref{fig:ysyb}(b), we compare the $y_b^+$ predicted by equation (\ref{eq:yb}) for each dataset in Table \ref{tab:cases} with the prediction of $y_b^+$ in the high-Reynolds-number limit. The data distribution exhibits three patterns relative to the curve, as explained below. 

For high-Reynolds-number cases (SV, SK21.2, and P1.4), the predicted $y_b^+$ values collapse perfectly onto the high-Reynolds-number asymptotic curve. This behavior arises from strong scale separation between the inner and outer regions at high Reynolds numbers. Meanwhile, the TSS reaches its high-Reynolds-number asymptotic state, leaving $y_b^+$ solely determined by $p^+$, consistent with the theoretical prediction.

For cases with low-to-moderate Reynolds numbers and weak APGs ($\beta<\beta_c$, $p^+<0.03$), the $y_b^+$ data consistently lie below the asymptotic curve. In these flows, the deviations are not induced by the 
finite-Reynolds-number effects on the TSS because $\alpha p^+y^+$ contributes only a small part to $\tau^+$ within $0<y^+<4y_c^+$. For the mixing-length profile, $y_{\rm log}^+\gg4y_c^+$ is generally satisfied because $\beta<\beta_c$. However, the outer-layer term $\frac{1-r^4}{m(1-r)}$, which is approximated by $1-\frac{m-1}{2}\frac{y^+}{\delta^+}$ near the wall, departs noticeably from unity near $y^+=4y_c^+$, indicating limited scale separation. This cross-layer interference reduces the near-wall mixing length. Since $y_b^+$ is solved self-consistently to match the integrated velocity, a smaller $y_b^+$ is required to compensate, resulting in a downward shift. 

For cases with strong APGs ($\beta>\beta_c$, $p^+>0.07$) and finite Reynolds numbers (e.g., L9 and K39), the $y_b^+$ data surprisingly collapse back onto the asymptotic curve. This results from a self-canceling interplay of two effects: the near-wall penetration of the outer-layer mixing-length term and the shrunken log-law region (since $\beta>\beta_c$) both act to reduce $y_b^+$, while the suppressed TSS under underdeveloped turbulence (since $\alpha p^+y^+$ contributes crucially) acts to increase it. Their near-perfect cancellation recovers the high-Reynolds-number scaling.

\subsection{Mixing length profile}\label{subsec:ML_full_valid}
\begin{figure}
    \centering
    \includegraphics[width=0.49\linewidth]{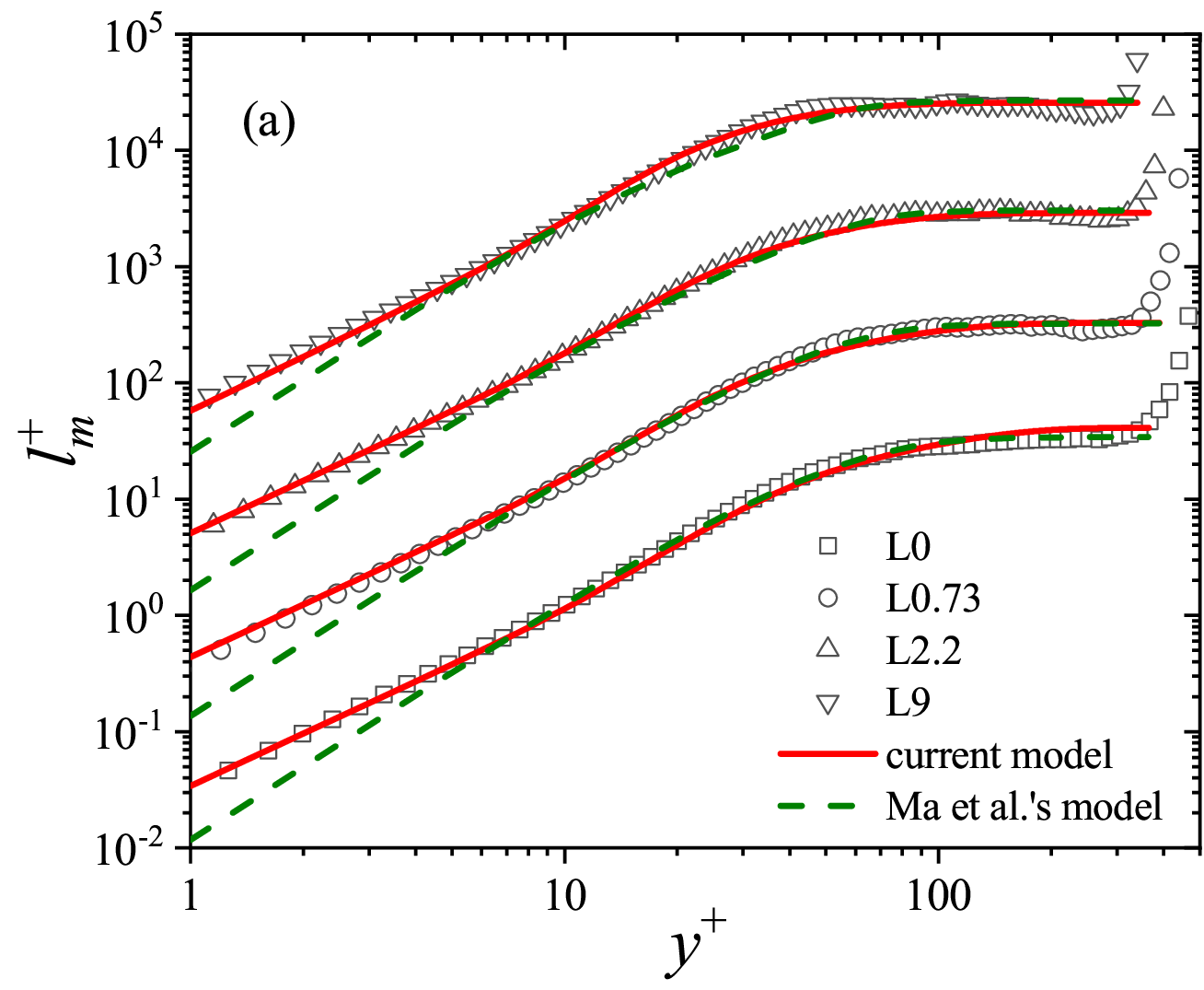}
    \includegraphics[width=0.5\linewidth]{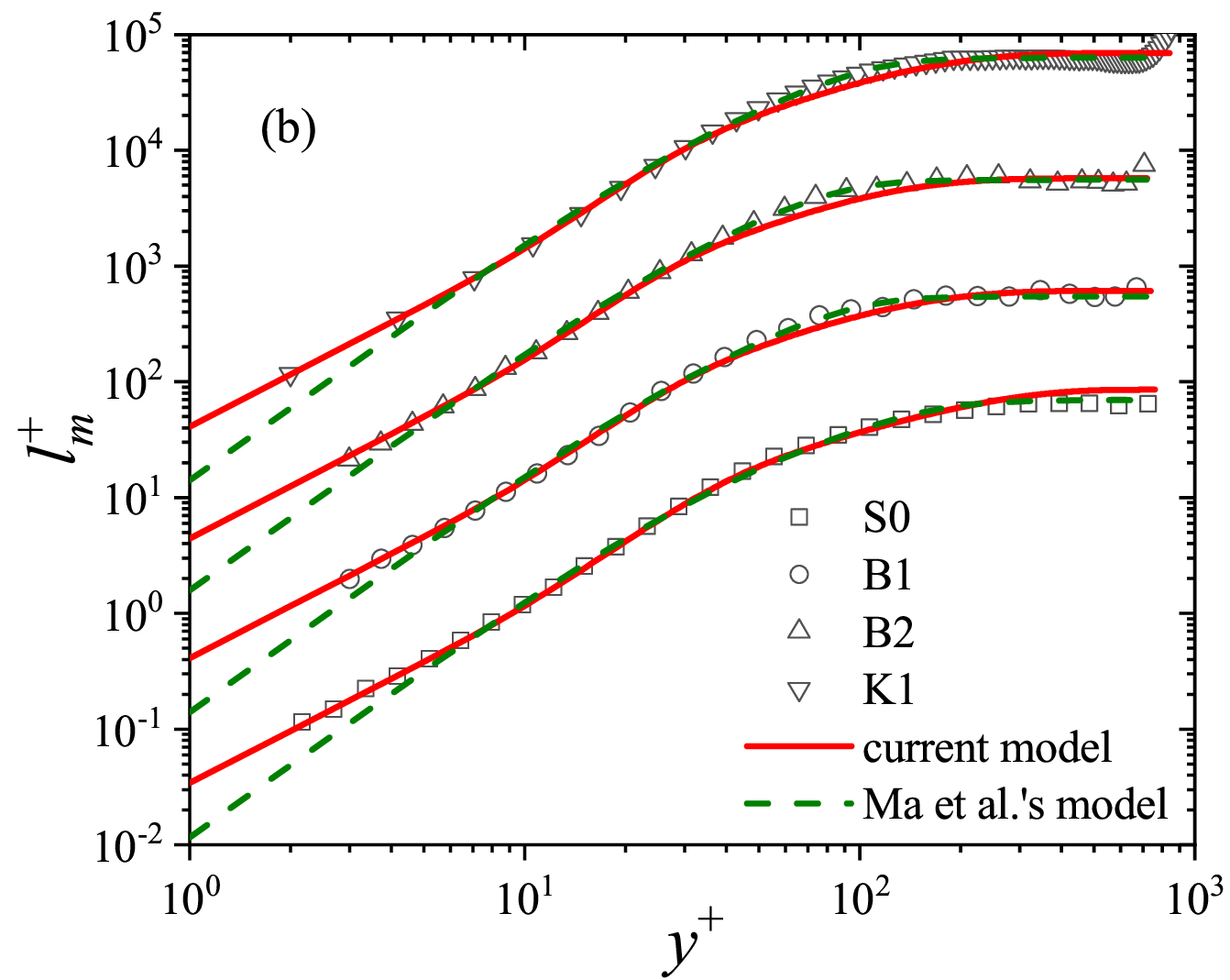}
     \includegraphics[width=0.49\linewidth]{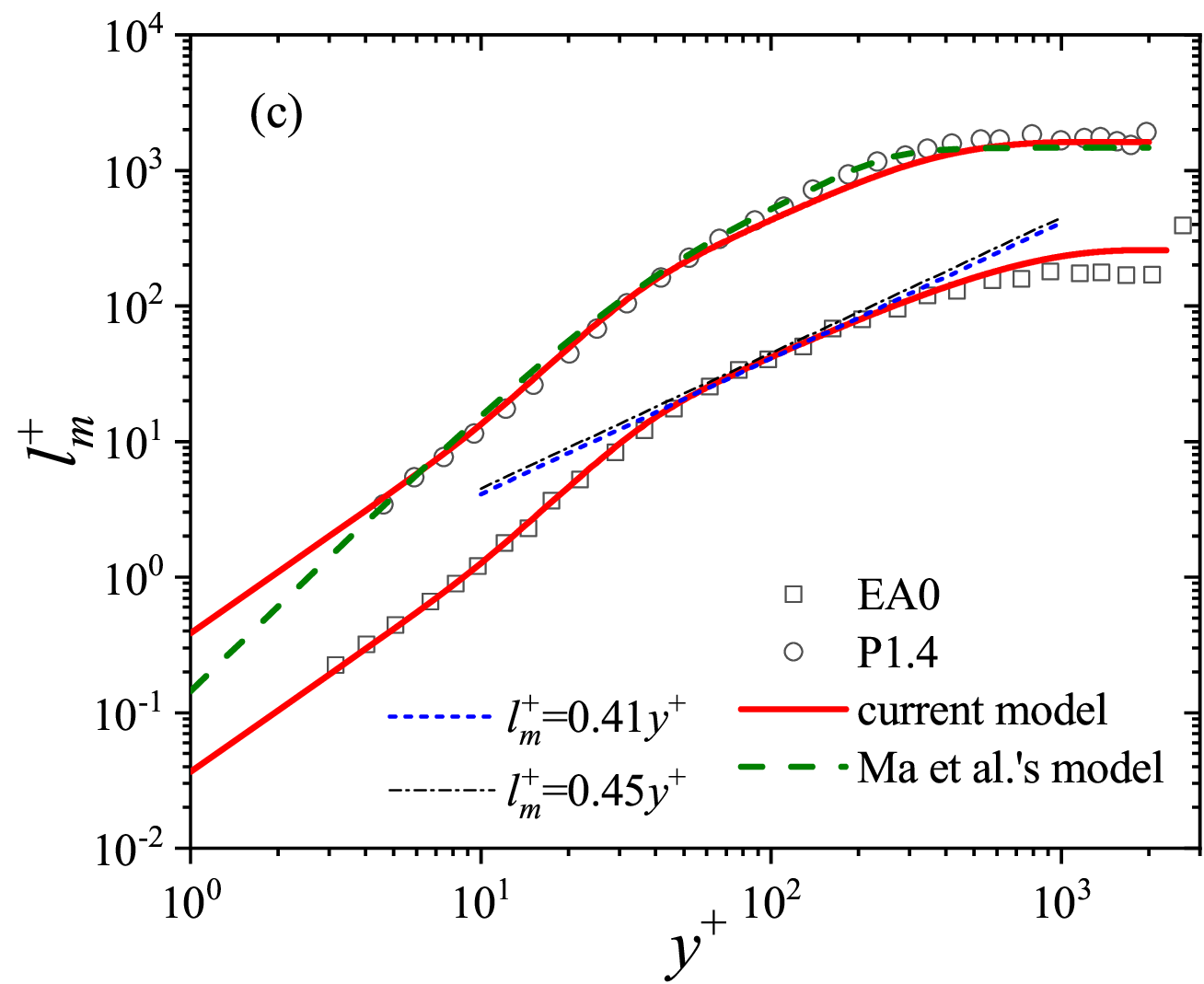}
    \includegraphics[width=0.5\linewidth]{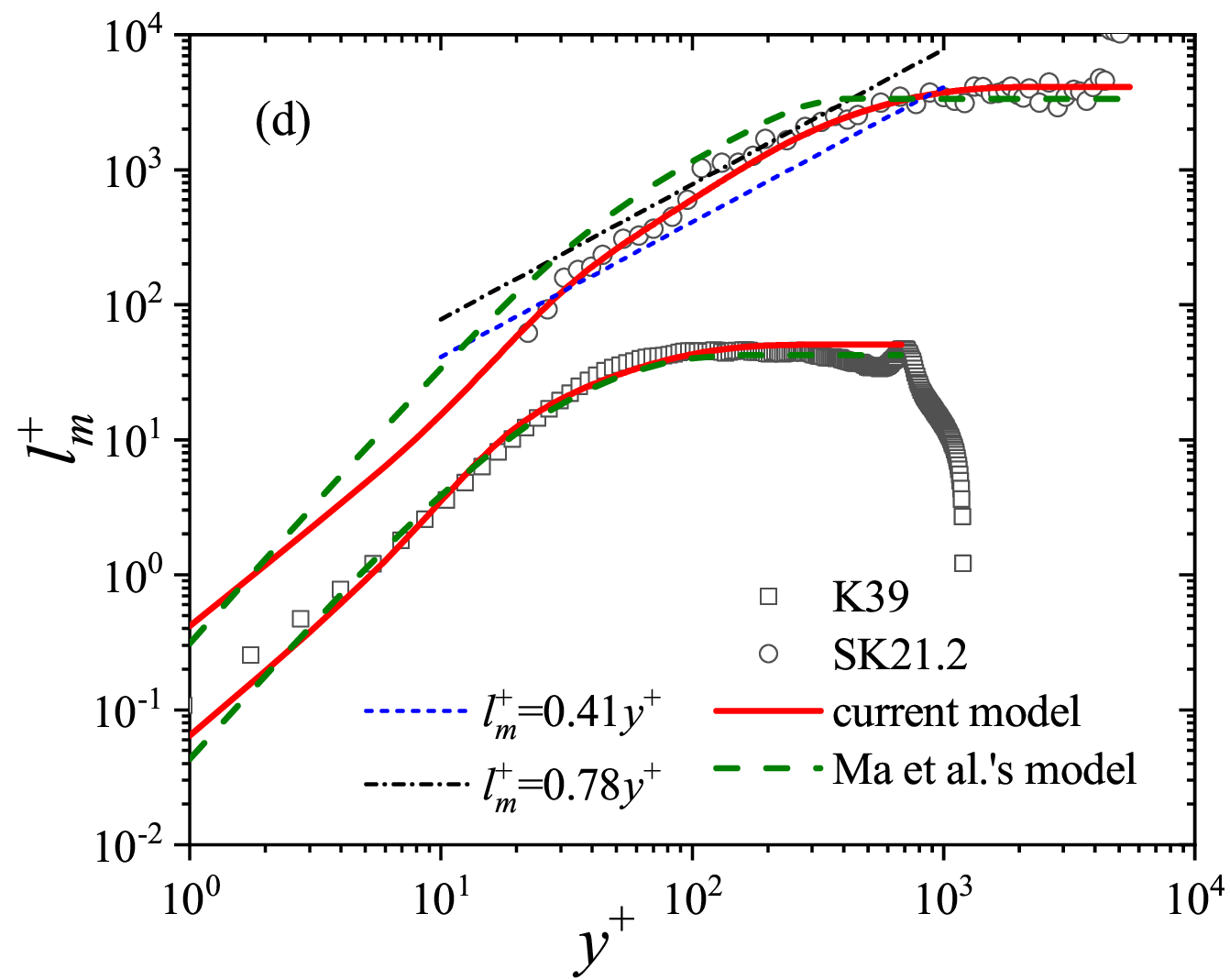}
  \caption{Profiles of mixing length in equilibrium APG TBLs. Symbols represent data from the experiments and numerical simulations in Table \ref{tab:cases}. Solid lines denote predictions from the current model (equation (\ref{eq:ell_APG})). Dashed lines denote predictions of Ma et al.'s model \cite{MaChen2026} (Appendix \ref{Append_Ma}). Curves and symbols are shifted along the vertical axis to avoid overlap.}
  \label{fig:ellm_valid}
\end{figure}

Figure \ref{fig:ellm_valid} compares the full-profile predictions of the present mixing-length model against DNS, LES, and experimental measurements across the entire dataset of equilibrium APG TBLs, spanning $Re_\theta=1250\sim50980$ and $\beta=0\sim39$. Overall, the present model yields excellent agreement with the reference data across the full parameter space, accurately capturing the multi-layer scaling behavior of the mixing length: from the viscous sublayer adjacent to the wall, through the buffer layer and log-law region, to the constant-$\ell_m$ wake region near the boundary layer edge.

For comparison, we evaluate the recent full-profile mixing-length model proposed by Ma et al. \cite{MaChen2026} (Appendix \ref{Append_Ma}), which was developed for both equilibrium and non-equilibrium APG TBLs. Three key characteristics of this model are observed across our test cases. First, it shows consistent deviations in the viscous sublayer, as its modified van Driest-type damping function yields a near-wall asymptotic scaling of $\ell_m^+\propto {y^+}^{2.1}$. This scaling corresponds to the buffer layer scaling in our multi-layer framework, rather than the required $\ell_m^+\propto {y^+}^{3/2}$ scaling for the viscous sublayer. Second, the model uses two flow-dependent fitting parameters, $b$ and $n$, to modulate the outer-region mixing length, achieving prediction accuracy comparable to our model in the outer region for low-to-moderate Reynolds-number cases. Third, it significantly overestimates the mixing length below the wake region for the high-Reynolds-number, strong-APG experimental case SK21.2, indicating its failure to describe high-Reynolds-number near-separation flows. In contrast, the present symmetry-based model maintains consistent, high-fidelity performance across the full parameter range without case-by-case empirical tuning, demonstrating its robust universality for equilibrium APG TBLs from ZPG to incipient separation conditions.

In addition to the inter-model comparison, Fig. \ref{fig:ellm_valid}(c) further interrogates the scaling behavior of the mixing length in the logarithmic region for the ZPG case EA0, where we overlay the canonical linear mixing-length scaling $\ell_m^+=\kappa y^+$ with $\kappa$ equaling, respectively, the canonical value 0.41 and the SED proposal 0.45 for reference. While the present model adopts $\kappa=0.45$ from the SED theory, both the inferred mixing-length data in the logarithmic region and our model predictions align more closely with the $\ell_m^+=0.41y^+$ scaling for this moderate-Reynolds-number case. This apparent discrepancy does not invalidate the $\kappa=0.45$ adopted in our framework. Instead, it demonstrates that $\kappa=0.45$ is a global, intrinsic flow parameter defined for the full boundary-layer profile in the SED theory, rather than a local fitting parameter restricted to the logarithmic region. The apparent fitted value of $\kappa=0.41$ is inevitably modulated by finite-Reynolds-number effects, and thus cannot be used to judge the validity of the intrinsic global K\'{a}rm\'{a}n constant. Only under asymptotically high-Reynolds-number conditions, where the scale separation between the inner and outer layers is fully developed, can the intrinsic global K\'{a}rm\'{a}n constant $\kappa=0.45$ be explicitly resolved in numerical or experimental observations.

For the high-Reynolds-number, strong-APG case SK21.2 shown in Fig. \ref{fig:ellm_valid}(d), we compare the measured mixing-length profile and predictions from the present model against two reference linear scalings: the canonical ZPG mixing-length law $\ell_m^+=0.41y^+$, and $\ell_m^+=0.78y^+$ — the fitted linear scaling for this specific case reported by Sk{\aa}re and Krogstad \cite{Skare1994exp}. Two distinct APG-induced modifications to the overlap-region mixing length are clearly resolved. First, the strong APG drastically elevates the apparent K\'{a}rm\'{a}n constant for the mixing-length linear law (denoted $\kappa_\ell$ in \cite{Skare1994exp}) from the canonical ZPG value of 0.41 to 0.78. Second, the wall-normal extent of the region where this linear scaling holds is markedly narrower than in the ZPG and weak-APG cases.

The present model fully and accurately captures these non-canonical behaviors under high-Reynolds-number, near-separation conditions via three coupled, physics-driven mechanisms: (1) the $\sqrt{\tau^+}$-scaling for the logarithmic-region mixing length, which preserves the invariant intrinsic K\'{a}rm\'{a}n constant $\kappa=0.45$ while accounting for APG-augamented TSS; (2) the critical Clauser parameter $\beta_c=6.2$, which quantifies the onset of logarithmic-layer shrinkage as APG intensifies; and (3) a symmetry-preserving smooth crossover from the log law to the half-power-law scaling, which characterizes the progressive breakdown of the logarithmic region in the near-separation limit.

These results for the mixing-length profile directly underpin the accurate predictions of the mean velocity and Reynolds shear stress profiles, which are presented in the following section.

\subsection{Mean velocity and Reynolds shear stress profiles}\label{subsec:MVP}

\begin{figure}
    \centering
    \includegraphics[width=0.49\linewidth]{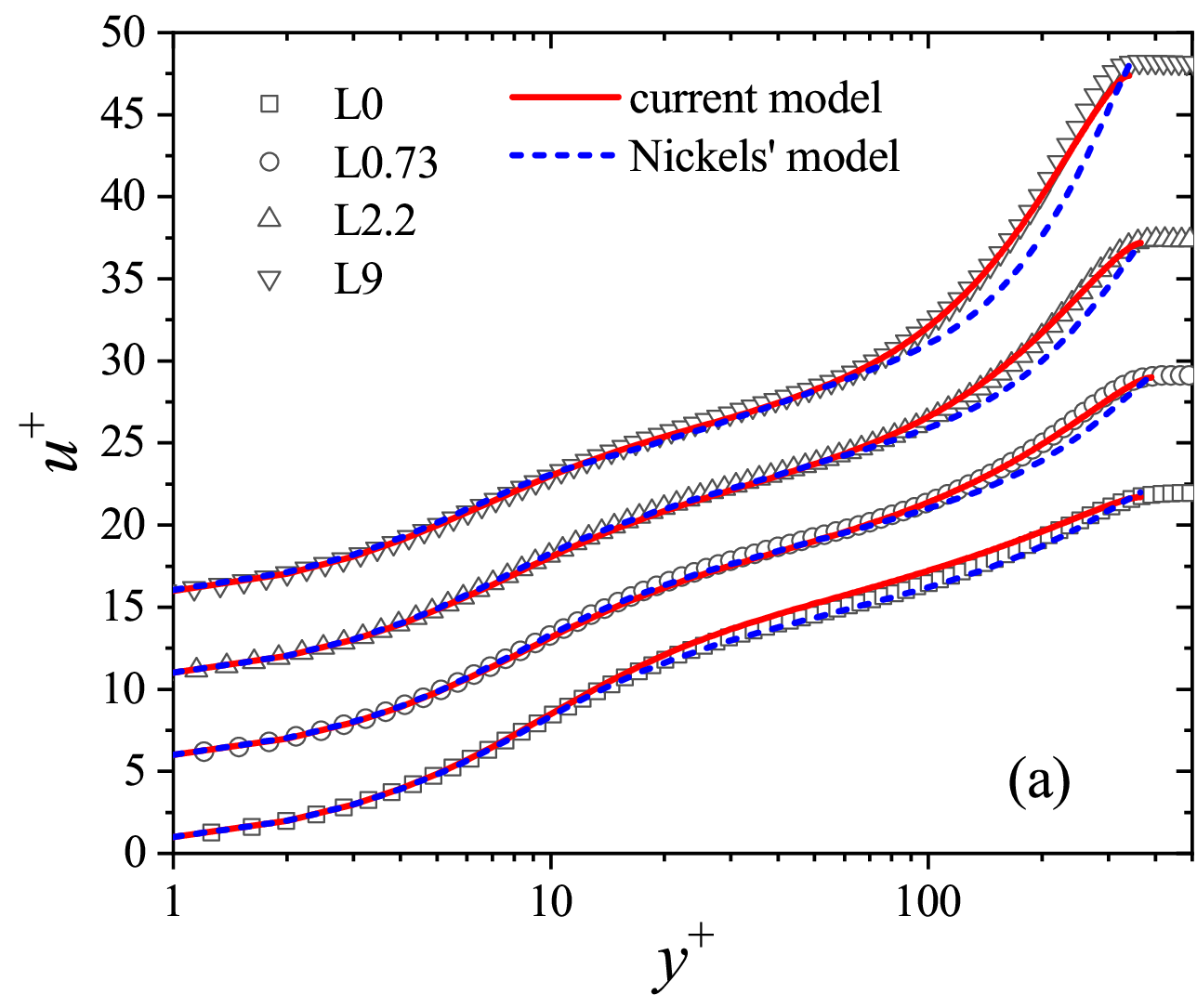}
    \includegraphics[width=0.5\linewidth]{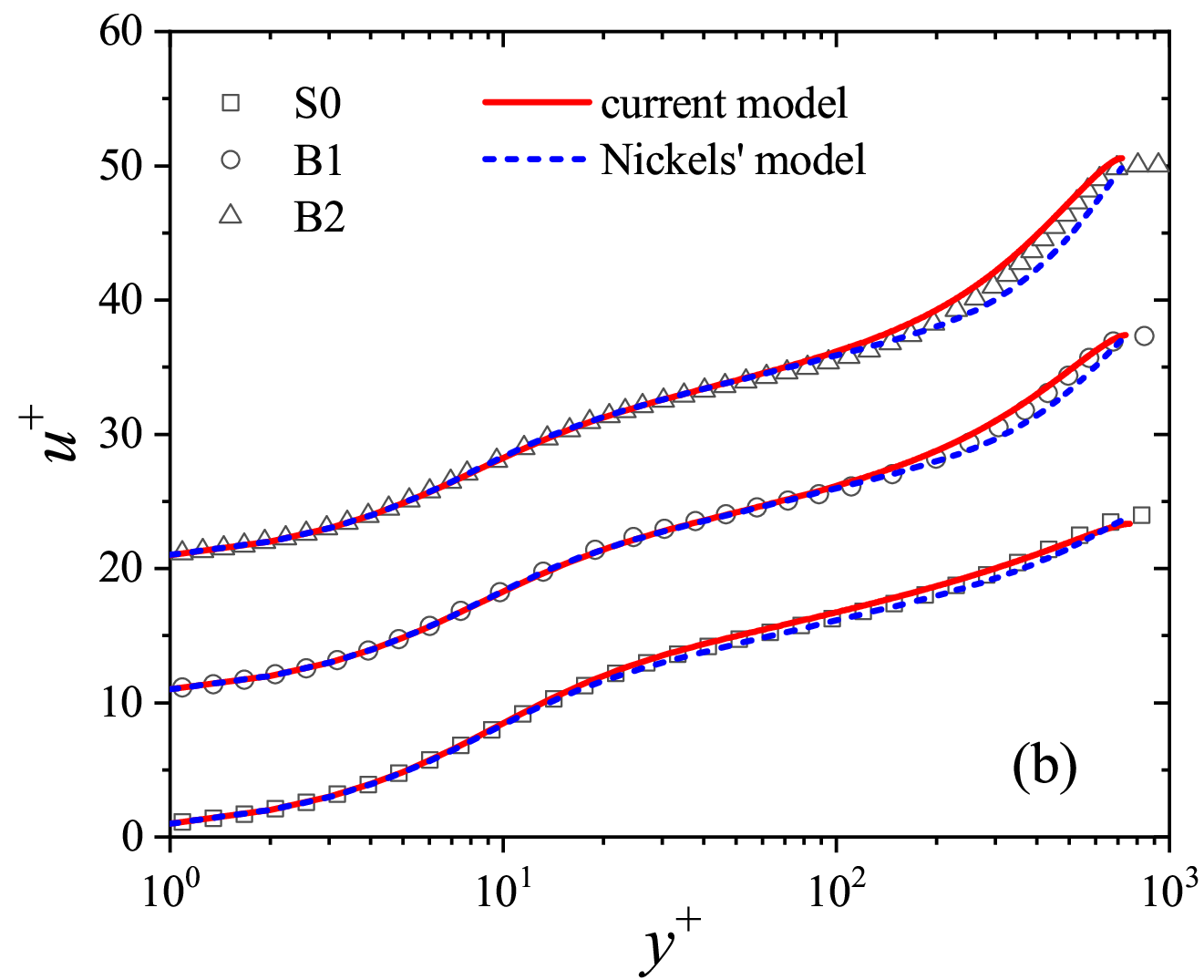}
     \includegraphics[width=0.49\linewidth]{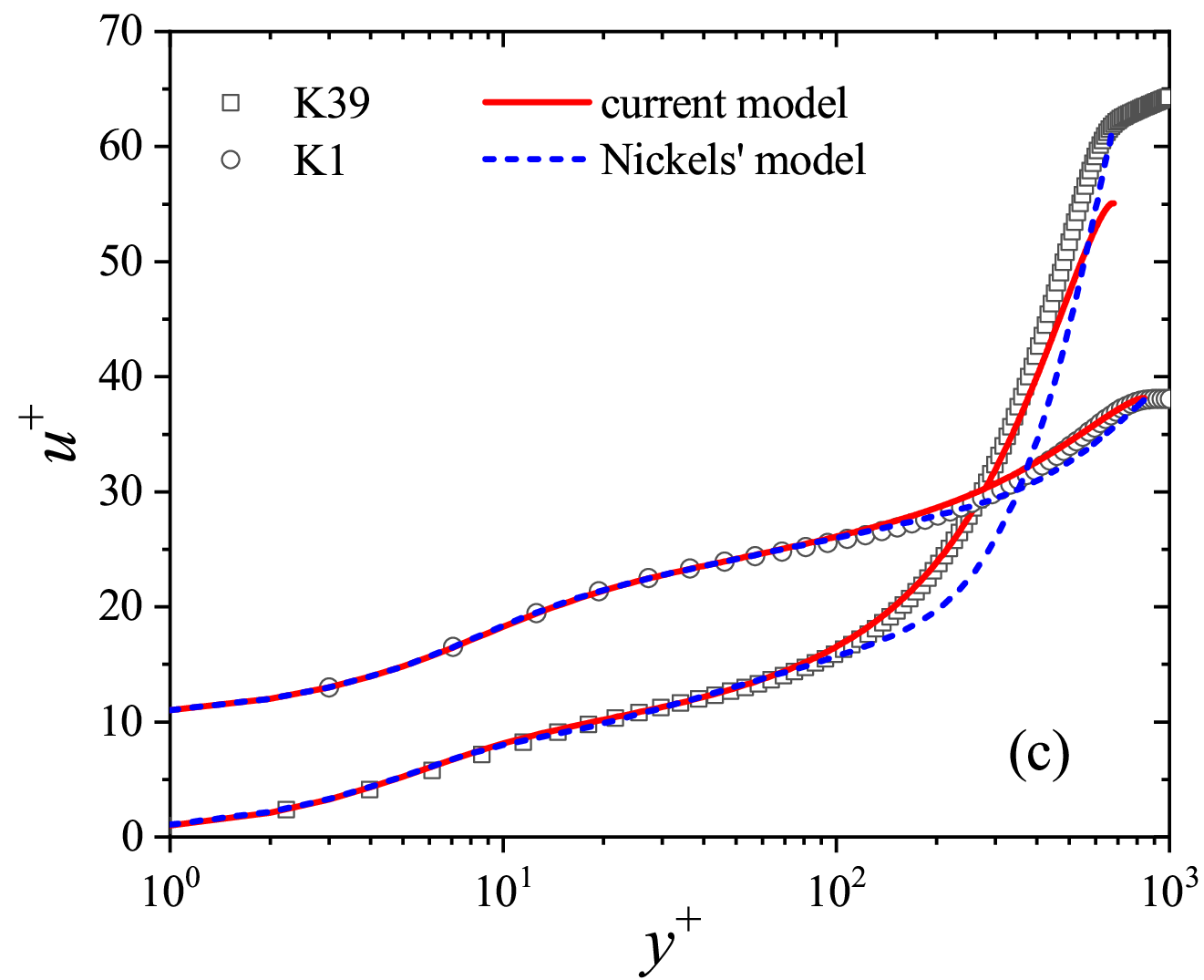}
    \includegraphics[width=0.49\linewidth]{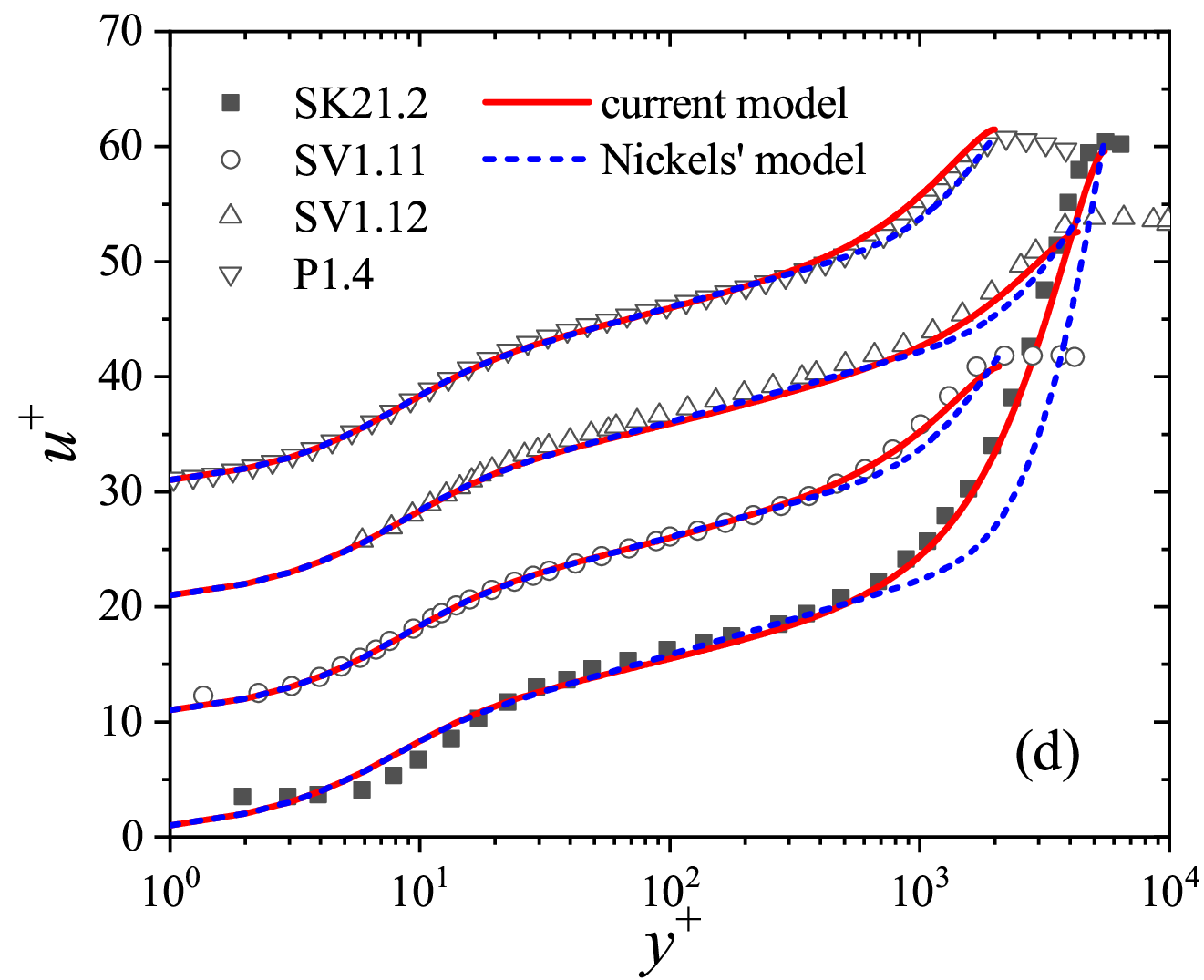}
  \caption{Profiles of streamwise mean velocity in equilibrium APG TBLs. Symbols represent data from the experiments and numerical simulations in Table \ref{tab:cases}. Solid lines denote predictions by equation (\ref{eq:u_model}), where $\tau^+$ and $\ell_m^+$ are predicted by (\ref{eq:tau_APG}) and (\ref{eq:ell_APG}), respectively. Short-dashed lines denote Nickels' model \cite{Nickels2004} (Appendix \ref{Append_Nickels}). Curves and symbols are shifted along the vertical axis to avoid overlap.}
  \label{fig:velocity_valid}
\end{figure}

Figure \ref{fig:velocity_valid} compares the full-profile mean-velocity predictions from the present model against DNS, LES, and experimental measurements across the entire equilibrium APG TBL dataset, alongside Nickels's mean-velocity model \cite{Nickels2004} (Appendix \ref{Append_Nickels}). Overall, the present model yields excellent agreement with the reference data across the full wall-normal domain for all test cases. The only notable systematic deviation occurs in the outer region of case K39, where the present model slightly underpredicts the mean velocity relative to the DNS data. This discrepancy can be fully explained by the finite-Reynolds-number effect on the wake-region mixing-length parameter $\lambda$. The K39 case features an extremely strong APG ($\beta=39$) but a limited Reynolds number ($Re_\theta=10000$). Under such conditions, the finite-Reynolds-number effect suppresses the development of large-scale turbulent eddies in the outer region, leading to an actual $\lambda$ significantly lower than the high-Reynolds-number asymptotic scaling adopted in the present model (Eq. [\ref{eq:Pi_model}]). The overestimation of the outer-region mixing length reduces the predicted mean velocity gradient ${\rm d}u^+/{\rm d}y^+$, and the wall-normal integration of this underestimated gradient ultimately yields a lower mean velocity profile in the outer region, consistent with the observed deviation.

Nickels' mean-velocity model exhibits robust performance across all test cases, especially in the inner region. In the present framework, the viscous sublayer thickness $y_s^+$ is derived from the critical sublayer thickness $y_c^+$ defined by Nickels, and the buffer layer thickness $y_b^+$ is solved self-consistently by matching the universal inner-layer scaling of Nickels' model at $y^+=4y_c^+$. Owing to this construction of the inner-layer parameters, the predictions of the present model and Nickels' model are nearly identical in the inner region. In the outer region, Nickels' model modulates the wake velocity distribution via an empirical fitting parameter $c$, which has a physical role analogous to Coles' wake parameter $\Pi$. For a fair and rigorous comparison, we determine the value of $c$ in Nickels' model by enforcing an exact match between the predicted velocity at the boundary-layer edge and the experimental or numerical data at that location. With this setup, Nickels' model systematically underestimates the mean velocity in the region above the overlap layer, as shown in Fig. \ref{fig:velocity_valid}.

\begin{figure}
    \centering
    \includegraphics[width=0.48\linewidth]{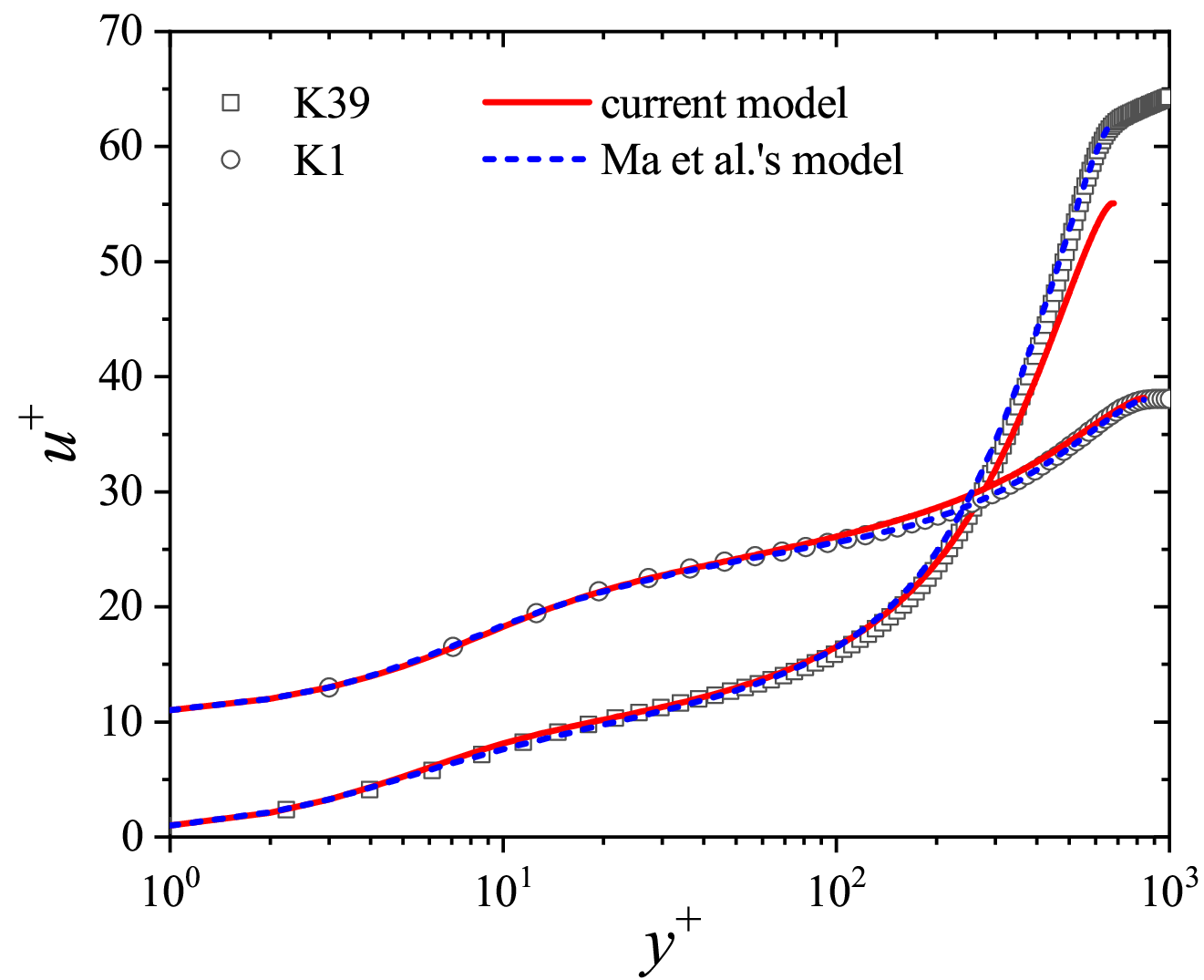}
    \includegraphics[width=0.48\linewidth]{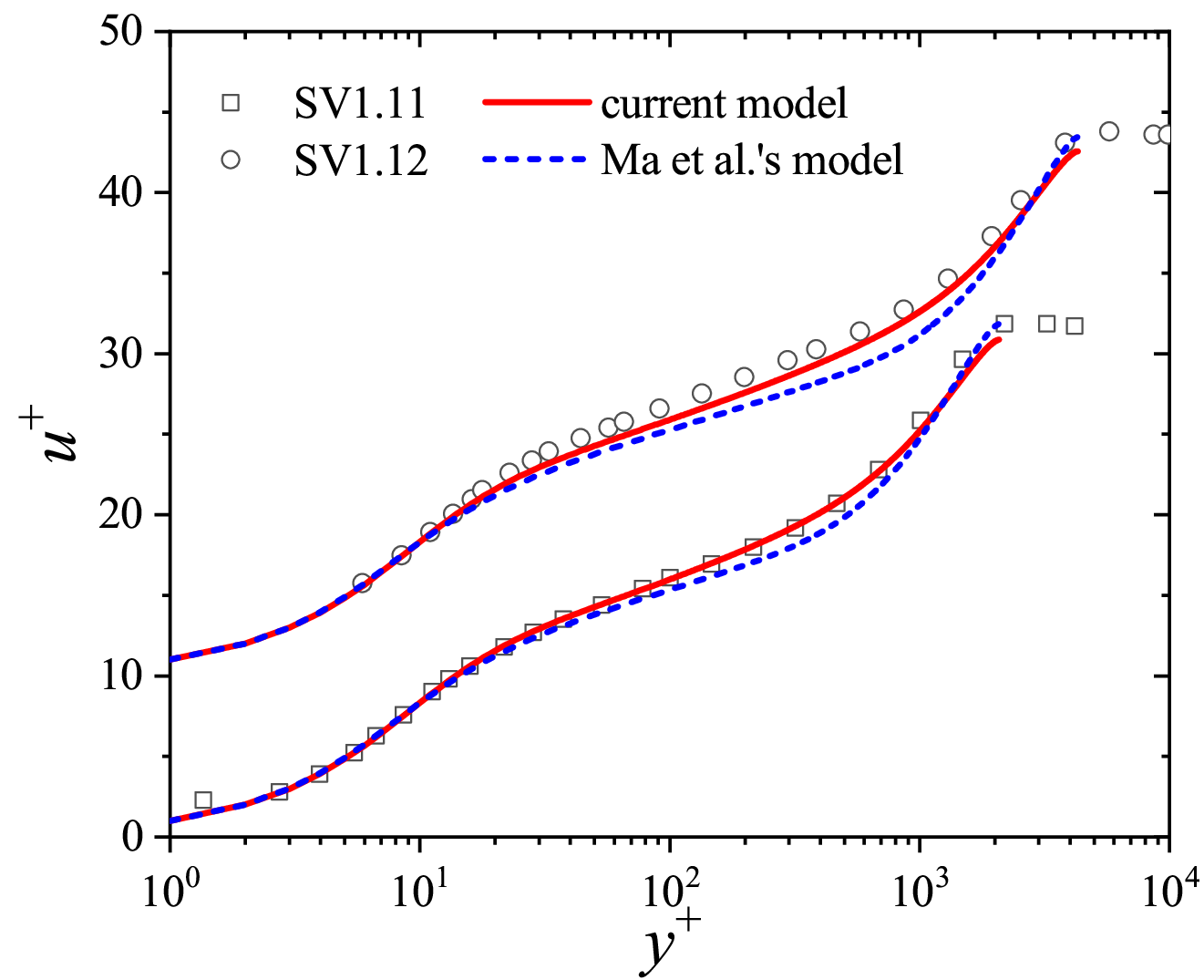}
    \\(a)\quad\quad\quad\quad\quad\quad\quad\quad\quad\quad\quad\quad\quad\quad\quad\quad\quad(b)\\
 \caption{Profiles of streamwise mean velocity for (a) cases K1 and K39, and (b) cases SV1.11 and SV1.12. Solid lines denote predictions by the current model. Short-dashed lines denote predictions by Ma et al.'s mixing-length model. The SK21.2 case is not shown, as Ma et al.'s model produces nonphysical results under this strong APG, high-Reynolds-number condition.}
 \label{fig:K_u_Ma}
\end{figure}

To further benchmark the predictive performance of the present model against the state-of-the-art mixing-length formulation for APG TBLs, Fig. \ref{fig:K_u_Ma} compares the mean velocity predictions of the present model with those from the mixing-length model proposed by Ma et al. \cite{MaChen2026} (Appendix \ref{Append_Ma}). To isolate the performance of the mixing-length formulation itself, we conduct a controlled comparison in which the mean velocity profile for  Ma et al.'s model is computed using the present symmetry-based TSS model (Eq. [\ref{eq:tau_APG}]), rather than their original TSS formulation. This setup ensures that differences in velocity predictions arise solely from the mixing-length closure. The original TSS model in Ma et al.'s framework is valid only for the inner region of the boundary layer, and imposes a nonphysical constant TSS in the outer region; it relies on tuning the two outer-region mixing-length parameters, $b$ and $n$, to compensate for this inherent defect in the TSS formulation.

As shown in Fig.~\ref{fig:K_u_Ma}(a), with case-by-case tuning of the flow-dependent parameters $b$ and $n$, the mixing-length model of Ma et al. achieves accurate predictions for the two finite-Reynolds-number cases: the weak-APG case K1 ($\beta=1$, $Re_\theta=3440$) and the extreme-APG case K39 ($\beta=39$, $Re_\theta=10000$), closely matching the DNS data across the full wall-normal profile. However, when applied to the higher-Reynolds-number experimental cases SV1.11 and SV1.12 ($\beta\approx1.1$, $Re_\theta=9060$ and $19780$, respectively) shown in Fig.~\ref{fig:K_u_Ma}(b), the model of Ma et al. systematically underestimates the mean velocity in the logarithmic overlap region and the overlying wake region, even with optimized tuning of parameters $b$ and $n$.

This underestimation becomes markedly more severe for the high-Reynolds-number strong-APG case SK21.2 ($\beta=21.2$, $Re_\theta=50980$), which is not shown in Fig.~10 as the model produces nonphysical velocity predictions. This failure is fully consistent with the behavior observed in Fig.~\ref{fig:ellm_valid}(d), where Ma et al.'s model significantly overestimates the mixing length in the inner and overlap regions of the SK21.2 case: an overestimated mixing length leads to an underestimated mean velocity gradient, which, upon wall-normal integration, results in unphysical, severely underpredicted velocity profiles in the overlap and wake regions.

Owing to the inherent uncertainty in the determination of the critical Clauser parameter $\beta_c$, which arises from the scatter in the half-power-law coefficient $C$ and the effective-APG coefficient $\alpha$ reported in the literature, we further investigate the influence of $\beta_c$ on the prediction of the mean velocity profile via three diagnostic functions for APG TBLs. These diagnostic functions are specifically designed to interrogate the scaling behavior in distinct regions of the boundary layer: (1) $y^+ \frac{{\rm d}u^+}{{\rm d}y^+}$, the canonical diagnostic for the logarithmic region, which exhibits a constant plateau equal to $1/\kappa$ when the log law holds; (2) $\sqrt{y^+} \frac{{\rm d}u^+}{{\rm d}y^+}$, the diagnostic for the half-power-law region, which displays a flat plateau under strong APG conditions where the half-power law dominates the overlap region; and (3) the mean velocity gradient $\frac{{\rm d}u^+}{{\rm d}y^+}$, which serves as the diagnostic for the quasi-linear velocity region in the wake. Near the peak of the TSS profile, both the TSS and the mixing length reach a near-constant plateau, leading to a broad region of approximately constant velocity gradient (quasi-linear mean velocity profile) in the wake, which can be clearly resolved via the $\frac{{\rm d}u^+}{{\rm d}y^+}$ diagnostic.

\begin{figure}
    \centering
    \includegraphics[width=0.4\linewidth]{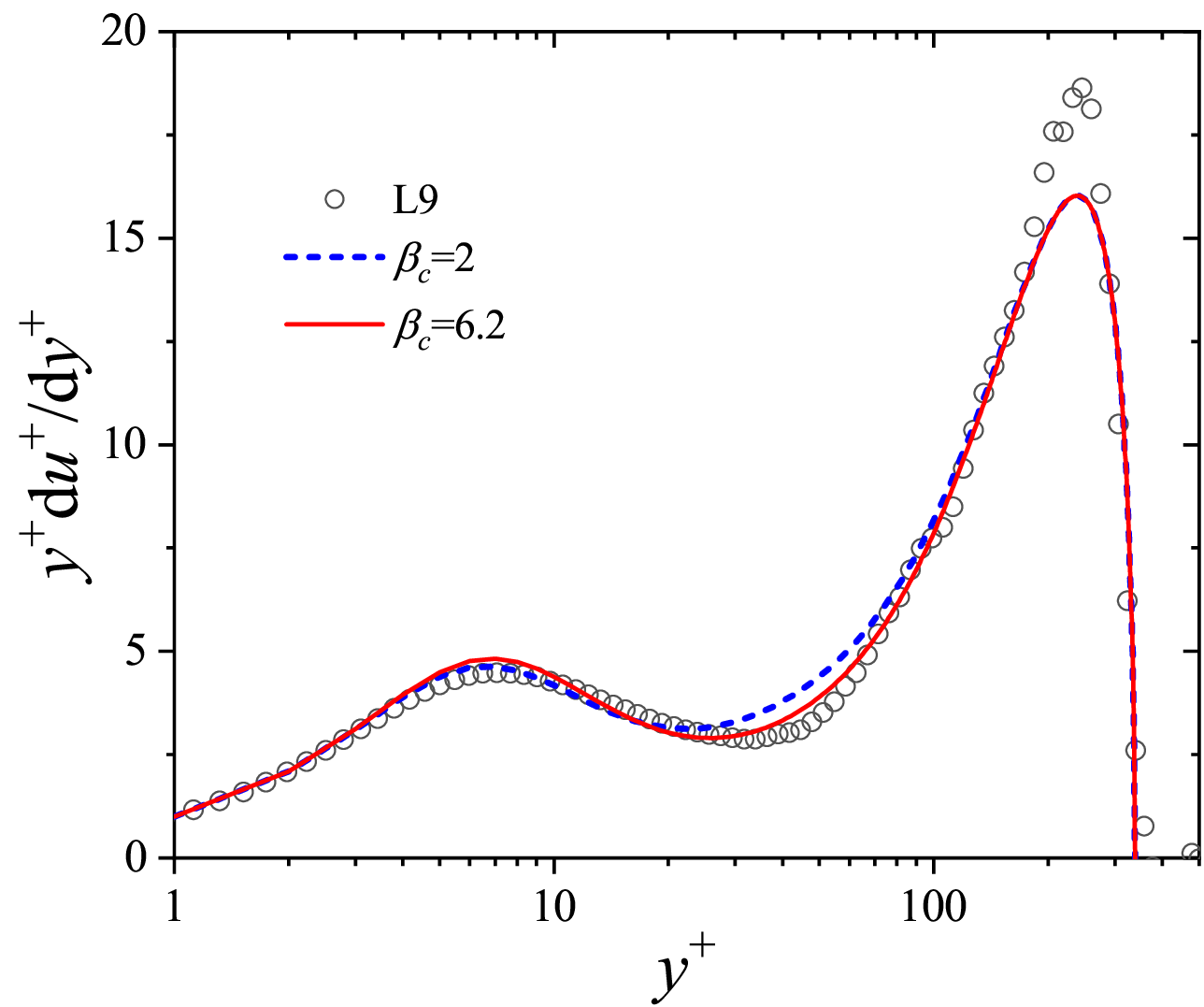}
    \includegraphics[width=0.4\linewidth]{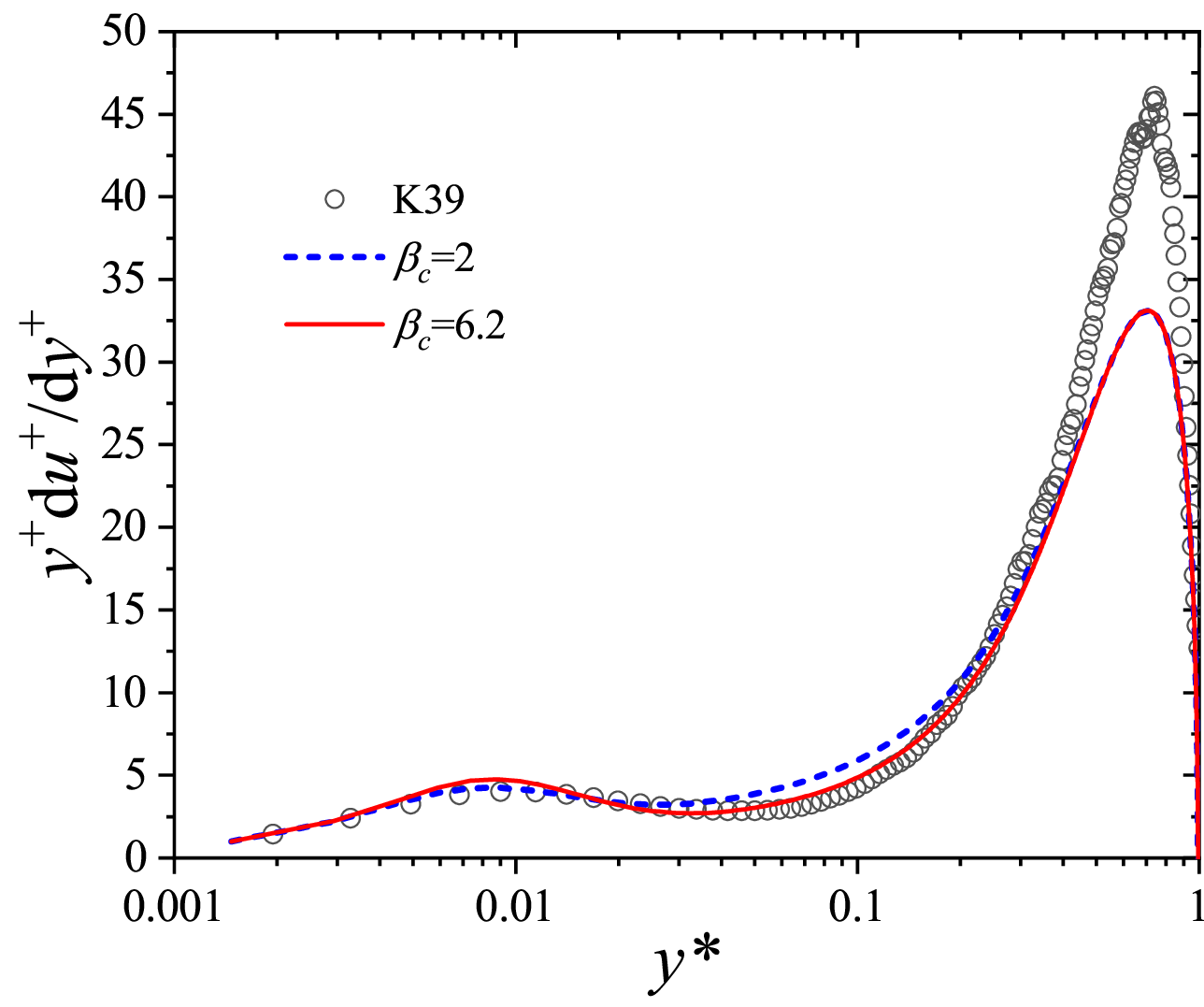}
    \\(a)\quad\quad\quad\quad\quad\quad\quad\quad\quad\quad\quad\quad\quad\quad\quad\quad\quad(b)\\
    \includegraphics[width=0.4\linewidth]{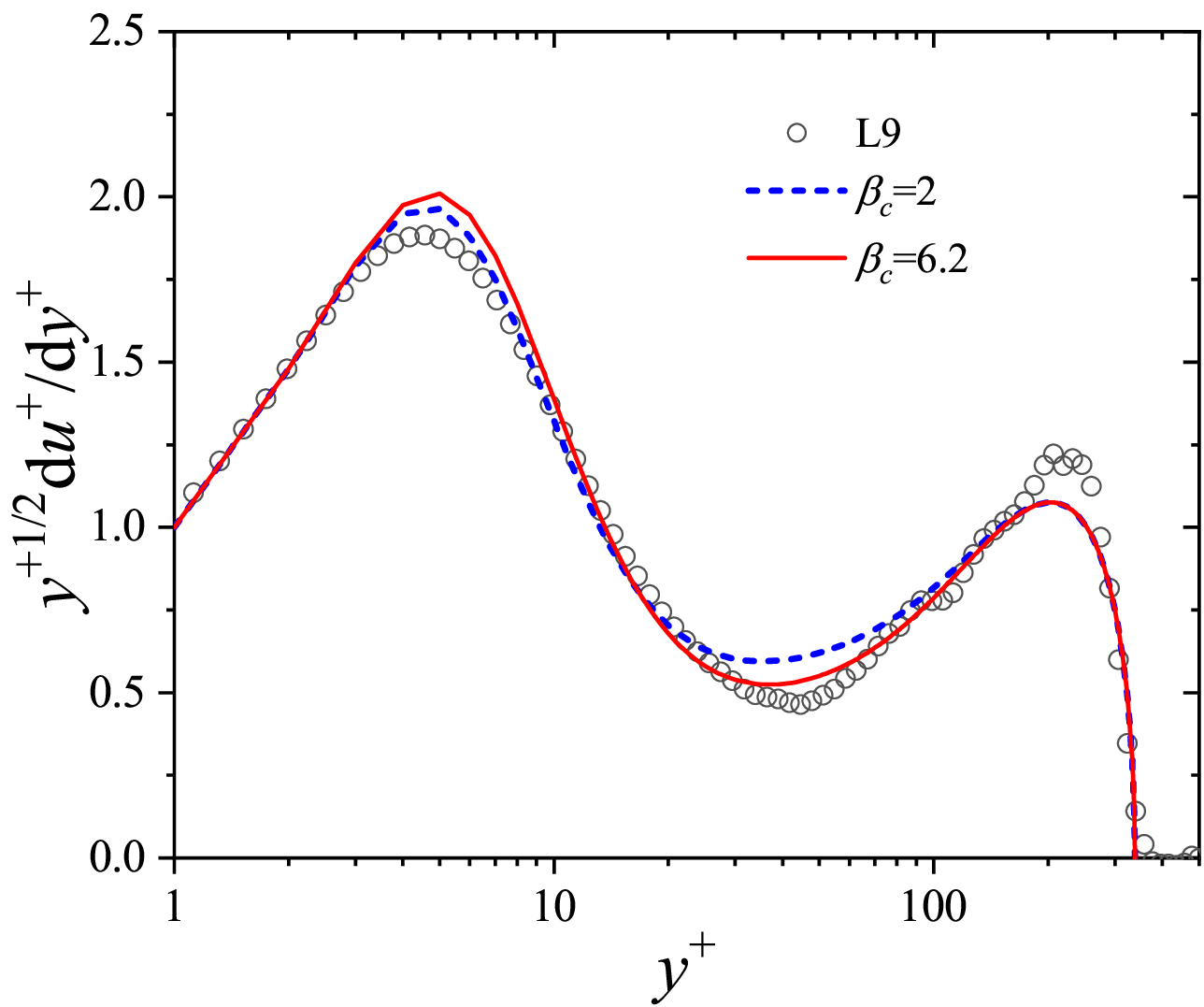}
    \includegraphics[width=0.4\linewidth]{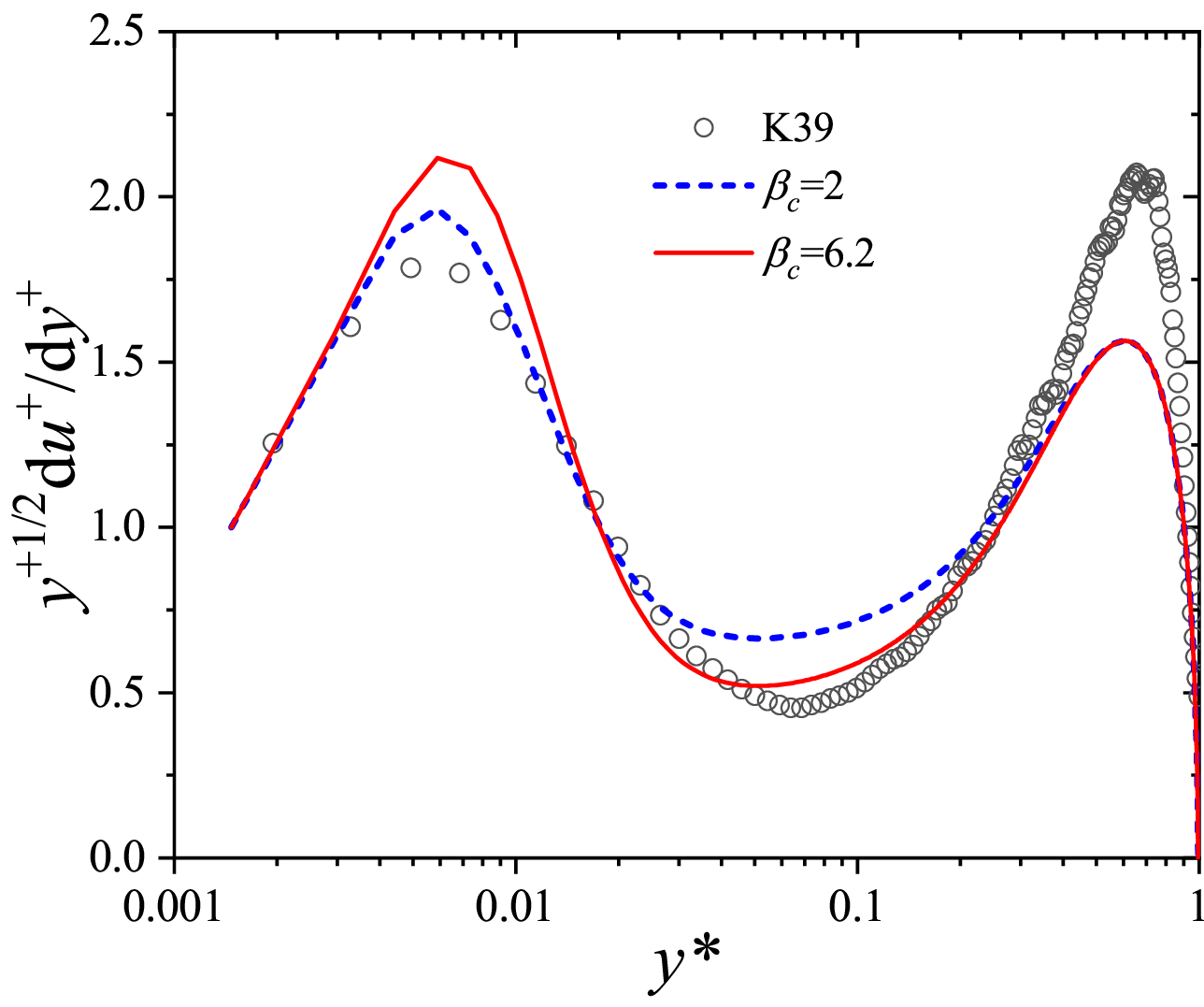}
    \\(c)\quad\quad\quad\quad\quad\quad\quad\quad\quad\quad\quad\quad\quad\quad\quad\quad\quad(d)\\
    \includegraphics[width=0.4\linewidth]{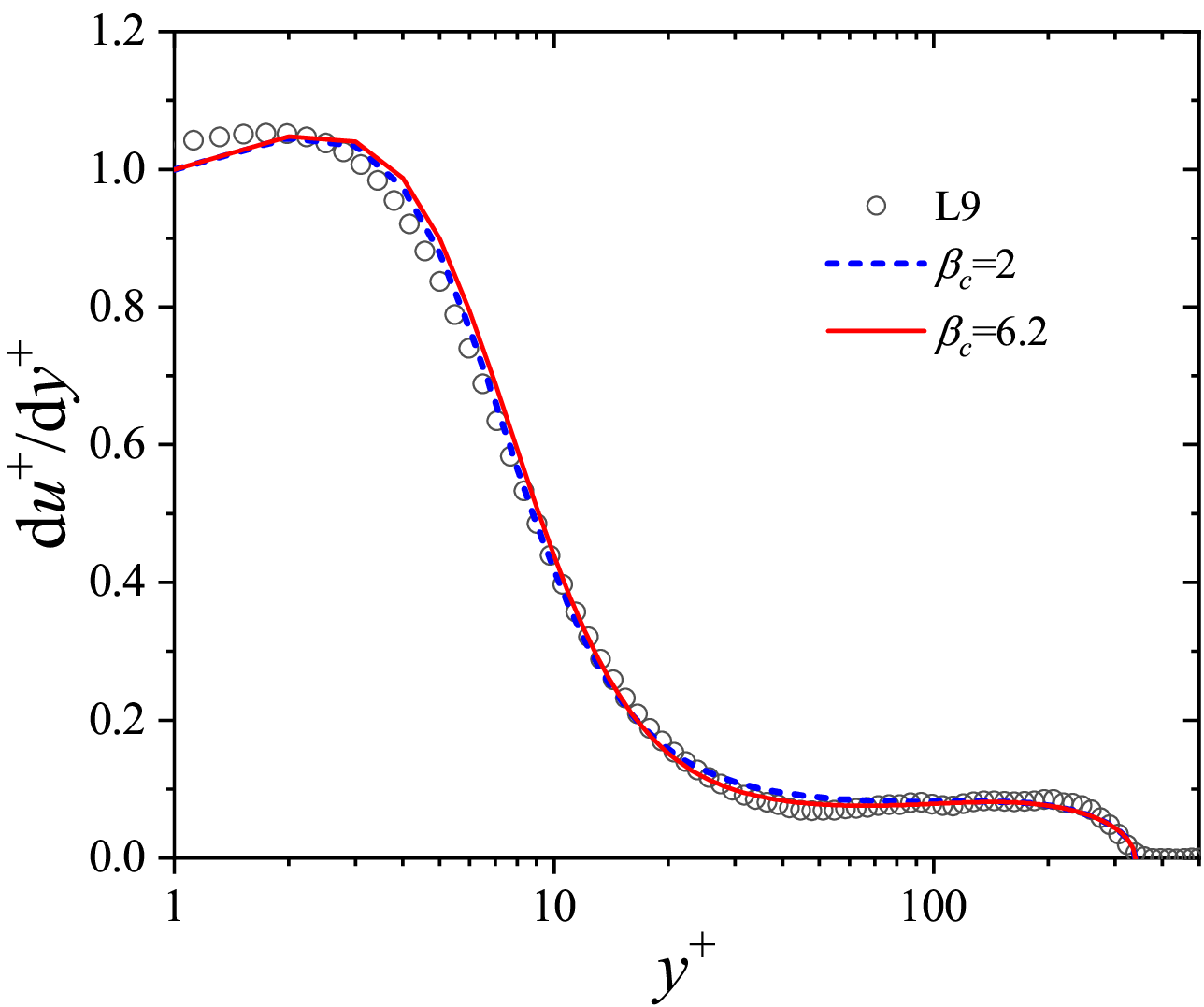}
    \includegraphics[width=0.4\linewidth]{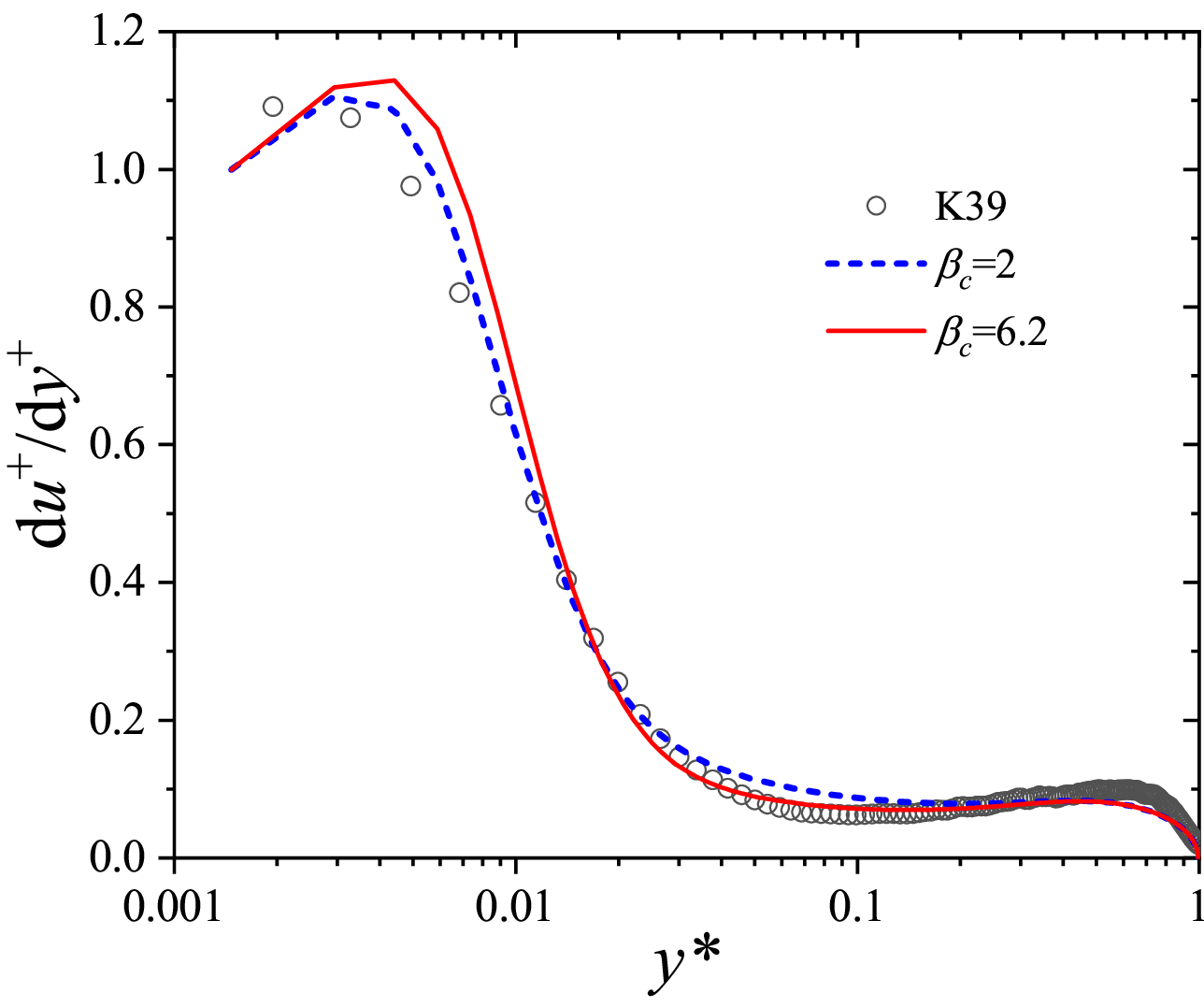}
    \\(e)\quad\quad\quad\quad\quad\quad\quad\quad\quad\quad\quad\quad\quad\quad\quad\quad\quad(f)\\
 \caption{Profiles of the diagnostic functions $y^+\frac{{\rm d}u^+}{{\rm d}y^+}$, $\sqrt{y^+}\frac{{\rm d}u^+}{{\rm d}y^+}$, and $\frac{{\rm d}u^+}{{\rm d}y^+}$ for cases L9 (panels (a), (c), and (e)) and K39 (panels (b), (d), and (f)). The solid lines denote the prediction of the current model employing $\beta_c=6.2$, and the short-dashed lines, $\beta_c=2$.}
 \label{fig:betac_valid}
\end{figure}

Figure~\ref{fig:betac_valid} presents the predictions of the present model with two representative values of $\beta_c$ (2 and 6.2) against the DNS data for cases L9 and K39. It is worth noting that the experimental case SK21.2, despite its higher Reynolds number which makes it an ideal candidate for validation, is not included in this analysis. The velocity gradient derived from experimental measurements inevitably contains significant noise and scatter, leading to excessive dispersion in the calculated diagnostic functions and preventing clear, reliable discrimination between predictions with different $\beta_c$ values.

As shown by the diagnostic functions in Fig.~\ref{fig:betac_valid}, the model with $\beta_c=2$ significantly underestimates the upper boundary of the logarithmic region, which is $y_{\mathrm{log}}^*=0.15\beta_c/\beta$ according to the current model. This leads to an artificially shortened $\mathrm{log}$-law region, with a premature transition from the $\mathrm{log}$ law to the half-power-law scaling, resulting in substantial deviations from the DNS data across the overlap region. In contrast, the model with $\beta_c=6.2$ yields predictions that are in good agreement with the DNS data for all three diagnostic functions: it accurately captures the length of the $\mathrm{log}$-law plateau for case L9, correctly reproduces the dominance of the half-power-law scaling for the extreme-APG case K39, and precisely matches the near-constant velocity gradient in the wake region for both cases. This quantitative comparison confirms that $\beta_c=6.2$ is the physically more reasonable value for the critical Clauser parameter, consistent with the asymptotic derivation in Section~\ref{subsubsec:half}.

\begin{figure}
    \centering
    \includegraphics[width=0.46\linewidth]{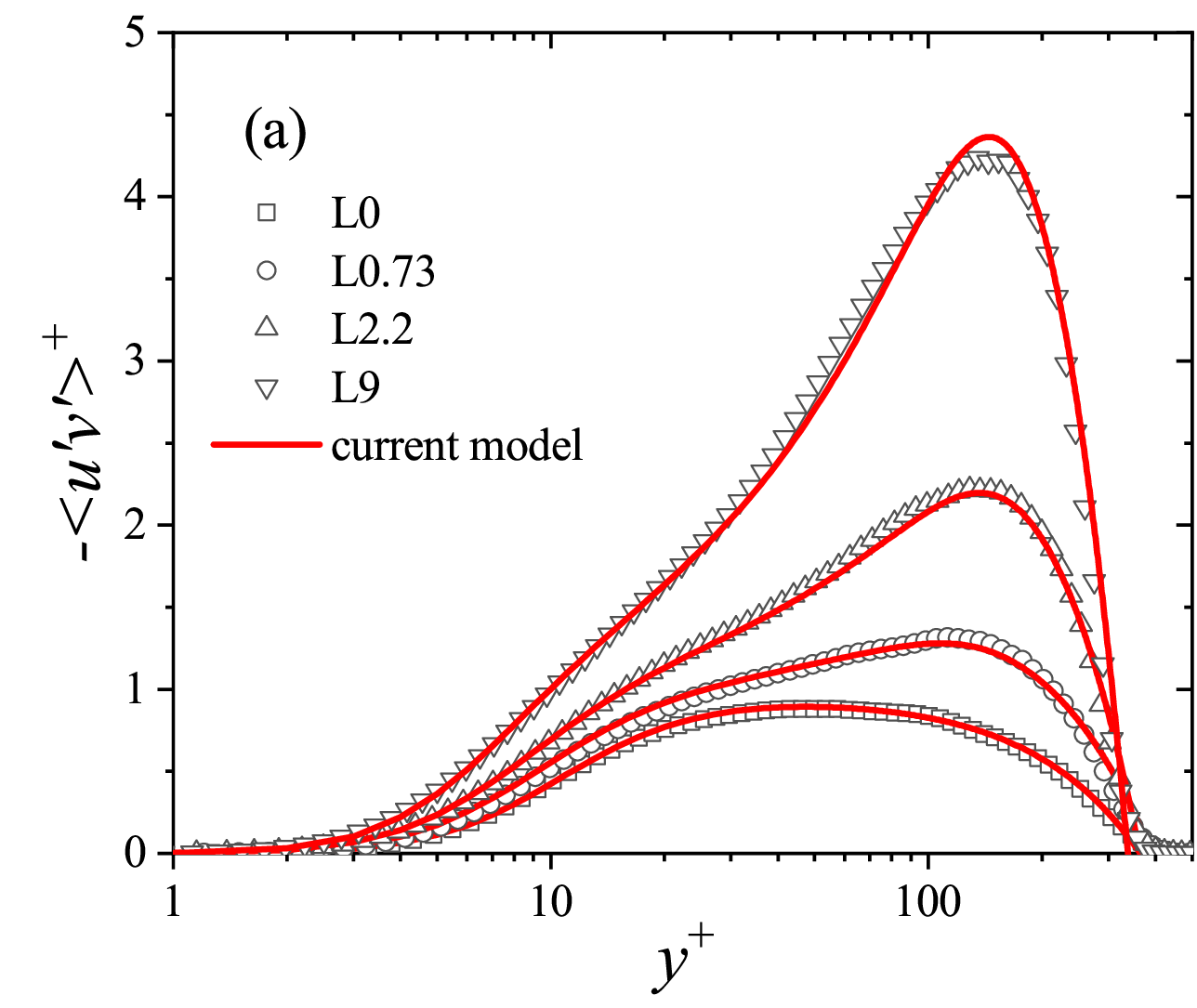}
    \includegraphics[width=0.49\linewidth]{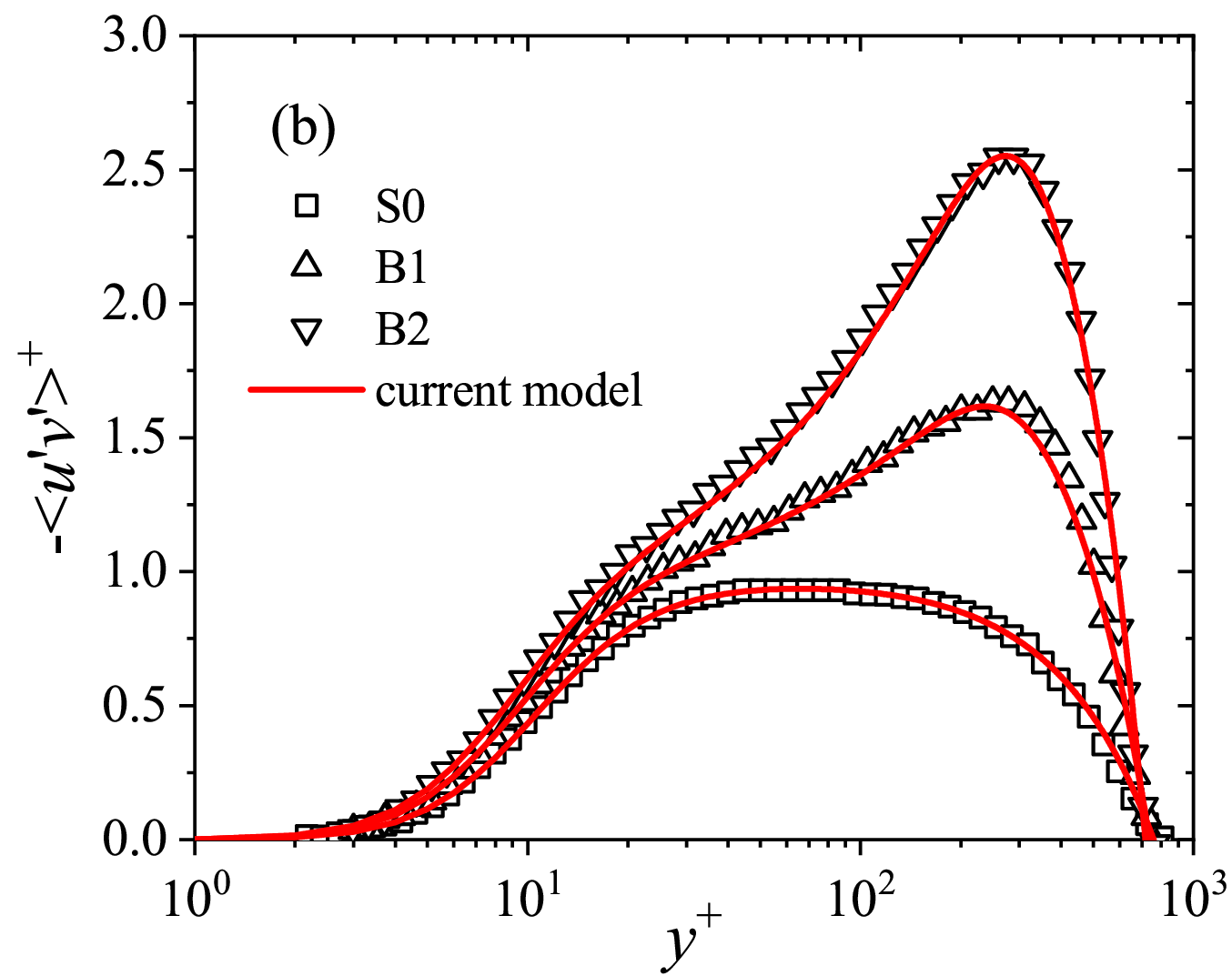}
     \includegraphics[width=0.49\linewidth]{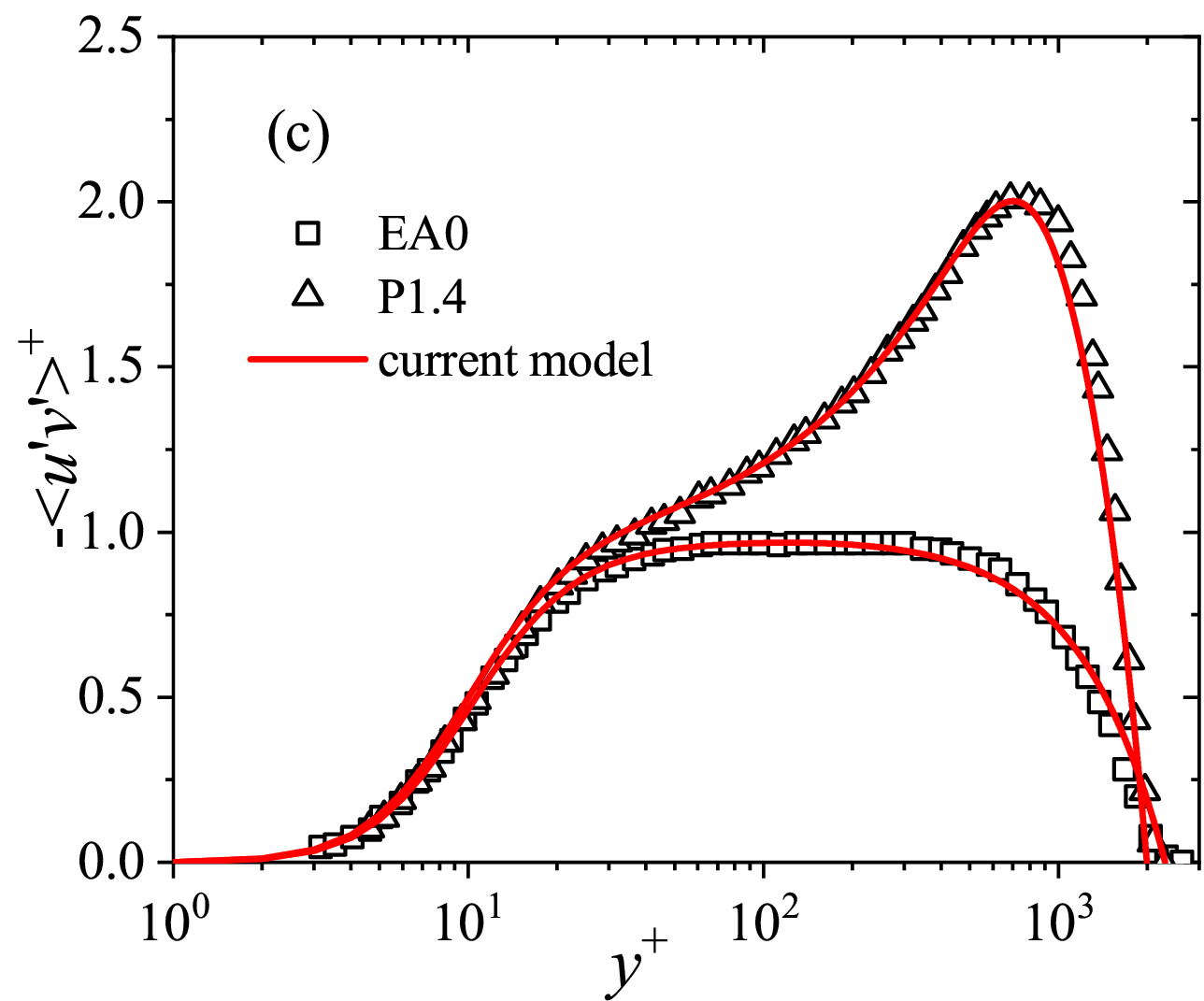}
    \includegraphics[width=0.49\linewidth]{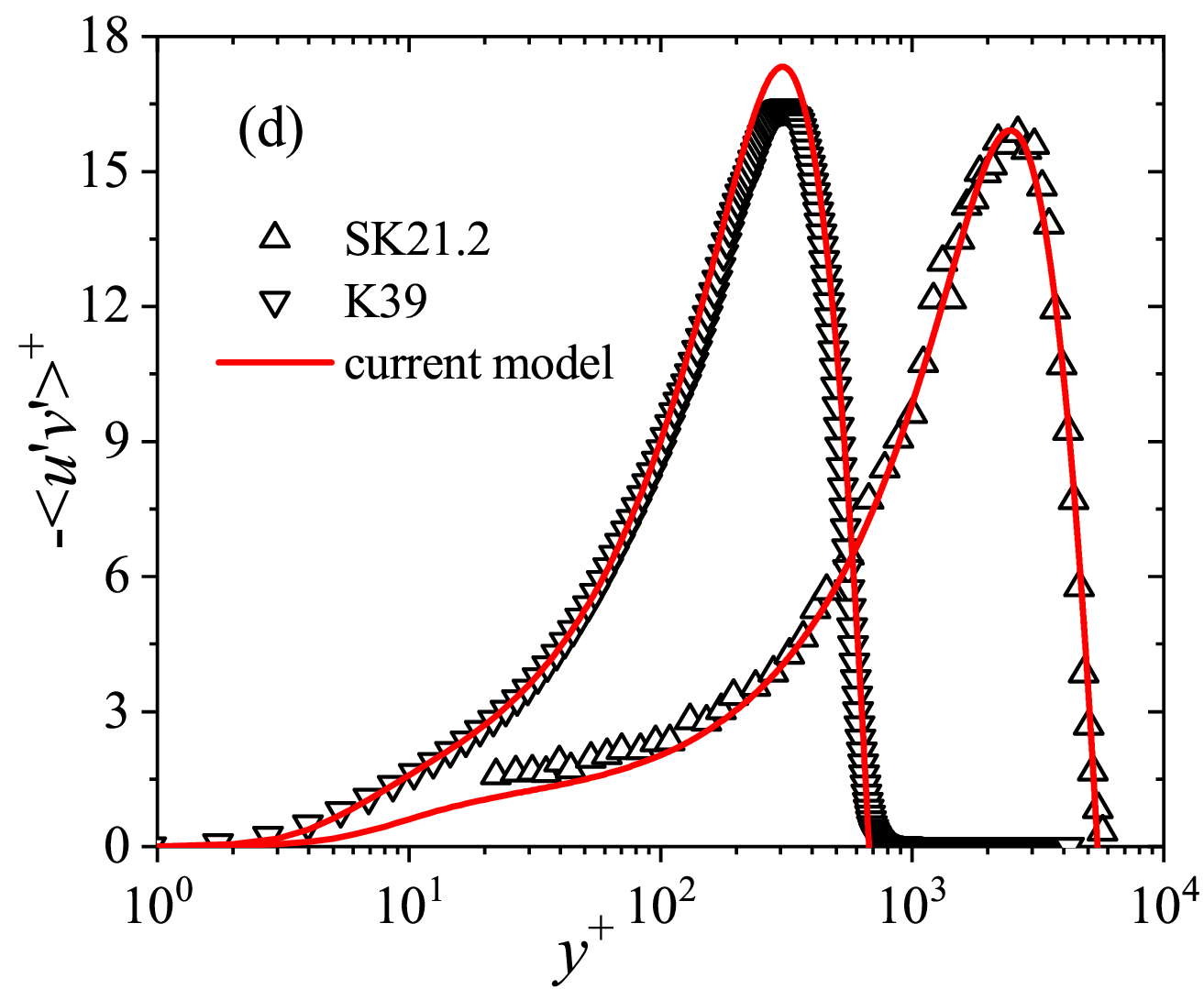}
  \caption{Profiles of Reynolds shear stress in equilibrium APG TBLs. Symbols represent data from the experiments and numerical simulations in Table \ref{tab:cases}. Solid lines denote predictions from the current model.}
  \label{fig:uv_valid}
\end{figure}

Finally, the Reynolds shear stress profiles predicted by the present model (calculated via $-\langle u'v' \rangle^+ = (\ell_m^+ \frac{\mathrm{d}u^+}{\mathrm{d}y^+})^2$) are validated against reference data in Fig.~\ref{fig:uv_valid}. For all test cases, the present model accurately captures the complete distribution of Reynolds shear stress: from zero at the wall, rising to a peak, then decaying back to zero at the boundary-layer edge. The model also precisely reproduces the key features induced by strong APG, including elevated peak shear stress and a shift of the peak location away from the wall as APG strengthens. The excellent agreement of the Reynolds shear stress predictions forms a closed validation loop with the mixing-length and mean-velocity profile results, fully demonstrating the self-consistency and accuracy of the present symmetry-based theoretical framework.

\section{Discussion}\label{sec:discuss}
\subsection{The symmetry-based approach}\label{subsec:discuss_symmetry}
The present work constructs a full-profile mixing-length formulation for APG TBLs using a symmetry-based framework, an approach that has been successfully validated for canonical wall-bounded turbulence within the SED theory \cite{she2017quantifying}. Unlike traditional asymptotic analysis or similarity analysis based on isolated flow features, this method postulates that the spatial evolution of key physical quantities (termed order functions in the SED, here the mixing length and TSS) obeys dilation symmetry --- the only allowable symmetry imposed by the solid wall constraint, which emerges as the macroscopic mean-field signature of stochastic turbulent fluctuations. The wall imposes spatially varying constraints on turbulent eddies at different wall-normal distances, manifested as shifts in the turbulent kinetic energy balance, which naturally gives rise to the multi-layer structure of TBLs \cite{chen2018quantifying}. The crossover scaling ansatz provides a rigorous analytical description of scaling transitions between adjacent layers, enabling the construction of a continuous full-profile formulation for the order function. In this work, we extend the SED framework from ZPG to APG flows via a coupled dual-order-function scheme, and show that APG acts on the TBL in two coupled ways: it deforms the existing multi-layer structure (e.g., modifying the viscous sublayer thickness $y_s^+$, buffer layer thickness $y_b^+$, and wake region extent), and induces new dilation-symmetry-breaking events, most notably the log-law breakdown and half-power-law strengthening above the critical Clauser parameter $\beta_c \approx 6.2$. Within this symmetry-based framework, order function modeling follows naturally from physical constraints, without ad hoc assumptions.

This approach, however, still has several open questions that warrant further investigation. First, its success relies on the intrinsic properties of the selected order function. While Townsend's attached eddy hypothesis provides a physical correspondence for the linear mixing-length scaling in ZPG TBLs, the direct link between coherent-structure evolution and mixing-length scaling shifts under APG conditions (e.g., the reduction of mixing-length parameter $\lambda$ and shrinkage of the log layer observed in this work) remains to be established. Second, despite the present framework containing only one measurable finite-Reynolds-number correction parameter (far fewer than empirical comparative models), the foundational parameters in the symmetry-based formulation (e.g., crossover steepness $\zeta=4$, log-law layer edge $0.15\delta$, critical Clauser PG parameter $\beta_c=6.2$, intrinsic K\'{a}rm\'{a}n constant $\kappa=0.45$, critical unstable Reynolds number $Re_c=12$) are currently determined via empirical constraints, with their first-principles physical roots not yet fully resolved. Verifying the postulated scaling laws also relies on high-fidelity, high-Reynolds-number data, which remains a long-standing challenge in wall turbulence research. Third, the extensibility of this approach to more complex flows (e.g., non-equilibrium TBLs and three-dimensional flows) remains unproven. In particular, flow-history effects and turbulence anisotropy pose fundamental challenges to the mixing-length description — which, with $\ell_m$ as the core modeling quantity, underpins the entire present framework — and its coupling with higher-order turbulence models.

\subsection{The log law of the wall and the K\'{a}rm\'{a}n constant}\label{subsec:discuss_loglaw}
The present framework is built on the hypothesis that the logarithmic law remains robust until its breakdown, and introduces the $\sqrt{\tau^+}$ correction to the mixing length in the logarithmic region accordingly. While this work identifies the critical Clauser parameter $\beta_c\approx6.2$ that governs the progressive breakdown of the log law under strong APG, it does not address the fundamental origin of the logarithmic law itself, which remains a long-standing open question in wall turbulence. However, the present full-profile mixing-length model opens a new avenue to assess the intrinsic K\'{a}rm\'{a}n constant from equilibrium PG TBLs (including canonical ZPG TBLs). Previously, investigating the K\'{a}rm\'{a}n constant via the mixing-length linear law was widely considered unreliable, as the apparent fitted $\kappa_\ell$ varies drastically with APG strength. In contrast, our framework enables the estimation of the global, intrinsic K\'{a}rm\'{a}n constant from the full mixing-length profile, independent of local fitting in the logarithmic region, providing a new path to interrogate the universality of the log law and the K\'{a}rm\'{a}n constant across a wider range of wall-bounded flows.

This work also has important implications for mixing-length scaling in internal flows, such as channel and pipe flows. Recent studies by Pirozzoli \cite{Pirozzoli2014} and Xu et al. \cite{XuYJ2025} suggested that the mixing length in the logarithmic region of channel flows should be corrected as $\ell_m^+=\kappa y^+\sqrt{1-y^+/\delta^+}$ to account for the wall-normal decay of TSS. Since the logarithmic region extends to $y^+/\delta^+\approx0.15$ in these flows, this correction has a non-negligible effect on the mixing-length profile. Adopting this correction means that the full-profile mixing-length formulation for internal flows in the SED theory needs to be revised in a manner similar to the present work, which may challenge the conclusion of a universal $\kappa=0.45$ across flow geometries. It is therefore necessary to systematically investigate the mixing-length scaling in different wall-bounded flows within the present symmetry-based framework, to further clarify the universality of the logarithmic law and the intrinsic K\'{a}rm\'{a}n constant.

\subsection{Joint effects of Reynolds number and pressure gradient}\label{subsec:discuss_RePG}
A key limitation of the present framework is that the TSS model contains a single parameter $\sigma_p$, which requires calibration via the measured peak TSS, and thus the model does not fully achieve a priori prediction independent of experimental/numerical data. Physically, $\sigma_p$ characterizes the underdevelopment of large-scale outer-layer eddies under finite-Reynolds-number conditions, which is also reflected in the deviation of the wake-region mixing-length parameter $\lambda$ from the high-Reynolds-number asymptotic scaling. This finite-Reynolds-number effect also propagates to the inner layer, as $\sigma_p$ influences the buffer layer thickness $y_b^+$ via the coupled TSS and mixing-length formulation (equation [\ref{eq:yb}]). Given the high accuracy of the TSS model based on $\sigma_p$, it is necessary to further establish the universal scaling of $\sigma_p$ with Reynolds number and APG strength.

For the inner-layer formulation, we adopt the universal inner-layer velocity scaling proposed by Nickels to determine the viscous sublayer thickness $y_s^+$ and buffer layer thickness $y_b^+$. While we validate the robustness of Nickels' inner scaling across a wide parameter range, its validity in a broader parameter space (e.g., extreme APG near separation, ultra-high Reynolds number conditions) remains unproven, which sets the current boundary of the model's universality. An interesting unresolved question is that Nickels' model postulates the inner velocity profile depends solely on $p^+$, while our results show that the inner-layer scaling of TSS and mixing length is also modulated by finite-Reynolds-number effects. The physical mechanism underlying this discrepancy warrants further investigation, along with the distribution of Reynolds normal stress components in the inner layer, which is closely linked to the near-wall turbulence dynamics that govern the inner-layer scaling.

\subsection{Extension to non-equilibrium flows}\label{subsec:discuss_extension}

As noted in Section \ref{subsec:discuss_symmetry}, extending the mixing-length closure framework to non-equilibrium TBLs remains a long-standing challenge. The present theory is developed exclusively for equilibrium APG TBLs, and is thus not directly applicable to general engineering flows dominated by strong non-equilibrium effects. However, recent work from our group has validated the extensibility of the present symmetry-based framework: we successfully extended the equilibrium TSS model to arbitrary two-dimensional non-equilibrium flows via the SED symmetry framework, accurately capturing the evolution of TSS in both weakly non-equilibrium flows and strongly non-equilibrium flows with an internal boundary layer developed \cite{ZhengBi2025}. In a separate study, we also established a symmetry-based description of TSS and mixing length for TBLs upstream and downstream of a two-dimensional separation bubble \cite{BiWT2024}. These works confirm the core SED hypothesis that new physical effects (here, non-equilibrium effects) induce deformation of the multi-layer boundary-layer structure and new dilation-symmetry-breaking events, demonstrating the feasibility of studying non-equilibrium boundary layers via the present symmetry-based approach.

\section{Conclusion}\label{sec:conclude}
This work develops a symmetry-based analytical model for the full-profile mixing-length scaling in equilibrium APG TBLs, extending the SED theory and coupling a two-layer TSS model to resolve long-standing debates on mixing-length scaling in APG TBLs.

The proposed framework unifies the viscous sublayer, buffer layer, logarithmic/half-power-law transition zone, and wake region with a globally invariant K\'{a}rm\'{a}n constant $\kappa=0.45$. A core finding is the identification of a critical Clauser parameter $\beta_c=6.2$, which governs the progressive breakdown of the logarithmic law: for $\beta < \beta_c$, the logarithmic layer retains its canonical upper bound at $0.15\delta$, while for $\beta > \beta_c$, the logarithmic region shrinks rapidly and transitions to the half-power-law scaling, with a smooth asymptotic limit to Stratford's square-root law at zero skin friction. This result clarifies the physical mechanism underlying the log-law-to-half-power-law transition under strong APG, reconciling conflicting observations in the existing literature.

We further derive an analytical formulation for the wake-region mixing-length parameter $\lambda$, which decreases monotonically from the ZPG value $\lambda_0=\kappa/4=0.1125$ to an asymptotic $\lambda_\infty=0.074$ in the strong APG limit, in excellent agreement with numerical and experimental data across a wide range of $\beta$. For the near-wall region, the thicknesses of the viscous sublayer ($y_s^+$) and buffer layer ($y_b^+$) are determined self-consistently via Nickels' universal inner scaling for PG TBLs, with no ad hoc fitting parameters, and the model accurately captures the compression of the inner layer with intensifying APG.

With only one finite-Reynolds-number correction parameter determined directly from the maximum TSS, the model achieves exceptional agreement with a comprehensive database of DNS, LES, and experimental measurements spanning $Re_\theta=1250\sim50980$ and $\beta=0\sim39$, accurately predicting full profiles of mixing length, mean streamwise velocity, and Reynolds shear stress. Unlike existing semi-empirical models that rely on flow-dependent tuning parameters, the present symmetry-based framework maintains robust predictive performance across the full parameter space, especially for high-Reynolds-number, strong-APG flows near incipient separation where existing models often fail.

This work provides a physically consistent, unified framework for mixing-length scaling in equilibrium APG TBLs, resolving the century-long challenge of extending Prandtl's mixing-length hypothesis to nonzero APG conditions. By distinguishing the global intrinsic K\'{a}rm\'{a}n constant from the apparent locally fitted value in the logarithmic region, the model reconciles the observed variation of the apparent K\'{a}rm\'{a}n constant with APG strength with the postulated universality of the intrinsic constant, offering a new path to assess the invariance of the logarithmic law across wall-bounded flows. The model also delivers a physics-based improvement to mixing-length closures in RANS turbulence models, with significant engineering value for aerodynamic and industrial flow applications involving PG TBLs. Preliminary work demonstrates the extensibility of the symmetry-based approach to non-equilibrium flows, which will be systematically explored in future studies to expand the model's applicability to more complex flow configurations.

\begin{acknowledgments}
\noindent We are grateful to Prof. J.-H. Lee, Prof. J. Soria, Dr. V. Kitsios, Dr. R. Vinuesa, and Dr. A. Bobke for sharing their numerical simulation data.
\end{acknowledgments}


\appendix
\section{Nickels' mean-velocity profile}\label{Append_Nickels}
\noindent Nickels \cite{Nickels2004} proposed the following model for the mean velocity profiles in PG TBLs:
\begin{align}
  u^+ =&y_c^+\left[1-\left(1+2(y^+/y_c^+)+\frac{1}{2}(3-p^+y_c^+)(y^+/y_c^+)^2-\frac{3}{2}p^+y_c^+(y^+/y_c^+)^3\right){\rm e}^{-3y^+/y_c^+}\right]\nonumber\\
&+\frac{\sqrt{1+p^+y_c^+}}{6\kappa_o}{\rm ln}\left(\frac{1+(0.6y^+/y_c^+)^6}{1+{y^*}^6}\right)+c\left(1-{\rm e}^{-\frac{5({y^*}^4+{y^*}^8)}{1+5{y^*}^3}}\right),
  \label{eq:Nickels}
\end{align}
where $y_c^+$ is found from equation (\ref{eq:yc}), $\kappa_o=0.39$, $y^*=y^+/\delta^+$, and $c$ is a measure of wake strength and is determined by fitting the full functional form to the data. Nickels stated that equation (\ref{eq:Nickels}) gives a good description of wall-bounded flows in the approximate range of $-0.02<p^+<0.06$. At $y^+=4y_c^+$, assuming that the Reynolds number is sufficiently large (i.e., $\delta^+\gg4y_c^+$), Nickels' model predicts that 
\begin{equation}
    u^+(4y_c^+)\approx y_c^++\frac{{\rm ln}2.4}{0.39}\sqrt{1+p^+y_c^+}.
    \label{eq:Nickels_4yc}
\end{equation}

\section{Ma et al.'s mixing-length model}\label{Append_Ma}
Based on the five-parameter formulation by Cantwell \cite{Cantwell2019}, Ma et al. \cite{MaChen2026} recently proposed a full-profile mixing-length model for APG TBLs, as follows
\begin{equation}
    \ell_m^+= \frac{\kappa y^+f(Y^+)}{[f(0.1\delta^+)]^{-1}+0.55\sqrt{\frac{p_0^+}{0.1\delta^+}}Y^+}\frac{1-{\rm e}^{-(y^+/a)^{1.1}}}{\left[1+(\frac{y^+}{b\delta^+})^n\right]^{1/n}},
    \label{eq:Ma_ellm}
\end{equation}
where $\kappa=0.41$, $a=25/\sqrt{1+p^+y_c^+}$, $p_0^+=p^+/(1+10{\rm e}^{-200(p^+-0.01)})$, and $b$ and $n$ are two flow-dependent parameters and are determined by fitting the full functional form to the data. $Y^+$ is defined as
\begin{equation}
    Y^+= y^+-\frac{0.1\delta^+}{10}{\rm ln}(1+{\rm e}^{10(\frac{y^+}{0.1\delta^+}-1)})+\frac{0.1\delta^+}{10}{\rm ln}(1+{\rm e}^{-10}).
    \label{eq:Ma_Y}
\end{equation}
The square of function $f$ is an inner-layer model for TSS $\tau^+$ \cite{MaYan2024}, expressed as
\begin{equation}
    [f(y^+)]^2=1+p^+y^+-A_p(y^+{\rm ln}y^+-y^+),
    \label{eq:Ma_f}
\end{equation}
where $A_p=\alpha y^+/{\rm ln}(0.1\delta^+)$, and $\alpha=\left(\frac{\partial \tau^+}{\partial y^+}|_w-\frac{\partial \tau^+}{\partial y^+}|_{0.1\delta}\right)/\frac{\partial \tau^+}{\partial y^+}|_w$.

\bibliography{MVPinPGTBL}

@article{JoTreview2024,
  title={On the physical structure, modelling and computation-based prediction of two-dimensional, smooth-wall turbulent boundary layers subjected to streamwise pressure gradients},
  author={Klewicki, J. and Sandberg, R. and Knopp, T. and Devenport, W. and Fritsch, D. and Vishwanathan, V. and Volino, R. and Toxopeus, S. and MeKeon, B. and Eca, L.},
  journal={J. Turbul.},
  volume={25},
  number={10--11},
  pages={345--368},
  year={2024},
  publisher={}
}

@article{Bradshaw1974,
  title={{Possible origin of Prandtl's mixing-length theory}},
  author={Bradshaw, P.},
  journal={Nature},
  volume={249},
  number={},
  pages={135--136},
  year={1974},
  publisher={}
}

@article{Marusic2013Rapid,
  title={On the logarithmic region in wall turbulence},
  author={Marusic, I. and Monty, J. P. and Hultmark, M.},
  journal={J. Fluid Mech.},
  volume={716},
  number={},
  pages={R3},
  year={2013},
  publisher={Cambridge University Press}
}

@article{Heisel2020,
  title={On the mixing length eddies and logarithmic mean velocity profile in wall turbulence},
  author={Heisel, M. and de Silva, C. M. and Hutchins, N. and Marusic, I. and Guala, M.},
  journal={J. Fluid Mech.},
  volume={887},
  number={},
  pages={R1},
  year={2020},
  publisher={Cambridge University Press}
}

@article{Knopp2021,
  title={Experimental analysis of the log law at adverse pressure gradient},
  author={Knopp, T. and Reuther, N. and Novara, M. and Schanz, D. and Schulein, E. and Schroder, A. and Kahler, C. J.},
  journal={J. Fluid Mech.},
  volume={918},
  number={},
  pages={A17},
  year={2021},
  publisher={Cambridge University Press}
}

@article{Cantwell2019,
  title={A universal velocity profile for smooth wall pipe flow},
  author={Cantwell, B. J.},
  journal={J. Fluid Mech.},
  volume={878},
  number={},
  pages={834--874},
  year={2019},
  publisher={Cambridge University Press}
}

@article{Cantwell2022,
  title={A universal velocity profile for turbulent wall flows including adverse pressure gradient boundary layers},
  author={Subrahmanyam, M. A. and Cantwell, B. J. and Alonso, J. J.},
  journal={J. Fluid Mech.},
  volume={933},
  number={},
  pages={A16},
  year={2022},
  publisher={Cambridge University Press}
}

@article{Coleman2017,
  title={Direct Numerical Simulation and Theory of a Wall-Bounded Flow with Zero Skin Friction},
  author={Coleman, G. N. and Pirozzoli, S. and Quadrio, M. and Spalart, P. R.},
  journal={Flow Turbul. Combust.},
  volume={99},
  number={},
  pages={553--564},
  year={2017},
  publisher={}
}

@article{Kitsios2017,
  title={Direct numerical simulation of a self-similar adverse pressure gradient turbulent boundary layer at the verge of separation},
  author={Kitsios, V. and Sekimoto, A. and Atkinson, C. and Sillero, J. A. and Borrell, G. and Gungor, A. G. and Jimenez, J. and Soria, J.},
  journal={J. Fluid Mech.},
  volume={829},
  number={2},
  pages={392--419},
  year={2017},
  publisher={Cambridge University Press}
}

@Article{Sanmiguel2020exp,
  title={Experimental realisation of near-equilibrium adverse-pressure-gradient turbulent boundary layers},
  author={Sanmiguel Vila, C. and Vinuesa, R. and Discetti, S. and Ianiro, A. and Schlatter, P. and {\"O}rl{\"u}, R.},
  journal={Exp. Thermal Fluid Sci.},
  volume={112},
  pages={109975},
  year={2020},
  publisher={}
}

@article{Eitel-Amor2014,
  title={{Simulation and validation of a spatially evolving turbulent boundary layer up to $Re_\theta=8300$}},
  author={Eitel-Amor, G. and {\"O}rl{\"u}, R. and Schlatter, P.},
  journal={Int. J. Heat Fluid Flow},
  volume={47},
  pages={57-69},
  year={2014}
}

@article{Pozuelo2022les,
  title={{An adverse-pressure-gradient turbulent boundary layer with nearly constant $\beta\approx1.4$ up to $Re_\theta\approx8700$}},
  author={Pozuelo, R. and Li, Q. and Schlatter, P. and Vinuesa, R.},
  journal={J. Fluid Mech.},
  volume={939},
  pages={A34},
  year={2022},
  publisher={Cambridge University Press}
}

@article{Schlatter2009dns,
  title={{Turbulent boundary layers up to $Re_\theta=2500$ studied through simulation and experiment}},
  author={Schlatter, P. and {\"O}rl{\"u}, R. and LI, Q. and Brethouwer, G. and Fransson, J. H. M. and Johansson, A. V. and Alfredsson, P. H. and Henningson, D. S.},
  journal={Phys. Fluids},
  volume={21},
  pages={051702},
  year={2009}
}

@article{MaPoF2024-2,
  title={Scaling laws for the mean velocity in adverse pressure gradient turbulent boundary layers based on asymptotic expansions},
  author={Ma, M. and Han, M. and Shao, Z. and Yan, C.},
  journal={Phys. Fluids},
  volume={36},
  pages={095175},
  year={2024}
}

@article{MaYan2024,
  title={Asymptotic expansions and scaling of turbulent boundary layers in adverse pressure gradients},
  author={Ma, M. and Bai, R. and Song, H. and Zhang, J. and Yan, C.},
  journal={Phys. Fluids},
  volume={36},
  pages={065139},
  year={2024}
}

@article{Sethna2022,
Author = {Sethna, J. P.},
Title = {Power laws in physics},
Journal = {Nature Rev. Phys.},
Year = {2022},
Volume = {4},
Number = {8},
Pages = {501-503},
DOI = {10.1038/s42254-022-00491-x},
}

@article{Gluzman1998,
Author = {Gluzman, S. and Yukalov, V. I.},
Title = {Unified approach to crossover phenomena},
Journal = {Phys. Rev. E},
Year = {1998},
Volume = {58},
Number = {4},
pages = {4197--4209},
}

@article{Reeves1974,
  title={Two-Layer Model of Turbulent Boundary Layers},
  author={Reeves, B. L.},
  journal={AIAA J.},
  volume={12},
  number={7},
  pages={932--939},
  year={1974}
}

@article{Knopp2022,
  title={{An empirical wall law for the mean velocity in an adverse pressure gradient for RANS turbulence modeling}},
  author={Knopp, T.},
  journal={Flow Turbul. Combust.},
  volume={109},
  number={},
  pages={571--601},
  year={2022},
  publisher={Springer}
}

@article{XuYJ2025,
  title={Extending the logarithmic velocity profile in turbulent channel flow},
  author={Xu, Y. J. and Schmidt, S. J. and Adams, N. A.},
  journal={Phys. Fluids},
  volume={37},
  pages={045109},
  year={2025},
  publisher={}
}

@article{BiWT2024,
  title={Mean-flow structures of the turbulent boundary layers bounding a two-dimensional separation bubble},
  author={Bi, W. T. and Du, T. T. and Chen, J. and She, Z. S.},
  journal={Phys. Fluids},
  volume={36},
  pages={085197},
  year={2024},
  publisher={}
}

@article{ZhengBi2025,
  title={A symmetry-based model for total shear stress of equilibrium pressure-gradient turbulent boundary layers},
  author={Zheng, K. X. and Bi, W. T. and Chen, J. and She, Z. S.},
  journal={Phys. Fluids},
  volume={37},
  pages={055126},
  year={2025},
  publisher={}
}

@article{Bi2025,
  title={Quantifying non-equilibrium pressure-gradient turbulent boundary layers through a symmetry-based framework},
  author={Bi, W. T. and Zheng, K. X. and Chen, J. and She, Z. S.},
  journal={AIP Adv.},
  volume={15},
  pages={095112},
  year={2025},
  publisher={}
}

@article{MaChen2026,
  title={A Generalized Mixing Length Model with Adverse-Pressure-Gradient Effects},
  author={Ma, M. Z. and Shi, Y. X. and Zhu, Y. L. and Han, A. X. and Chen, X},
  journal={Symmetry},
  volume={18},
  pages={105},
  year={2026},
  publisher={}
}

@article{LiRong2025,
  title={Scaling crossovers in non-equilibrium critical dynamics},
  author={Li, R. and Ding, Q. R. and Cui, W. C.},
  journal={I. Phys. A: Math. Theor.},
  volume={58},
  pages={385004},
  year={2025},
  publisher={}
}

@article{Pirozzoli2014,
  title={Revisiting the mixing-length hypothesis in the outer part of turbulent wall layers: mean flow and wall friction},
  author={Pirozzoli, S.},
  journal={J. Fluid Mech.},
  volume={745},
  pages={378--397},
  year={2014},
  publisher={Cambridge University Press}
}

@article{Granville1989,
  title={{A modified van Driest formula for the mixing length of turbulent boundary layers in pressure gradients}},
  author={Granville, P. S.},
  journal={J. Fluids Engn. Trans. ASME},
  volume={111},
  pages={94--97},
  year={1989},
  publisher={}
}

@article{Thomas1989,
  title={Supplementary boundary-layer approximations for turbulent flow},
  author={Thomas, L. C. and Hasani, S. M. F.},
  journal={J. Fluids Engn. Trans. ASME},
  volume={111},
  pages={420--427},
  year={1989},
  publisher={}
}

@article{Lee2008,
  title={Effects of an adverse pressure gradient on a turbulent boundary layer},
  author={Lee, J. H. and Sung, H. J.},
  journal={Int. J. Heat Fluid Flow},
  volume={29},
  number={3},
  pages={568--578},
  year={2008},
  publisher={Elsevier}
}

@article{Bobke2017les,
  title={History effects and near equilibrium in adverse-pressure-gradient turbulent boundary layers},
  author={Bobke, A. and Vinuesa, R. and {\"O}rl{\"u}, R. and Schlatter, P.},
  journal={J. Fluid Mech.},
  volume={820},
  number={},
  pages={667--692},
  year={2017},
  publisher={Cambridge University Press}
}

@article{Bernard2003,
  title={Decelerating Boundary Layer: A New Scaling and Mixing Length Model},
  author={Bernard, A. and Foucaut, J. M. and Dupont, P. and Stanislas, M.},
  journal={AIAA J.},
  volume={41},
  number={2},
  pages={248--255},
  year={2003}
}

@article{Galbraith1977,
Author = {Galbraith, R. A. and Sjolander, S. and Head, M. R.},
Title = {Mixing Length in the Wall Region of Turbulent Boundary Layers},
Journal = {Aeronaut. Q.},
Year = {1977},
Volume = {28},
Number = {},
DOI = {},
pages = {97--110},
}

@book{Voyage,
  title={A Voyage Through Turbulence},
  author={Davidson, P. A. and Kaneda, Y. and Moffatt, K. and Sreenivasan, K. R.},
  volume={},
  year={2011},
  publisher={Cambridge University Press}
}

@book{Wilcox2006,
  title={Turbulence modeling for CFD, 3rd ed},
  author={Wilcox, D. C.},
  volume={},
  year={2006},
  publisher={DCW Industries, California}
}

@article{Buschmann2005,
  title={New Mixing-Length Approach for the Mean Velocity Profile of Turbulent Boundary Layers},
  author={Buschmann, M. H. and Gad-el-Hak, M.},
  journal={J. Fluids Engn. Trans. ASME},
  volume={127},
  pages={393--396},
  year={2005},
  publisher={Elsevier}
}

@incollection{prandtl1932,
  author    = {Prandtl, L.}, 
  title     = {Zur turbulenten Str\"omung in Rohren und l\"angs Platten}, 
  booktitle = {Ergebnisse der Aerodynamischen Versuchsanstalt zu G\"ottingen},
  volume    = {4},
  pages     = {18--29},
  year      = {1932},
  note      = {Also published in LPGA 2, 632--648}
}

@article{Skare1994exp,
  title={A turbulent equilibrium boundary layer near separation},
  author={Skare, P. E. and Krogstad, P. A.},
  journal={J. Fluid Mech.},
  volume={272},
  pages={319--348},
  year={1994},
  publisher={Cambridge University Press}
}

@Article{Lee2017direct,
  title={Large-scale motions in turbulent boundary layers subjected to adverse pressure gradients},
  author={Lee, J. H.},
  journal={J. Fluid Mech.},
  volume={810},
  pages={323--361},
  year={2017},
  publisher={Cambridge University Press}
}

@article{prandtl1925ausgebildete,
  title={Bericht {\"u}ber Untersuchungen zur ausgebildeten Turbulenz},
  author={Prandtl, L.},
  journal={Z. Angew. Math. Mech.},
  volume={5},
  pages={136--139},
  year={1925}
}

@article{Karman1948,
  title={Progress in the Statistical Theory of Turbulence},
  author={von Karman, T.},
  journal={PANS},
  volume={34},
  number={11},
  pages={530--539},
  year={1948}
}

@TechReport{Klebanoff1954,
   author = {Klebanoff, P. S.},
   title = {Characteristics of turbulence in a boundary layer with zero pressure gradient},
   institution = {NACA},
   year = {1954},
   number = {TN 3178}
}

@TechReport{Bradshaw1965,
   author = {Bradshaw, P. and Ferriss, D. H.},
   title = {The response of a retarded equilibrium turbulent boundary layer to the sudden removal of pressure gradient},
   institution = {NPL Aero. Rep.},
   year = {1965},
   number = {1145}
}

@article{clauser1954turbulent,
  title={Turbulent boundary layers in adverse pressure gradients},
  author={Clauser, F. H.},
  journal={J. Aeronaut. Sci.},
  volume={21},
  number={2},
  pages={91--108},
  year={1954}
}

@article{van1956turbulent,
  title={On turbulent flow near a wall},
  author={Van Driest, E. R.},
  journal={J. Aeronaut. Sci.},
  volume={23},
  number={11},
  pages={1007--1011},
  year={1956}
}

@article{Nickels2004,
  title={Inner scaling for wall-bounded flows subject to large pressure gradients},
  author={Nickels, T. B.},
  journal={J. Fluid Mech.},
  volume={521},
  number={},
  pages={217--239},
  year={2004},
  publisher={Cambridge University Press}
}

@article{resilience,
  title={The resilience of the logarithmic law to pressure gradients: evidence from direct numerical simulation},
  author={Johnstone, R. and Coleman, G. N. and Spalart, P. R.},
  journal={J. Fluid Mech.},
  volume={643},
  number={},
  pages={163--175},
  year={2010},
  publisher={Cambridge University Press}
}

@inproceedings{Das1987,
  title={A Numerical Study of Turbulent Separated Flows},
  author={Das, DK},
  booktitle={Am. Soc. Mech. Eng. Forum on Turbulent Flows, FED Vol. 51},
  pages={85--90},
  year={1987},
  organization={}
}

@inproceedings{ColesHirst1968,
  title={Computation of Turbulent Boundary Layers},
  author={Coles, DE and Hirst, EA},
  booktitle={Proc. 1968 AFOSR-IFP Stanford Conf., vol. 2},
  pages={},
  year={1968},
  organization={Stanford Univ., Stanford, Calif.}
}

@article{coles1956law,
  title={The law of the wake in the turbulent boundary layer},
  author={Coles, D.},
  journal={J. Fluid Mech.},
  volume={1},
  number={2},
  pages={191--226},
  year={1956},
  publisher={Cambridge University Press}
}

@Article{stratford1959prediction,
  title={The prediction of separation of the turbulent boundary layer},
  author={Stratford, B. S.},
  journal={J. Fluid Mech.},
  volume={5},
  number={1},
  pages={1--16},
  year={1959},
  publisher={Cambridge University Press}
}

@Article{townsend1961equilibrium,
  title={Equilibrium layers and wall turbulence},
  author={Townsend, A. A.},
  journal={J. Fluid Mech.},
  volume={11},
  number={1},
  pages={97--120},
  year={1961},
  publisher={Cambridge University Press}
}

@article{bradshaw1967turbulence,
  title={The turbulence structure of equilibrium boundary layers},
  author={Bradshaw, P.},
  journal={J. Fluid Mech.},
  volume={29},
  number={4},
  pages={625--645},
  year={1967},
  publisher={Cambridge University Press}
}

@Article{skote2002direct,
  title={Direct numerical simulation of a separated turbulent boundary layer},
  author={Skote, M. and Henningson, D. S.},
  journal={J. Fluid Mech.},
  volume={471},
  pages={107--136},
  year={2002},
  publisher={Cambridge University Press}
}

@article{she2010new,
  title={New perspective in statistical modeling of wall-bounded turbulence},
  author={She, Z. S. and Chen, X. and Wu, Y. and Hussain, F.},
  journal={Acta Mech. Sinica},
  volume={26},
  number={6},
  pages={847--861},
  year={2010},
  publisher={Springer}
}

@article{Perry1973,
  title={Mean velocity and shear stress distributions in turbulent boundary layers},
  author={Perry, A. E. and Schofield, W. H.},
  journal={Phys. Fluids},
  volume={16},
  number={},
  pages={2068--2074},
  year={1973},
  publisher={AIP Publishing LLC}
}

@article{chen2016analytic,
  title={Analytic prediction for planar turbulent boundary layers},
  author={Chen, X. and She, Z. S.},
  journal={Sci. China Phys., Mech. \& Astron.},
  volume={59},
  pages={114711},
  year={2016},
  publisher={Springer}
}

@article{chen2018quantifying,
  title={{Quantifying wall turbulence via a symmetry approach. Part 2. Reynolds stresses}},
  author={Chen, X. and Hussain, F. and She, Z. S.},
  journal={J. Fluid Mech.},
  volume={850},
  pages={401--438},
  year={2018},
  publisher={Cambridge University Press}
}

@article{she2017quantifying,
  title={{Quantifying wall turbulence via a symmetry approach: A Lie group theory}},
  author={She, Z. S. and Chen, X. and Hussain, F.},
  journal={J. Fluid Mech.},
  volume={827},
  pages={322--356},
  year={2017},
  publisher={Cambridge University Press}
}

@article{devenport2022equilibrium,
  title={Equilibrium and non-equilibrium turbulent boundary layers},
  author={Devenport, W. J. and Lowe, K. T.},
  journal={Prog. Aerosp. Sci.},
  volume={131},
  pages={100807},
  year={2022},
  publisher={Elsevier}
}

@article{ShuXu2025, 
title={Mean velocity profiles and total shear stress profiles in adverse-pressure-gradient turbulent boundary layers considering history effect}, 
volume={}, 
 journal={arXiv}, 
 author={Shu, Z. Q. and Xu, C. X.}, 
 year={2025}, 
 pages={2508.07663v2}}

@article{MarusicPoF2010,
Author = {Marusic, I. and McKeon, B. J. and Monkewitz, P. A. and Nagib, H. M. and Smits, A. J. and Sreenivasan, K. R.},
Title = {{Wall-bounded turbulent flows at high Reynolds numbers: Recent advances and key issues}},
Journal = {Phys. Fluids},
Year = {2010},
Volume = {22},
Number = {6},
DOI = {10.1063/1.3453711},
pages = {065103},
}

@book{White2005,
  title={Viscous Fluid Flow, 3rd ed},
  author={White, F. M.},
  volume={},
  year={2006},
  publisher={McGraw-Hill}
}

@book{Schlichting,
  title={Boundary Layer Theory, 9th ed},
  author={Schlichting, H. and Gersten, K.},
  volume={},
  year={2017},
  publisher={Springer}
}

\end{document}